\documentclass[12pt,thmsa]{article}
\usepackage{amsfonts}
\usepackage{setspace}
\usepackage{graphicx}
\usepackage{subfig}
\usepackage{supertabular}
\usepackage{amssymb}
\usepackage{amsmath}
\usepackage[margin=1.5in]{geometry}
\usepackage{latexsym}
\usepackage{paralist}
\usepackage{appendix}
\usepackage{amsthm}
\usepackage{color}
\usepackage{url}
\usepackage{hyperref}
\usepackage{upgreek}

\newtheorem{Assumption}{Assumption}
\newtheorem{lemma}{Lemma}
\newtheorem{proposition}{Proposition}
\newtheorem{theorem}{Theorem}

\input{tcilatex}
\newtheorem{innercustomgeneric}{\customgenericname}
\providecommand{\customgenericname}{}
\newcommand{\newcustomtheorem}[2]{  \newenvironment{#1}[1]
  {   \renewcommand\customgenericname{#2}   \renewcommand\theinnercustomgeneric{##1}   \innercustomgeneric
  }
  {\endinnercustomgeneric}
}
\newcustomtheorem{customassumption}{Assumption}

\begin{document}

\title{Doubly Robust Identification of Causal Effects of a Continuous
Treatment using \\
Discrete Instruments}
\date{{\small February 2024}}
\author{Yingying Dong and Ying-Ying Lee{\small \thanks{%
Yingying Dong and Ying-Ying Lee, Department of Economics, University of
California Irvine, yyd@uci.edu and yingying.lee@uci.edu.}}}
\maketitle

\begin{abstract}
Many empirical applications estimate causal effects of a continuous
endogenous variable (treatment) using a binary instrument. Estimation is
typically done through linear 2SLS. This approach requires a mean treatment
change and causal interpretation requires the LATE-type monotonicity in the
first stage. An alternative approach is to explore distributional changes in
the treatment, where the first-stage restriction is treatment rank
similarity. We propose causal estimands that are doubly robust in that they
are valid under either of these two restrictions. We apply the doubly robust
estimation to estimate the impacts of sleep on well-being. Our new estimates
corroborate the usual 2SLS estimates.

\ \ \ \ \ \ \ \ \ \ \ \ \ \ \ \ \ \ \ \ \ \ \ \ \ \ \ \ \ \ \ \ \ \ \ \ \ \
\ \ \ \ \ \ \ 

\noindent \textbf{JEL codes}: C14, C21, I30

\noindent \textbf{Keywords}: Doubly Robust Identification, Non-separable
Model, Treatment effect heterogeneity, Continuous treatment, Monotonicity,
Rank similarity, Sleep time, Well-being
\end{abstract}


\onehalfspacing

\section{Introduction}

Many empirical applications estimate causal effects of a continuously
distributed endogenous variable (treatment), such as air pollution
concentration, poverty rate, income, price, birth weight, and time use, etc.
A common approach is to apply two-stage least squares (2SLS) estimation,
using a binary or discrete instrumental variable (IV). See, for recent
examples, Chay and Greenstone (2005), Kling et al.\ (2007), Goda et al.
(2011), Angrist et al.\ (2000), Maruyama and Heinesen (2020), Giaccherini et
al. (2021), Aggeborn and Ohman (2021), and Bessone et al.\ (2021). In the
case of a binary instrument, the 2SLS estimator essentially estiamtes the
Wald ratio. In its basic form without considering covariates, the Wald ratio
estimand is given by 
\begin{equation}
\tau ^{Wald}\text{:=}\frac{\mathbb{E}\left[ Y|Z=1\right] -\mathbb{E}\left[
Y|Z=0\right] }{\mathbb{E}\left[ T|Z=1\right] -\mathbb{E}\left[ T|Z=0\right] }%
,  \label{Wald}
\end{equation}%
where $Y$ is the outcome of interest, $Z$ is the binary instrument and $T$
is the treatment. The above estimand $\tau ^{Wald}$ requires a mean change
in the treatment variable, as the denominator cannot be zero. In addition,
when treatment effect is heterogenous and individuals select treatment
intensity based on idiosyncratic gains, causal interpretation of the Wald
ratio relies on a monotonicity assumption, which restricts treatment to
change in one direction when the IV\ changes. This monotonicity assumption
is originally proposed in Imbens and Angrist (1994) to show that in the case
of a binary treatment and a binary IV, $\tau ^{Wald}$ identifies a local
average treatment effect (LATE).

One drawback of the above approach is that causal identification may be weak
or may even fail. Frequently, policy instruments aim to shift one or two
tails of the treatment distribution or change other features, such as the
variance, of the treatment. As a result, treatment changes may concentrate
at some selected quantiles, say lower or upper quantiles. By solely focusing
on the mean treatment change, one may miss where the true changes are in the
treatment distribution. Examples of such policies include minimum wage,
minimum capital requirements, and the pollution ceiling set by Environmental
Protection Agency (EPA). In this paper, we consider an alternative approach,
which explores the distributional change in the first stage for causal
identification. This idea has been proposed and explored in the
non-separable IV model literature. The commonly employed restriction in the
first stage is treatment rank invariance or more generally treatment rank
similarity. Rank invariance in the first stage is typically stated as the
condition that the treatment function is monotonic in a scalar disturbance.
See, e.g., Imbens and Newey (2002), and Chesher (2001, 2002) for early
papers exploring this condition in non-separable models. Note that both the
LATE-type monotonicity and treatment rank similarity are restrictions on the
first-stage instrument effect heterogeneity. In general neither assumption
implies the other. Also, these assumptions are not verifiable, in the sense
that one may at best test their testable implications, which are necessary
but not sufficient conditions of these assumptions.\footnote{%
See, e.g., Angrist and Imbens (1995) and Fiorini and Stevens (2021) for the
discussion of the testable implication of the LATE-type monotonicity when
treatment is multi-valued. See Dong and Shu (2018) and Frandsen and Lefgren
(2018) for tests of the testable implication of rank similarity.}

This paper takes a novel nonparametric doubly robust (DR) identification
approach to identify causal effects of a continuous treatment using a binary
or discrete instrument. We consider the two alternative restrictions on the
first-stage instrument effect heterogeneity: the LATE-type monotonicity or
treatment rank similarity. Either of these assumptions can be consistent
with certain treatment choice behaviors and has been used extensively to
identify causal treatment effects. See, e.g., Imbens and Newey (2009) and
the reference therein\ for justifications of rank invariance (a stronger
version of rank similarity) and Angrist and Imbens (1995) for a
justification of monotonicity when treatment is multi-valued. We focus on
discrete instruments since discrete instruments are widely used.

The parameters of interest include 1) the average effect at a given
treatment quantile, which captures treatment effect heterogeneity at
different treatment intensities and 2) weighted average effects for the
largest subpopulation that respond to the IV change. Identification of the
former parameter requires treatment rank similarity to hold. In contrast,
for the latter parameter, we develop doubly robust estimands that are valid
under either monotonicity or treatment rank similarity. When monotonicity
holds, these estimands reduce to the LATE-type estimands and individuals who
respond to the IV change can be labeled as compliers, since they change
treatment in a monotonic way (similar to compliers in the classic LATE\
model with a binary treatment and binary IV); otherwise, these estimands
continue to be valid under treatment rank similarity, i.e., they continue to
identify weighted averages of the average treatment effects for all
individuals that change their treatment values when the IV\ changes, even
though these individuals no longer respond in a monotonic way. Instead,
their treament changes are subject to the rank restriction, i.e., the
probability distribution of their treatment ranks stay the same, which is a
slight generalization of requiring that their treatment rank to be exactly
the same. Since these first-stage restrictions are not verifiable, our
doubly robust estimands\ identify causal effects for the largest
subpopulation while allowing either of these two assumptions holds true.

Our identification is nonparametric in that we consider non-separable models
for both the first-stage and the outcome equation. Non-separable models
allow for treatment effect heterogeneity and individuals self-selection of
different treatment levels based on idiosyncratic gains, both of which are
important features of the data as supported by economic theory and empirical
evidence. For estimation, we opt for convenient semiparametric estimators to
avoid cumbersome fully nonparametric estimation. We establish consistency
and asymptotic normality of our proposed estimators. Lastly we apply our
proposed approach to estimate the impacts of sleep time on individuals'
well-being using data from a recent field experiment by Bessone et al.\
(2021). We show that our doubly robust approach can serve as a valuable tool
to corroborate the IV/2SLS estimates.

This paper's identification approach builds upon two strands of literature -
the LATE literature and the non-separable IV model literature. The LATE\
model is proposed in the seminal work of Imbens and Angrist (1994) and is
further extended in Angrist and Imbens (1995), Angrist, Imbens and Rubin
(1996), Angrist, Graddy and Imbens (2000), Abadie (2003), Fr\"{o}lich
(2007), de Chaisemartin (2017), Dahl, Huber, and Mellace (2023), etc. The
LATE model relies on the monotonicity assumption mentioned previously or
some weaker versions of it for causal identification. Many studies in the
nonseparable IV model literature explore rank invariance or rank similarity
in the first stage for causal identification. See, e.g., Chesher (2001,
2002, 2003, 2005), Imbens and Newey (2002, 2009), Florens et al.\ (2008) and
more recently Torgovitsky (2015), and D'Haultfoeuille and F\'{e}vrier
(2015), among others. In particular, Torgovitsky (2015), and D'Haultfoeuille
and F\'{e}vrier (2015) provide detailed discussions of the identifying power
of rank restrictions in the treatment and/or in the outcome equation. In
addition, Masten and Torgovitsky (2016) consider a random correlated
coefficients model and utilize treatment rank invariance to identify the
average partial effect of continuous treatment variables, using binary or
discrete instruments. For the DR identification approach, a few existing
studies take this approach, see, e.g., Dong, Lee, and Gou (2023) and
Arkhangelsky and Imbens (2022). Both papers are set in different frameworks
than the current one. Dong, Lee, and Gou (2023) study the regression
discontinuity design, while Arkhangelsky and Imbens (2022) investigate the
panel data model.\footnote{%
The current paper extends the regression discontinuity setup of Dong, Lee,
and Gou (2023) in multiple directions, including allowing the IV
independence and treatment rank similarity to hold conditional on a vector
of continuous and/or discrete covariates, allowing for a multi-valued IV or
a vector of discrete IVs and completely different estimation and inference
procedures.}

The rest of the paper proceeds as follows. Section~\ref{SecId} presents the
DR identification results for the basic setup with a binary IV and without
covariates. Section~\ref{SecGen} extends the identification results to the
general setup with covariates. Section~\ref{SecEstInf} proposes convenient
partial linear estimators and establishes their consistency and asymptotic
normality. Section~\ref{SecExt} discusses extensions to the case with a
multi-valued IV or a vector of discrete IVs, with or without covariates;
Section~\ref{SecEmp} presents our empirical analysis. Short concluding
remarks are provided in Section~\ref{SecCon}.

\section{\protect\large Doubly Robust Identification in the Basic Setup}

\label{SecId}Let $Y\in \mathcal{Y}\subset \mathcal{R}$ be the outcome of
interest, e.g., a measure of well-being. $Y$ can be continuous or discrete.
Let $T\in \mathcal{T}\subset \mathcal{R}$ be a continuous treatment
variable, e.g., sleep time. Let $Z\in \{0,1\}$ be a binary IV for $T$, e.g.,
an indicator for being randomly assigned to a group receiving encouragement
or financial incentives to increase night sleep.

To present the core ideas, we subsume all the covariates in this section.
The general setup with covariates is presented in the next section. Assume
that $Y$ and $T$ are generated as 
\begin{eqnarray}
Y &=&g\left( T,\varepsilon \right),  \label{Y_eq} \\
T &=&h\left( Z,U\right),  \label{T_eq}
\end{eqnarray}%
where $\varepsilon $ captures all the other factors other than $T$ that
affect $Y$, and similarly $U$ captures all the other reduced-form factors
other than $Z$ that affect $T$. The outcome disturbance $\varepsilon \in 
\mathcal{E}\subset \mathcal{R}^{d_{\varepsilon }}$ is allowed to be of
arbitrary dimension, so $d_{\varepsilon }$ does not need to be finite.
Without loss of generality, rewrite eq.\ (\ref{T_eq})\ as%
\begin{equation}
T=ZT_{1}\left( U_{1}\right) +\left( 1-Z\right) T_{0}\left( U_{0}\right),
\label{T_eq2}
\end{equation}%
where $T_{z}\left( \cdot \right),z=0,1$ are some unknown functions, and the
reduced-form disturbance $U_{z}\in \mathcal{U}_{z}\subset \mathcal{R}$, $%
z=0,1$. Later we impose an assumption that essentially requires $T_{z}\left(
\cdot \right) $ to be the quantile functions and $\mathcal{U}_{z}$ to be the
rank variables. Note by construction $U=ZU_{1}+\left( 1-Z\right) U_{0}$.

Define $Y_{t}$:=$g\left( t,\varepsilon \right) $ as the potential outcome
when $T$ is exogenously set to be $t$. Further define $T_{z}$:=$T_{z}\left(
U_{z}\right) $, $z=0,1$, as the potential treatment when $Z$ is exogenously
set to be $z$. Denote the support of $T_{z}$ as $\mathcal{T}_{z}$. The
observed treatment is then $T=ZT_{1}+\left( 1-Z\right) T_{0}$. We use $%
F_{\cdot }\left( \cdot \right) $ and $F_{\cdot |\cdot }\left( \cdot |\cdot
\right) $ to denote the unconditional cumulative distribution function (CDF)
and conditional CDF, respectively.

\begin{Assumption}[Treatment quantile representation]
$T_{z}(u)$ is strictly increasing in $u$, and $U_{z}\sim Unif\left(
0,1\right) $, $z=0,1$.\label{Cont_T}
\end{Assumption}

Assumption \ref{Cont_T} requires that the potential treatment $T_{z}$ is
continuous with a strictly increasing CDF. The condition $U_{z}\sim
Unif\left( 0,1\right) $ involves a normalization. This kind of normalization
is necessary, since the identification results hold up to a monotonic
transformation of $U_{z}$, as long as $U_{z}$ is continuous with a strictly
increasing CDF. See discussions in Matzkin (2003) and more recently
Torgovitsky (2015). By Assumption \ref{Cont_T}, $T_{z}(u) $ is the $u$
quantile of $T_{z}$, and $U_{z}=F_{T_{z}}(T_{z})$ is the rank of the
potential treatment. Further, $U=ZU_{1}+\left( 1-Z\right) U_{0}$ is the
observed treatment rank.

\begin{Assumption}[Independence]
$Z\perp \left( U_{z},\varepsilon \right) $, $z=0,1$.\label{Inde}
\end{Assumption}

Assumption \ref{Inde} essentially requires $Z$ to be randomly assigned. More
generally, we can allow the independence condition to hold only after
conditioning on relevant pre-determined covariates, which we will discuss in
the next session. Assumptions \ref{Cont_T} and \ref{Inde} imply $U\perp Z$,
because for $z=0,1$, $\Pr \left( U\leq \tau |Z=z\right) =\Pr \left(
U_{z}\leq \tau |Z=z\right) =\Pr \left( U_{z}\leq \tau \right) =\tau $, where
the last equality follows\ the condition $Z\perp U_{z}$ as implied by
Assumption \ref{Inde}.

\begin{Assumption}[First-stage]
$T_{1}(u)\neq T_{0}(u)$ for at least some $u\in \left( 0,1\right) $. \label%
{1st_stage}
\end{Assumption}

Assumption~\ref{1st_stage} requires that the distribution of $T$ changes
with $Z$. Assumption~\ref{1st_stage} is strictly weaker than the standard
first-stage assumption of the LATE model, which requires $\mathbb{E}\left[
T_{1}\right] \neq \mathbb{E}\left[ T_{0}\right] $. For example, when the
policy instrument $Z$ affects the variance or shifts the tails of the
treatment distribution but otherwise leaves the average treatment level
unaffected, we have the standard LATE first-stage assumption fails, but the
above Assumption~\ref{1st_stage} holds.

\begin{Assumption}[Monotonicity]
$\Pr \left( T_{1}\geq T_{0}\right) =1$.\label{Mono}
\end{Assumption}

Assumption \ref{Mono} requires that treatment can only change in one
direction when $Z$ changes - without loss of generality, we normalize it to
be non-decreasing. For example, this assumption holds in the usual linear
regression model of $T$ with a constant coefficient on $Z$.

Assumption~\ref{Mono} can not be tested directly, but it has testable
implications. It implies $T_{1}(u)-T_{0}(u)\geq 0$ for all $u\in \left(
0,1\right) $, i.e., $T_{1}$ stochastically dominates $T_{0}$. Since
stochastic dominance is a necessary but not sufficient condition for
Assumption~\ref{Mono}, rejecting stochastic dominance could mean
monotonicity does not hold, but failing to reject does not necessarily mean
that monotonicity holds. That is, in a given empirical scenario, even we see
that the quantile curve of $T_{1}$ does not cross the quantile curve of $%
T_{0}$, it does not necessarily mean that Assumption~\ref{Mono} holds.
Assumption \ref{Mono} is essentially not verifiable. Assumptions \ref{Inde}
- \ref{Mono} together imply $\mathbb{E}\left[ T|Z=1\right] -\mathbb{E}\left[
T|Z=0\right] >0$.

For convenience of exposition, we generalize the standard definition of
compliers, which is defined for a binary treatment (Angrist, Imbens, and
Rubin, 1996). Let $\mathcal{T}_{c}=\{(t_{0},t_{1})\in \mathcal{T}_{0}\times 
\mathcal{T}_{1}\text{ : }t_{1}-t_{0}>0\}$ be the set of compliers. Define\ $%
LATE(t_{0},t_{1}):=\mathbb{E}\Big[\frac{Y_{t_{1}}-Y_{t_{0}}}{t_{1}-t_{0}}%
|T_{1}=t_{1},T_{0}=t_{0}\Big]$ for any $(t_{0},t_{1})\in \mathcal{T}_{c}$. $%
LATE(t_{0},t_{1})$ is the local average treatment effect for complier type $%
(t_{0},t_{1})\in \mathcal{T}_{c}$. For example, in the case of a binary
treatment, $\tau ^{Wald}=LATE(0,1)$. More generally when treatment is
continuous as in our setup, $\tau ^{Wald}$ is a weighted average of $%
LATE(t_{0},t_{1})$ for all $(t_{0},t_{1})\in \mathcal{T}_{c}$. We formalize
this result in the following lemma.

\begin{lemma}
\textit{Let Assumptions \ref{Cont_T}-\ref{Mono} hold. Then} 
\begin{align*}
\tau ^{Wald}=\iint_{\mathcal{T}_{c}}w_{t_{0},t_{1}}LATE\left(
t_{0},t_{1}\right) F_{T_{0},T_{1}}\left( dt_{0},dt_{1}\right)
\end{align*}%
where $w_{t_{0},t_{1}}=\left( t_{1}-t_{0}\right) /\iint_{\mathcal{T}%
_{c}}\left( t_{1}-t_{0}\right) F_{T_{0},T_{1}}\left( dt_{0},dt_{1}\right) $. %
\label{L_LATE}
\end{lemma}

The above lemma states that under Assumptions \textit{\ \ref{Cont_T}-\ref%
{Mono}}, $\tau ^{Wald}$ in eq.\ (\ref{Wald}) identifies a weighted average
of the average treatment effects for different compliers, where the weights
are proportional to their treatment intensity change $\left(
t_{1}-t_{0}\right) $. Fr\"{o}lich (2007) gives a comparable expression when
treatment is multi-valued. Angrist and Imbens (1995) provide a slightly
different expression than that of Fr\"{o}lich (2007), but as pointed out by
Fr\"{o}lich (2007), these expressions are equivalent.\footnote{%
Unlike in Fr\"{o}lich (2007) and here, Angrist and Imbens (1995) present the
weighted average effect in terms of overlapping subpopulations.}

When $g(T,\varepsilon )$\ is continuously differentiable in its first
argument, the identified causal parameter can be further expressed as a
weighted average derivative of $Y$\ w.r.t. $T$, following Angrist et al.
(2000, Theorem 1). The exact form of the weighted average derivative is
provided in the proof of Lemma \ref{L_LATE} in the online supplementary
appendix.\footnote{%
Our weighted average derivative appears to be different than that of Angrist
et al.\ (2000). Similar to the point made in Fr\"{o}lich (2007), both are
equivalent and the difference lies in that they express their weighted
average derivative in terms of overlapping subpopulations.}

In the following, we provide an alternative assumption to Assumption~\ref%
{Mono}, which allows us to identify causal effects at different treatment
quantiles and further a convexly weighted average effect. This convexly
weighted average effect is in contrast to $\tau ^{Wald}$, which is also a
convexly weighted average effect under Assumption~\ref{Mono} monotonicity.

\begin{Assumption}[Treatment Rank Similarity]
$U_{0}|\varepsilon \sim U_{1}|\varepsilon $.\label{RS}
\end{Assumption}

Assumption \ref{RS} assumes that conditional on $\varepsilon $, $U_{0}$ and $%
U_{1}$ follow the same distribution. Without conditioning on $\varepsilon $, 
$U_{0}$ and $U_{1}$ both follow a uniform distribution over the unit
interval due to normalization, so $F_{U_{0}}(u) =F_{U_{1}}\left( u\right) $
by construction. Assumption \ref{RS} implies $\varepsilon |U_{0}=u\sim
\varepsilon |U_{1}=u$ by Bayes' theorem, so $\varepsilon $ has the same
distribution at the same rank of the potential treatment.

A slightly stronger assumption is rank invariance, which is the condition $%
U_{0}=U_{1}$. Rank invariance essentially requires that the joint
distribution of $T_{0}$ and $T_{1}$ are degenerate. Rank invariance holds
trivially when the treatment model is additively separable in a scalar
disturbance, but this assumption does not require additive separability in
general. Rank invariance is frequently imposed in the non-separable IV
literature. For example, Imbens and Newey (2009) propose a control variable
approach to identify various causal parameters for the non-separable IV
model. They assume that in the treatment model $T=h\left( Z,U\right) $, $U$
is a scalar unobservable, and that $h\left( Z,u\right) $ is strictly
increasing in $u$ with probability 1. Monotonicity in a scalar disturbance
implies rank invariance, because under this assumption, $U_{z}$:=$%
F_{T_{z}}\left( T_{z}\right) =F_{U}(u)$ for any $z$ in the support of $Z$.
When $Z\in \{0,1\}$, it means $U_{0}=U_{1}$. In addition to rank invariance,
Imbens and Newey (2009) assume $Z\perp \left( U,\varepsilon \right) $, which
is equivalent to Assumption \ref{Inde} when rank invariance holds.

Rank similarity in Assumption \ref{RS} relaxes rank invariance - instead of
assuming the ranks of the potential treatments to be the same, it only
assumes that they have the same conditional probability distribution for any
given $\varepsilon $, and thereby permits random deviations from the common
rank level between the potential treatments. For example, if the common rank
level for night sleep (the actual time one is in sleep as measured by
actigraphy) is determined by individuals' biological clock (possibly after
conditioning on observable covariates as disscussed in our general setup),
which does not change with $Z$, then rank similarity permits that the
increase in night sleep is subject to some random factors. Rank similarity
was proposed by Chernozhukov and Hansen (2005, 2006) to identify quantile
treatment effects in IV models. Note that they impose rank similarity on the
ranks of potential outcomes, instead of ranks of potential treatments.

\begin{lemma}
Under Assumptions \textit{\ref{Cont_T}, }\ref{Inde} and \ref{RS}, $T\perp
\varepsilon |U$. \label{L_CV}
\end{lemma}

Lemma \ref{L_CV} suggests that $U$ is a control variable as defined by
Imbens and Newey (2009), i.e., conditional on the observed treatment rank $U$%
, $T$ is exogenous to $Y$. Intuitively, under Assumptions \ref{Inde} and \ref%
{RS} and holding $U$ fixed, the only variation in $T$ is the exogenous
variation induced by $Z$.\footnote{%
This result is closely related to Theorem 1 of Imbens and Newey (2009),
except that we assume rank similarity instead of rank invariance and that we
focus on a binary IV instead of an IV that may have a large support. The
large support is required to identify structural parameters, like the
average structural function (Blundell and Powell, 2003), when the outcome
disturbance is of arbitrary dimension.}

Based on Lemma \ref{L_CV}, one may condition on $U$ in the outcome equation
to estimate the causal effect of $T$ on $Y$. Let $q_{z}\left( u\right)
=F_{T|Z}^{-1}\left( u|z\right) $ be the conditional $u$ quantile of $T$
given $Z=z$, and further $\Delta q\left( u\right) =q_{1}\left( u\right)
-q_{0}\left( u\right) $. In addition, let $\mathcal{U=}\left\{ u\in \left(
0,1\right) \text{: }\Delta q\left( u\right) \neq 0\right\} $. By
conditioning on $U=u$, for any $u\in \mathcal{U}$, the resulting IV estimand
can be written as 
\begin{equation}
\tau (u)\text{:=}\frac{\mathbb{E}\left[ Y|Z=1,U=u\right] -\mathbb{E}\left[
Y|Z=0,U=u\right] }{\mathbb{E}\left[ T|Z=1,U=u\right] -\mathbb{E}\left[
T|Z=0,U=u\right] }.  \label{estimand_tau_u}
\end{equation}
The numerator captures the reduced-form effect of $Z$ on $Y$ given $U=u$,
while the denominator captures the first-stage treatment change given $U=u$.
The corresponding estimator (by replacing the population means and ranks by
their sample analogues)\ is analogous to the indirect least square estimator
in the linear IV model setting.

Intuitively, conditional on $U=u$, with a binary instrument, $T$ potentially
can take two values $T_{0}\left( u\right) $ and $T_{1}(u)$. When $T$ changes
exogenously from $T_{0}(u)$ and $T_{1}(u)$, the corresponding average effect
on the outcome $\mathbb{E}\big[ Y_{T_{1}(u)}-Y_{T_{0}(u)}|U=u\big] $ can be
identified, as we show in the following Theorem \ref{Thm_LATE_u}. For
notational convenience, let $\Delta T(u)=T_{1}\left( u\right) -T_{0}(u)$.

\begin{theorem}
Let Assumptions \ref{Cont_T}-\ref{1st_stage} and Assumption \ref{RS} hold.
Then for any $u\in \mathcal{U}$, 
\begin{eqnarray}
\tau (u) &=&\mathbb{E}\left[ \frac{Y_{T_{1}(u) }-Y_{T_{0}(u) }}{\Delta T(u) }%
|U=u\right]  \label{QLATE_1} \\
&=&\int \left\{ g\left( T_{1}(u) ,e\right) -g\left( T_{0}\left( u\right)
,e\right) \right\} \frac{1}{\Delta T(u) }F_{\varepsilon |U}\left(
de|u\right) .  \label{Q-LATE_2}
\end{eqnarray}%
\label{Thm_LATE_u}
\end{theorem}

To see the above results, note%
\begin{eqnarray*}
\mathbb{E}\left[ Y|Z=1,U=u\right] &=&\mathbb{E}\left[ g\left( T_{1}\left(
u\right),\varepsilon \right) |Z=1,U=u\right] \\
&=&\mathbb{E}\left[ g\left( T_{1}(u),\varepsilon \right) |U=u\right] \\
&=&\mathbb{E}\left[ Y_{T_{1}(u)}|U=u\right]
\end{eqnarray*}%
where the first equality follows from the models of $Y$ and $T$, (\ref{Y_eq}%
) and (\ref{T_eq2}), respectively, the second equality follows from the
condition $Z\perp \varepsilon |U$\ as shown in the proof of Lemma \ref{L_CV}%
, and the last equality is by the definition of the potential outcome. One
can similarly show $\mathbb{E}\left[ Y|Z=0,U=u\right] =\mathbb{E}\left[
Y_{T_{0}(u)}|U=u\right] $.\textbf{\ }That is, we can identify $\mathbb{E}%
\left[ Y_{T_{z}(u)}|U=u\right] $ for $z=0,1$ and $u\in \mathcal{U}$. Ideally
one may wish to recover $\mathbb{E}\left[ Y_{t}\right] $ for any $t\in 
\mathcal{T}$, which is known as the average dose-response function or the
average structural function. However, identifying $\mathbb{E}\left[ Y_{t}%
\right] $ for any $t\in \mathcal{T}$ is not possible in our setup, because
we have a binary instrument and we do not restrict the dimensionality of the
outcome disturbance, i.e., we do not impose rank invariance in the outcome
model.

Theorem \ref{Thm_LATE_u} shows that $\tau (u)$ identifies an average (per
unit) treatment effect at the $u$ quantile of the treatment. \ $\tau (u)$
measures treatment effect heterogeneity at different treatment intensities,
which can be useful.\ The denominator in eq.\ (\ref{QLATE_1}) reflects the
fact that $T_{z}\left( u\right) \notin \left\{ 0,1\right\} $ in general.
Inside of the integral in eq.\ (\ref{Q-LATE_2}), $T$ exogenously changes
from $T_{0}(u)$ to $T_{1}(u)$ while holding $\varepsilon $ fixed at $e$, so $%
\tau (u)$ is causal from a \textit{ceteris paribus }point of view.

By eq.\ (\ref{T_eq2}) and Assumption \ref{Cont_T}, $U$ and $T$ follow a
one-to-one mapping given $Z=z$, i.e., conditioning on $U=u$ is the same as
conditioning on $T=T_{z}\left( u\right) $.\footnote{%
The $\sigma $-algebra is the same.} Further by Assumption \ref{Inde}, $%
T_{z}\left( u\right) =q_{z}\left( u\right) $. 
Let the conditional mean function of $Y$ given $Z$ and $T$ be $m_{z}(t)=%
\mathbb{E}\left[ Y|Z=z,T=t\right] $, $z=0,1$. Then $\tau (u)$ in eq.\ (\ref%
{estimand_tau_u})\ can be re-written as 
\begin{equation}
\tau (u)=\frac{m_{1}(q_{1}(u))-m_{0}(q_{0}(u))}{q_{1}\left( u\right)
-q_{0}\left( u\right) }.  \label{estimand_tau_u_2}
\end{equation}%
Later our estimation is directly based on eq.\ (\ref{estimand_tau_u_2}).

Oftentimes, researchers or policy makers are interested in some summary
measure of the overall treatment effect. With $\tau (u) $, one can further
identify and estimate a weighted average of $\tau \left( u\right) $, i.e., 
\begin{align*}
\tau ^{RS}\left( w\right) \text{:=}\int_{\mathcal{U}}\tau (u) w(u) du
\end{align*}%
for any known or estimable weighting function $w(u) $ such that $w(u) \geq 0$
and $\int_{\mathcal{U}}w(u) du=1$. The weighting function $w(u) $ is
required to be non-negative; otherwise, $\tau ^{RS}\left( w\right) $ can be
a weighted difference of the average treatment effects for units. For
example, if $\mathcal{U=}\left( 0,1\right) $, and one chooses $w(u) =1$,
then $\tau ^{RS}\left( w\right) =\mathbb{E}\left[ \tau(U)\right] $.

$\tau ^{RS}\left( w\right) $ is a weighted average of the average treatment
effects at all treatment quantiles where treatment changes under Assumption %
\ref{RS}, treatment rank similarity. By Lemma \ref{L_LATE}, $\tau ^{Wald}$
is a weighted average of the average treatment effects for all compliers
under Assumption \ref{Mono},\ monotonicity. Note that both assumptions
impose restrictions on the first-stage IV effect heterogeneity -
monotonicity imposes a sign restriction, while treatment rank similarity
imposes a rank restriction. Neither assumption implies the other. Neither
assumption is verifiable. In practice, it is not ideal to have to choose one
versus the other estimand based on some pre-testing results. We therefore
consider a weighting function that leads to a DR property of the resulting
estimand, i.e., the estimand is valid under either of the two alternative
identifying assumptions.

\begin{proposition}
Let Assumptions \ref{Cont_T}-\ref{1st_stage} hold. If either Assumption \ref%
{Mono} or Assumption \ref{RS} holds, then $\tau ^{DR}$:=$\int_{\mathcal{U}%
}\tau (u)w^{DR}(u)du$ for $w^{DR}(u)=|\Delta q\left( u\right) |/\int_{%
\mathcal{U}}|\Delta q\left( u\right) |du$ identifies a weighted average of
the average treatment effects among units for which $T_{1}\neq T_{0}$.\label%
{Prop_DR}
\end{proposition}

Proposition \ref{Prop_DR} combines the results of Lemma \ref{L_LATE} and
Theorem \ref{Thm_LATE_u}. It shows that under either first-stage restriction
on the IV effect heterogeneity, $\tau ^{DR}$ identifies a weighted average
of the average effects for all the units that respond to the IV\ change.
These units represent the largest subpopulation one can identify causal
effects for without any further restrictions. The two alternative
first-stage assumptions specify exactly how these units respond - either
they change treatment in a monotonic way or they change treatment such that
the probability distribution of their treatment ranks remains the same.

When Assumption \ref{Mono} monotonicity holds, $w^{DR}(u)=\Delta q\left(
u\right) /\int_{\mathcal{U}}\Delta q\left( u\right) du$. Then 
\begin{eqnarray*}
\tau ^{DR} &=&\frac{\int_{0}^{1}\left\{ \mathbb{E}\left[ Y|Z=1,U=u\right] -%
\mathbb{E}\left[ Y|Z=0,U=u\right] \right\} du}{\int_{0}^{1}\Delta q\left(
u\right) du} \\
&=&\frac{\mathbb{E}\left[ Y|Z=1\right] -\mathbb{E}\left[ Y|Z=0\right] }{%
\mathbb{E}\left[ T|Z=1\right] -\mathbb{E}\left[ T|Z=0\right] } \\
&=&\tau ^{Wald}
\end{eqnarray*}%
That is, $\tau ^{DR}$ reduces to the standard LATE estimand $\tau ^{Wald}$
given by eq.\ (\ref{Wald}) when monotonicity holds. By Lemma \ref{L_LATE},
in this case $\tau ^{DR}$ identifies a weighted average of the average
treatment effects for different compliers. Otherwise, when Assumption \ref%
{Mono} monotonicity does not hold, but Assumption \ref{RS} rank similarity
holds, $\tau ^{DR}$ is a weighted average of $\tau (u)$ for $u\in \mathcal{U}
$, and by Theorem \ref{Thm_LATE_u}, $\tau (u)$ captures the average
treatment effects at the $u$ quantile of treatment. Either way, $\tau ^{DR}$
identifies a weighted average of the average treatment effects for all the
units that change their treatment levels in response to the IV changes. The
weights are proportional to the magnitude of their treatment changes.

The weighting function in Proposition \ref{Prop_DR} allows $\Delta q\left(
u\right) $ to change signs, which in fact indicates that the LATE\
monotonicity condition does not hold. As a result, $\tau ^{DR}$ may average
over two different types of units, those who increase their treatment levels
and those who decrease their treatment levels when the IV changes.\footnote{%
This issue is not unique to our setting. This issue would arise whenever a
research estimates some average effects, but does not assume that the
treatment change cannot switch signs.} Ideally one may want to separately
consider these two types of units. However, individual types are not point
identified, so point identification of causal effects over different
individual types is not possible. As a mitigation measure, if desired, one
may separately consider treatment quantiles where $\Delta q\left( u\right)
>0 $ and those where $\Delta q\left( u\right) <0$.

Let $\mathcal{U}_{+}\mathcal{=}\{u\in \mathcal{U}\text{ : }\Delta q\left(
u\right) >0\}$. Define $\tau _{+}^{DR}$:=$\int_{\mathcal{U}_{+}}\tau
(u)w_{+}(u)du$, where $w_{+}(u)=\Delta q\left( u\right) /\int_{\mathcal{U}%
_{+}}\Delta q\left( u\right) du$. $\tau _{+}^{DR}$ can be rewritten as the
ratio of the mean outcome difference over $u\in \mathcal{U}_{+}$ to the mean
treatment difference over $u\in \mathcal{U}_{+}$, i.e.,%
\begin{equation*}
\tau _{+}^{DR}=\frac{\int_{\mathcal{U}_{+}}\int \left\{ g\left( T_{1}\left(
u\right) ,e\right) -g\left( T_{0}(u),e\right) \right\} F_{\varepsilon
|U}\left( de|u\right) du}{\int_{\mathcal{U}_{+}}\Delta T(u)du}\text{.}
\end{equation*}%
$\tau _{+}^{DR}$\ carries a similar interpretation as that of $\tau ^{DR}$
but is only for the subset of treatment quantiles $u\in \mathcal{U}_{+}$.%
\footnote{%
Under treatment rank invariance, monotonicity holds automatically if
treatment quantile changes do not switch signs. This is not true in general.
Intuitively under rank invariance, $U_{1}=u$ implies $U_{0}=u$ and vise
versa; however, under treatment rank similarity (but not rank invariance),
individuals counterfactual treatment rank is not point identified, and hence
monotonic treatment quantile changes do not gurantee individual level
monotonicity.} In particular, when either monotonicity or treatment rank
similarity holds over $\mathcal{U}_{+}$, $\tau _{+}^{DR}$ identifies a
weighted average of the average treatment effects for all the responding (to
IV changes) units associated with this subset of quantiles. Similarly, one
can define $\tau _{-}^{DR}$:=$\int_{\mathcal{U}_{-}}\tau (u)w_{-}(u)du$,
where $\mathcal{U}_{-}\mathcal{=}\left\{ u\in \mathcal{U}\text{ : }\Delta
q\left( u\right) <0\right\} $ and $w_{-}(u)=\Delta T(u)/\int_{\mathcal{U}%
_{-}}\Delta q\left( u\right) du$.

So far, we have focused our discussion on (weighted) average effects. One
may extend the above identification results to identify distributional
effects at a given treatment quantile $u\in \mathcal{U}$. In particular,
under Assumptions \ref{Cont_T}-\ref{1st_stage} and Assumption \ref{RS}, for
any $u\in \mathcal{U}$, $F_{Y_{T_{z}(u)}|U}\left( y|u\right) =\mathbb{E}[%
\mathbf{1}(Y\leq y)|U=u,Z=z]$, $z=0,1$. Let the conditional quantile
function of $Y_{T_{z}(u)}$ given $U=u$ be $F_{Y_{T_{z}(u)}|U=u}^{-1}\left( 
\widetilde{u}\right) $ for $\widetilde{u}\in \left( 0,1\right) $ and $z=0,1$%
. The corresponding reduced-form quantile treatment effect is given by $%
F_{Y_{T_{1}(u)}|U=u}^{-1}\left( \widetilde{u}\right)
-F_{Y_{T_{0}(u)}|U=u}^{-1}\left( \widetilde{u}\right) $ for any $\widetilde{u%
}\in \left( 0,1\right) $ and $u\in \mathcal{U}$. Replacing $Y$ by $\mathbf{1}%
\left( Y\leq y\right) $ in Theorem \ref{Thm_LATE_u} and further in
Proposition \ref{Prop_DR} leads to the DR\ estimand for the weighted average
effect of $T$ on $\mathbf{1}\left( Y\leq y\right) $ for all $y\in \mathcal{Y}
$. It is worth emphasizing that our goal is to develop robust identification
results in the presence of\ both treatment effect heterogeneity and
instrument effect heterogeneity. Whether any of these proposed weighted
averages are of interest depends on empirical scenarios.

\section{\protect\large Doubly Robust Identification with Covariates}

\label{SecGen}The previous section presents our core idea without
considering covariates. IV\ independence and rank similarity may be more
plausible when conditioning on relevant pre-determined covariates. For this
claim on rank similarity, see e.g., discussion in Chernozhukov and Hansen
(2005, 2006). If covariates enter the non-separable models for $Y$\ and $T$,
i.e., (\ref{Y_eq}) and (\ref{T_eq}), and all the previous assumptions hold
conditional on covariates, then it follows readily that all the previous
results hold conditional on covariates. However, such conditional results
may not be very useful in practice, as it can be unwieldy to present all the
conditional results if there are many covariates and worse many continuous
covariates. In this section, we seek to directly identify unconditional
weighted average effects as before while allowing for covariates.

Let $X\in \mathcal{X}\subset \mathcal{R}^{d_{X}}$ denote the vector of
covariates. We do not require $X$ to be exogenous. We consider the following
models for $Y$ and $T$:%
\begin{eqnarray}
Y &=&G\left( T,X,\epsilon \right),  \label{Y_general_eq} \\
T &=&H(Z,X,V)  \notag \\
&=&ZT_{1}\left( X,V_{1}\right) +\left( 1-Z\right) T_{0}\left( X,V_{0}\right)
,  \label{T_general_eq}
\end{eqnarray}%
where by construction $V=V_{1}Z+V_{0}\left( 1-Z\right) $.

As before, $T_{z}$:=$T_{z}\left( X,V_{z}\right) $ is the potential treatment 
$Z$ is exogenously set to be $z\in \left\{ 0,1\right\} $ and $Y_{t}$:=$%
G\left( t,X,\epsilon \right) $ is the potential outcome when $T$ is
exogenously set to be $t\in \mathcal{T}\subset \mathcal{R}$. We extend
Assumptions \ref{Cont_T}, \ref{Inde}, \ref{1st_stage} and \ref{RS} to
condition on covariates $X$ as follows.\footnote{%
We do not extend Assumption \ref{Mono} monotonicity, as little will be
changed from the identification perspective. For example, one way to relax
Assumption \ref{Mono} is to assume that either $\Pr \left( T_{1}\geq
T_{0}|X=x\right) =1$ or $\Pr \left( T_{1}\leq T_{0}|X=x\right) =1$ for any $%
x\in \mathcal{X}$. Since the sign of the first-stage change is identified
from the data given the assumptions here, for those covariate values at
which treatment change is negative when $Z$ changes from $0$ to $1$
(consistent with $\Pr \left( T_{1}\leq T_{0}|X=x\right) =1$), one may change
the observed value $Z=1$ to $Z=0$ and similarly $Z=0$ to $Z=1$, so that
after the switch, the condition $\Pr \left( T_{1}\geq T_{0}|X=x\right) =1$
for any $x\in \mathcal{X}$ holds, which then is essentially the same as
having $\Pr \left( T_{1}\geq T_{0}\right) =1$. Our identification results in
this section would go through with this rearranging values of $Z$.}

\begin{customassumption}{C1}
[{\small Conditional Treatment Quantile}]
For any $x\in \mathcal{X}$, $T_{z}(x,v) $, $z=0,1$, is strictly
increasing in $v$, and $V_{z}\sim Unif\left( 0,1\right) $.
\label{Cont_T_extend}
\end{customassumption}

By Assumption \ref{Cont_T_extend}, $T_{z}\left( x,v\right) $ is the
conditional quantile function of $T_{z}$ given $X$, and $V_{z}=F_{T_{z}|X}%
\left( T_{z}|X\right) $ is the conditional rank of $T_{z}$ given $X$.

\begin{customassumption}{C2}
[{\small Conditional Independence}]
$Z\perp \left( V_{z},\epsilon \right) |X$, $z=0,1$.\label{Inde_extend}
\end{customassumption}

\begin{customassumption}{C3}
[{\small Conditional First-stage}]
$T_{z}\left( x,v\right) \neq T_{z}\left( x,v\right) $ for at least some $%
x\in \mathcal{X}$ and $v\in \left( 0,1\right) $.\label{1st_extend}
\end{customassumption}

\begin{customassumption}{C4}
[{\small Common Support}]
$\Pr \left( Z=1|X=x\right) \in \left( 0,1\right) $ for any $x\in \mathcal{X}$%
.\label{CS}
\end{customassumption}

\begin{customassumption}{C5}
[{\small Conditional Treatment Rank Similarity}]
$V_{1}|( \epsilon , X) \sim V_{0}|(\epsilon ,X) $.%
\label{RS_extend}
\end{customassumption}

Assumption \ref{Inde_extend} requires $Z$ to be unconfounded, instead of
being randomly assigned as required by Assumption \ref{Inde}. Assumption \ref%
{CS} is a common support assumption to ensure our parameters are
well-defined. In addition, Assumption \ref{RS_extend}\ requires that
treatment rank similarity holds only among the subgroup of units with the
same observed covariate values, which is weaker than Assumption~\ref{RS}.%
\footnote{%
Note that $V_{z}$ is defined conditionally on $X$, while $U_{z}$ is defined
unconditionally. Given that $X$ are determinants of $Y$, one can let $X$ be
an observable sub-vector of $\varepsilon $\ in $Y=g\left( T,\varepsilon
\right) $. That is, $\varepsilon =\left( X,\epsilon \right) $. Assumption %
\ref{RS} $U_{1}|\varepsilon \sim U_{0}|\varepsilon $ implies $U_{1}|X\sim
U_{0}|X$, so $F_{U_{1}|X}\left( u|x\right) =F_{U_{0}|X}\left( u|x\right) $
for any $u\in \left( 0,1\right) $ and $x\in \mathcal{X}$. It follows that $%
F_{V_{0}|X,\epsilon }\left( v|X=x,\epsilon =e\right) =\mathbb{E}\left[
1\left( V_{0}\leq v\right) |X=x,\epsilon =e\right] =\mathbb{E}\left[ 1\left(
F_{U_{0}|X}\left( U_{0}|x\right) \leq v\right) |X=x,\epsilon =e\right] =%
\mathbb{E}\left[ 1\left( F_{U_{1}|X}\left( U_{1}|x\right) \leq v\right)
|X=x,\epsilon =e\right] =F_{V_{1}|X,\epsilon }\left( v|X=x,\epsilon
=e\right) $ for any $v$, $x$, and $e$ in their support, where the second
equality follows from $V_{z}=F_{Tz|X}\left( T_{z}|X\right) $ by Assumption %
\ref{Cont_T_extend}, which can be further written as $V_{z}=F_{Uz|X}\left(
U_{z}|X\right) $, $z=0,1$, since $T_{z}$ and $U_{z}$ follow a one-to-one
mapping by Assumption \ref{Cont_T}. Therefore, $V_{0}|X,\epsilon \sim
V_{1}|X,\epsilon $.}. The following Lemma extends Lemma \ref{L_CV} to allow
for covariates.

\begin{lemma}
Under Assumptions \ref{Cont_T_extend}-\ref{1st_extend} and \ref{RS_extend}, $%
T\perp \epsilon |\left( V,X\right) $.\label{L_CCV}
\end{lemma}

Lemma \ref{L_CCV} is a conditional (on $X$)\ version of Lemma \ref{L_CV}. So 
$V$ is a control variable given $X$. Let $q_{z}\left( v,x\right)
=F_{T|Z,X}^{-1}\left( v|z,x\right) $ be the conditional $v$ quantile of $T$
given $Z=z$ and $X=x$. Let $\Delta q(x,v)=q_{1}\left( v,x\right)
-q_{0}\left( v,x\right) $. Assumptions \ref{Cont_T_extend} and \ref{CS}
ensure that $\Delta q(x,v)$ is well defined for all $x\in \mathcal{X}$ and $%
v\in (0,1)$. Further let $\mathcal{S}=\{(x,v)\in \mathcal{X\times }(0,1)$: $%
\Delta q(x,v)\neq 0\}$. The resulting IV estimand by conditioning on $X=x$
and $V=v$ can be defined as 
\begin{equation}
\pi (x,v)\text{:=}\frac{\mathbb{E}\left[ Y|Z=1,X=x,V=v\right] -\mathbb{E}%
\left[ Y|Z=0,X=x,V=v\right] }{\mathbb{E}\left[ T|Z=1,X=x,V=v\right] -\mathbb{%
E}\left[ T|Z=0,X=x,V=v\right] }  \label{estimand_pi_x_v}
\end{equation}%
for any $(x,v)\in \mathcal{S}$. By eq.\ (\ref{T_general_eq}) and Assumption %
\ref{Cont_T_extend}, $T$ and $V$ follow a one-to-one mapping given $Z=z$ and 
$X=x$, i.e., conditioning on $V=v$ is the same as conditioning on $%
T=T_{z}\left( x,v\right) $ in (\ref{estimand_pi_x_v}). Further by Assumption %
\ref{Inde_extend}, $T_{z}\left( x,v\right) =q_{z}\left( x,v\right) $. Then $%
\pi (x,v)$ can be re-written as 
\begin{equation*}
\pi (x,v)=\frac{m_{1}(x,q_{1}(x,v))-m_{0}(x,q_{0}(x,v))}{q_{1}\left(
x,v\right) -q_{0}\left( x,v\right) },
\end{equation*}%
where $m_{z}(x,t)=\mathbb{E}\left[ Y|Z=z,X=x,T=t\right] $.


Let $\Delta T(x,v)=T_{1}(x,v)-T_{0}(x,v)$. We have the following Theorem \ref%
{Thm_LATE_v_extend}, which extends Theorem \ref{Thm_LATE_u}.

\begin{theorem}
Let Assumptions \ref{Cont_T_extend}-\ref{RS_extend} hold. Then for any $%
(x,v)\in \mathcal{S}$, 
\begin{align}
\pi (x,v)& =\mathbb{E}\left[ \frac{Y_{T_{1}(x,v)}-Y_{T_{0}(x,v)}}{\Delta
T(x,v)}|X=x,V=v\right]  \label{QLATE_X_1} \\
& =\int \left\{ G\left( T_{1}(x,v),x,e\right) -G\left( T_{0}(x,v),x,e\right)
\right\} \frac{F_{\epsilon |X,V}\left( de|x,v\right) }{\Delta T(x,v)}.
\label{QLATE_X_2}
\end{align}%
\label{Thm_LATE_v_extend}
\end{theorem}

By Theorem \ref{Thm_LATE_v_extend}, $\pi (x,v)$ identifies a conditional
weighted average treatment effect at the conditional $v$ quantile of the
treatment given $X=x$. By eq.\ (\ref{QLATE_X_2}), it is clear that $\pi
(x,v) $ represents the causal effect of an exogenous change in treatment
from $T_{0}\left( x,v\right) $ to $T_{1}(x,v)$, while holding $X$ and $%
\epsilon $ fixed at $x$ and $e$.

If desired, one may average $\pi (x,v)$ over the distribution of $X$ to
obtain a weighted average effect at the conditional $v$ quantile of the
treatment. For notational convenience, in the following, we assume $\pi
(x,v)=0$ when $\Delta q(x,v)=0$, so that $\pi (x,v)$ is defined for all $%
(x,v)\in \mathcal{X\times }\left( 0,1\right) $. For example, for any $v\in
\left( 0,1\right) $ such that $\Pr \left( \Delta q(X,v)\neq 0\right) >0$,\
one can define%
\begin{equation*}
\pi \left( v\right) \text{:=}\int_{\mathcal{X}}\pi (x,v)w_{v}(x)dx,
\end{equation*}%
where $w_{v}(x)=|\Delta q(x,v)|f_{X}\left( x\right) /\int_{\mathcal{X}%
}|\Delta q(x,v)|f_{X}(x)dx$. $\pi \left( v\right) $ identifies a weighted
average effect at the conditional $v$ quantile of the treatment. In
contrast, $\tau (u)$ identifies an average effect at the unconditional $u$
quantile of the treatment. $\pi \left( v\right) $ can be useful in
investigating treatment effect heterogeneity at the conditional $v$ quantile
of the treatment.

Consider now constructing a DR\ estimand for the overall unconditional
weighted average effect based on $\pi (x,v)$. Since $Z$ is valid only after
conditioning on pre-determined covariates, $\tau ^{Wald}$ is no longer a
valid causal estimand. Define 
\begin{equation}
\tau ^{Wald\_X}\text{:=}\frac{\int_{\mathcal{X}}\left\{ \mathbb{E}\left[
Y|Z=1,X=x\right] -\mathbb{E}\left[ Y|Z=0,X=x\right] \right\} f_{X}\left(
x\right) dx}{\int_{\mathcal{X}}\left\{ \mathbb{E}\left[ T|Z=1,X=x\right] -%
\mathbb{E}\left[ T|Z=0,X=x\right] \right\} f_{X}(x)dx}.  \label{Wald_X}
\end{equation}%
The numerator of eq.\ (\ref{Wald_X}) does not reduce to $\mathbb{E}\left[
Y|Z=1\right] -\mathbb{E}\left[ Y|Z=0\right] $ and similarly the denominator
of eq. (\ref{Wald_X})\ does not reduce to $\mathbb{E}\left[ T|Z=1,X=x\right]
-\mathbb{E}\left[ T|Z=0,X=x\right] $, as $X$ is not required to be
independent of $Z$, and hence the distribution of $X$ given $Z=0$ and that
given $Z=1$ are different in general. Nevertheless, the following lemma
shows that $\tau ^{Wald\_X}$ identifies the same unconditional effect as
what $\tau ^{Wald}$ would if $Z$ were valid without conditioning on
covariates.

\begin{lemma}
\textit{Let Assumptions \ref{Cont_T_extend} - \ref{CS} and further
Assumption \ref{Mono} hold. Then} 
\begin{equation*}
\tau ^{Wald\_X}=\iint_{\mathcal{T}_{c}}w_{t_{0},t_{1}}LATE\left(
t_{0},t_{1}\right) F_{T_{0},T_{1}}\left( dt_{0},dt_{1}\right)
\end{equation*}%
where $w_{t_{0},t_{1}}=\left( t_{1}-t_{0}\right) /\iint_{\mathcal{T}%
_{c}}\left( t_{1}-t_{0}\right) F_{T_{0},T_{1}}\left( dt_{0},dt_{1}\right) $. %
\label{L_LATE_X}
\end{lemma}

Fr\"{o}lich (2007) presents a comparable result for a binary or discrete
treatment. Our DR estimand below incorporates $\tau ^{Wald\_X}$ (instead of
its invalid counterpart $\tau ^{Wald}$)\ as a special case.

\begin{proposition}
Let Assumptions \ref{Cont_T_extend} -\textit{\ \ref{CS}} hold. When further
either \ref{Mono} or \ref{RS_extend} holds, 
\begin{equation*}
\pi ^{DR}\text{:=}\iint_{\mathcal{S}}\pi (x,v)w\left( x,v\right) dvdx
\end{equation*}%
for $w(x,v)=|\Delta q(x,v)|f_{X}(x)/\iint_{\mathcal{S}}|\Delta
q(x,v)|f_{X}(x)dvdx$ identifies a weighted average of the average treatment
effects among all the units for which $T_{1}\neq T_{0}$.\label%
{Prop_DR_extend}
\end{proposition}

When Assumption \ref{RS_extend} conditional rank similarity holds, $\pi
^{DR} $ is a weighted average of $\pi \left( x,v\right) $ for $(x,v)\in 
\mathcal{S} $, which by Theorem \ref{Thm_LATE_v_extend}, is a causal
estimand; Otherwise, when Assumption \ref{Mono} monotonicity holds, $\pi
^{DR}=\tau ^{Wald\_X}$, which we show in Lemma \ref{L_LATE_X} identifies a
weighted average of $LATE(t_{0},t_{1})$ for $(t_{0},t_{1})\in \mathcal{T}%
_{c} $. Either way, $\pi ^{DR}$ identifies a weighted average of the average
treatment effects for all the units responding to the IV\ change, the
largest subpopulation one can identify treatment effects without further
assumptions. The weights are proportional to both the magnitude of the
treatment change and the density of $X$.

Let $\mathcal{S}_{+}=\{(x,v)\in \mathcal{X}\times (0,1)\text{: }\Delta
q(x,v)>0\}$ and $\mathcal{S}_{-}=\{(x,v)\in \mathcal{X}\times (0,1)\text{: }%
\Delta q(x,v)<0\}$. Define 
\begin{equation}
\pi _{+}^{DR}\text{:=}\iint_{\mathcal{S}_{+}}\pi (x,v)w_{+}(x,v)dvdx,
\label{pi+_DR}
\end{equation}%
where $w_{+}(x,v)=\Delta q(x,v)f(x)/\iint_{\mathcal{S}_{+}}\Delta
q(x,v)f(x)dvdx$. $\pi _{+}^{DR}$ identifies a weighted average of the
average treatment effects for all the responding units with $(x,v)\in S_{+}$%
, when either monotonicity or conditional treatment rank similarity holds
for $\mathcal{S}_{+}$. $\pi _{-}^{DR}$ can be analogously defined by
replacing $w_{+}(x,v)$ with $w_{-}(x,v)$ and $\mathcal{S}_{+}$ with $%
\mathcal{S}_{-}$ in eq.\ (\ref{pi+_DR}) respectively. $\pi _{-}^{DR}$\
identifies a weighted average of the average treatment effects for units
experiencing negative treatment changes, regardless of whether they stay at
the same treatment rank or not.

\section{Estimation and Inference}

\label{SecEstInf} For estimation and inference, we focus on the general
setup with covariates. To avoid cumbersome fully nonparametric estimation,
we assume that covariates enter linearly and propose convenient
semi-parametric estimation. We briefly discuss the practical implications of
the additional functional form assumptions required in our estimation toward
the end of this section. The estimation without covariates can be seen as a
special case of that with covariates.

\subsection{Estimation}

\label{SecEst}We assume a linear quantile regression model for the
conditional $v$ quantile of $T$ given $Z=z$ and $X=x$, i.e., $q_{z}\left(
x,v\right) =a_{0}(v)+x^{\prime }a_{1}(v)+za_{2}(v)+zx^{\prime }a_{3}(v)$; we
further assume a partially linear model for the conditional mean function of 
$Y$ given $Z$, $X$ and $T$, i.e., $m_{z}(x,t)=x^{\prime
}b_{0}+g_{0}(t)+zx^{\prime }b_{1}+zg_{1}(t)$, where $g_{z}$, $z=0,1$, are
some unknown functions. Given a sample of $i.i.d.$\ observations $\{\left(
Y_{i},T_{i},X_{i},Z_{i}\right) \}_{i=1}^{n}$ for $(Y,T,X,Z)$, we propose the
following estimation procedure.

\begin{itemize}
\item[Step 1.] Estimate the first-stage conditional treatment quantiles $%
q_{z}(x,v)$:

\item $\widehat{q}_{z}\left( x,v\right) =\widehat{a}_{0}(v)+x^{\prime }%
\widehat{a}_{1}(v)+z\hat{a}_{2}(v)+zx^{\prime }\widehat{a}_{3}(v)$

for $v\in {V}^{(l)}$, where ${V}^{(l)}=\{v_{1},v_{2},...,v_{l}\}$ is the set
of equally spaced quantiles over $(0,1)$.

Then $\Delta \widehat{q}(x,v)=\hat{a}_{2}(v)+x^{\prime }\widehat{a}_{3}(v)$.


\item[Step 2.] 

Estimate the conditional mean function $m_{z}(x,t)$ by a partially linear
series estimator:

\item $\widehat{m}_{z}(x,t)=x^{\prime }\widehat{b}_{0}+\widehat{g}%
_{0}(t)+zx^{\prime }\widehat{b}_{1}+z\widehat{g}_{1}(t)$

Let $\Delta \hat{m}(X_{i},v)=\widehat{m}_{1}(X_{i},\widehat{q}_{1}(X_{i},v))-%
\widehat{m}_{0}(X_{i},\widehat{q}_{0}(X_{i},v))$.

\item[Step 3.] Assume that the trimming parameter $\varrho _{n}$ is a
positive sequence that goes to $\varrho =0$ as $n\rightarrow \infty $. For $%
v\in V^{(l)}$ and $i=1,...,n$, the plug-in estimator of $\pi (X_{i},v)$ is $%
\hat{\pi}(X_{i},v)=\Delta \hat{m}(X_{i},v)/\Delta \widehat{q}(X_{i},v)$ when 
$|{\Delta }\widehat{q}(X_{i},v)|\geq \varrho _{n}$. Let $\widehat{\pi }%
(X_{i},v)=0$ when $|{\Delta }\widehat{q}(X_{i},v)|<\varrho _{n}$.

\item Estimate $\pi (v)$ for $v\in V^{(l)}$ such that $\sum_{i}1\big(|{%
\Delta }\widehat{q}(X_{i},v)|\geq \varrho _{n}\big)\neq 0$: \newline
$\widehat{\pi }(v)=\sum_{i}\widehat{\pi }(X_{i},v)\widehat{w}_{v}\left(
X_{i}\right) $, where $\widehat{w}_{v}\left( X_{i}\right) =\frac{|{\Delta }%
\widehat{q}(X_{i},v)|1\left( |{\Delta }\widehat{q}(X_{i},v)|\geq \varrho
_{n}\right) }{\sum_{i}|{\Delta }\widehat{q}(X_{i},v)|1\left( |{\Delta }%
\widehat{q}(X_{i},v)|\geq \varrho _{n}\right) }$

\item Estimate $\pi ^{DR}$: $\widehat{\pi }^{DR}=\sum_{v\in V^{(l)}}\sum_{i}%
\widehat{\pi }(X_{i},v)\widehat{w}\left( X_{i},v\right) $, \newline
where $\widehat{w}\left( X_{i},v\right) =\frac{|{\Delta }\widehat{q}%
(X_{i},v)|1\left( |{\Delta }\widehat{q}(X_{i},v)|\geq \varrho _{n}\right) }{%
\sum_{v\in V^{(l)}}\sum_{i}|{\Delta }\widehat{q}(X_{i},v)|1\left( |{\Delta }%
\widehat{q}(X_{i},v)|\geq \varrho _{n}\right) }.$
\end{itemize}

One may estimate $\pi _{+}^{DR}$ or $\pi _{-}^{DR}$ analogously by replacing 
$|{\Delta }\widehat{q}(X_{i},v)|$ with ${\Delta }\widehat{q}(X_{i},v)$ or $-{%
\Delta }\widehat{q}(X_{i},v)$, respectively. The following provides details
on the partial linear series estimator in Step 2. Let $\{\psi _{J1},...,\psi
_{JJ}\}$ be a collection of basis functions of $t$ for approximating the
nonparametric component $g_{z}(t)$. Let $\psi ^{J}(x,t,z)=\big(x^{\prime
},\psi _{J1}(t),...,\psi _{JJ}(t),zx^{\prime },z\psi _{J1}(t),...,z\psi
_{JJ}(t)\big)^{\prime }$, a $2(d_{x}+J)\times 1$ vector. Let $\Psi = (\psi
^{J}(X_{1},T_{1},Z_{1}),...,\psi ^{J}(X_{n},T_{n},Z_{n}))^{\prime }$, a $%
n\times 2(d_{x}+J)$ matrix. Then the series coefficient estimate is $\hat{c}%
=[\Psi ^{\prime -}\Psi ^{\prime }(Y_{1},...,Y_{n})^{\prime }$, and a series
least squares estimator of $m_{z}(x,t)$ is $\hat{m}_{z}(x,t)=\psi
^{J}(x,t,z)^{\prime }\hat{c}$.

For the trimming parameter $\varrho _{n}$, one may choose $\varrho
_{n}=1.96\times \min_{v\in {V}^{(l)},\{X_{i}\}_{i=1}^{n}}$\newline
$se({\Delta }\widehat{q}(X_{i},v))/\log (n)$. This $\varrho _{n}$ satisfies
the rate condition required by our asymptotic theory as that given in
Theorems \ref{Tpiv} and \ref{TpiDR} in Section \ref{SecAsy}.%

\subsection{Asymptotic Theory}

\label{SecAsy}This section presents inference results for $\pi (v)$ and $\pi
^{DR}$. Inference results for the other parameters $\pi (x,v)$ and $\pi
_{\pm }^{DR}$ are presented in Section \ref{Secpixv} and Section \ref%
{ASecInfPf}, respectively, in the online supplementary appendix.

We derive the asymptotic theory based on the literature of quantile
regression and sieve estimation. The main complication here is that we need
to account for the variation from the Step 1 quantile regression and Step 2
sieve estimation, as well as the trimming function. Let $%
a(v)=(a_{0}(v),a_{1}^{\prime }(v),a_{2}(v),a_{3}^{\prime }(v))^{\prime }$ be
the quantile coefficients in Step 1. For the quantile regression estimator $%
\hat{a}(v)$, we apply the results of Angrist, Chernozhukov, and Fern\'{a}%
ndez-Val (2006). They show that $\hat{a}(v)$ converges uniformly over $v$ in
a closed subset of $\left( 0,1\right) $ to a zero mean Gaussian process
indexed by $v$. For the partially linear estimation in Step 2, we apply the
results of Chen and Christensen (2018).{\ They establish uniform inference
for nonlinear functionals of nonparametric IV regression.} We apply their
results for a special case of exogenous regressors and linear functionals.
Our assumptions for asymptotics collect the assumptions in these two papers.

Assumption~\ref{AQR} collects the conditions in Theorem 3 in Angrist,
Chernozhukov, and Fern\'{a}ndez-Val (2006).

\begin{customassumption}{A1}
The conditional density $f_{T|X,Z}(t|x,z)$ is bounded and uniformly
continuous in $t$, uniformly for $x\in \mathcal{X}$, $z=0,1$.
$\mathbb{E}\left[\Vert X\Vert ^{3}\right]<\infty $.
Let $\vartheta (v)$:=$\mathbb{E}[f_{T|X,Z}(S^{\prime }a(v)|X,Z)SS^{\prime }]$, where $S$:=$(1,X^{\prime },Z,ZX^{\prime })^{\prime }$, be positive
definite for all $v\in \mathcal{V}$ which is a closed subset of $(0,1)$.
\label{AQR}
\end{customassumption}Let $e=Y-\mathbb{E}\left[ Y|Z,X,T\right] $. 
Let $G=\mathbb{E}\left[ \psi ^{J}(X,T,Z)\psi ^{J}(X,T,Z)^{\prime }\right] =%
\mathbb{E}\left[ \Psi ^{\prime }\Psi /n\right] $ be positive definite for
each $J$. Let $\Omega =\mathbb{E}\left[ e^{2}\psi ^{J}(X,T,Z)\psi
^{J}(X,T,Z)^{\prime }\right] $ and $\mho =G^{-1}\Omega G^{-1}$. 

Let $L^{\infty }(T)$ denote the set of all bounded measurable functions $g:%
\mathcal{T}\rightarrow \mathcal{R}$ endowed with the sup-norm $\Vert g\Vert
_{\infty }=\sup_{t}|g(t)|$. Let $\Vert \cdot \Vert _{\ell ^{q}}$ denote the
vector $\ell ^{q}$-norm when applied to vectors and the operator norm
induced by the vector $\ell ^{q}$-norm when applied to matrices. If $%
\{a_{n}\}$ and $\{b_{n}\}$ are sequences of positive numbers, then we say $%
a_{n}\lesssim b_{n}$ if $\lim \sup_{n\rightarrow \infty }a_{n}/b_{n}<\infty $%
.

Consider a collection of linear functionals $\{L_{\ell }:\ell \in \mathcal{L}%
\}$ with an index set $\mathcal{L}$. For example, for the conditional mean
function $m_{z}(x,t)$, one can let $L_{\ell }(m_{z})=m_{z}(x,t)$ with $\ell
=(x,t)\in \mathcal{L}=\mathcal{X}\times \mathcal{T}$, for $z=0,1$.
Assumptions~\ref{ACC} and~\ref{Astep1} below collect the assumptions in Chen
and Christensen (2018).

\begin{customassumption}
{A2}

\begin{enumerate}
\item
\begin{enumerate}
\item[(i)] $(X,T)$ have compact rectangular support $\mathcal{XT}\subset
\mathcal{R}^{d_x + 1}$ and the density of $(X, T)$ is uniformly bounded away
from $0$ and $\infty$ on $\mathcal{XT}$.

\item[(ii)] For $z=0,1$, $m_z \in \mathcal{H} \subset L^\infty(X,T)$. The
sieve space for $(X,T)$ is the closed linear span $\Psi_J =
clsp\{\psi_{J1},...,\psi_{JJ}\}\subset L^2(X,T)$, and $\cup_J \Psi_J$ is
dense in $(\mathcal{H}, \|\cdot\|_{L^\infty(X,T)})$.
\end{enumerate}

\item
\begin{enumerate}
\item[(i)]
$\mathbb{E}\big[|e_i|^{2+\delta}\big] < \infty$ for some $\delta > 0$.

\item[(ii)] $\mathbb{E}\big[|e_i|^3|Z_i=z, X_i=x, T_i=t\big] <\infty$ and
$\mathbb{E}\big[e_i^2|Z_i=z, X_i = x, T_i = t\big] \in [\underline{%
\sigma}^2, \bar{\sigma}^2]$ for some finite and positive constants $(%
\underline{\sigma}^2, \bar{\sigma}^2)$, uniformly for $(x,t)\in\mathcal{XT}$%
, for $z=0,1$.
\end{enumerate}

\item
\begin{enumerate}
\item[(i)] $\Psi _{J}$ is H\"{o}lder continuous: there exist finite
constants $C\geq 0$, $\tilde{C}>0$ such that
$\Vert G_{{}}^{-1/2}\{\psi ^{J}(x,t,z)-\psi ^{J}(\tilde{x},\tilde{t}%
,z)\}\Vert _{\ell ^{2}}\lesssim J^{C}\Vert (x,t)-(\tilde{x},\tilde{t})\Vert
_{\ell ^{2}}^{\tilde{C}}$ for $t,\tilde{t}\in \mathcal{T}$, $x,\tilde{x}\in
\mathcal{X}$, $z=0,1$.

\item[(ii)] Let $\zeta $:=$\sup_{x,t,z}\Vert G^{-1/2}\psi ^{J}(x,t,z)\Vert
_{\ell ^{2}}$ satisfy $\zeta ^{2}/\sqrt{n}=O(1)$ and $\zeta ^{(2+\delta
)/\delta }$\\$\sqrt{(\log n)/n}=o(1)$.
\end{enumerate}

\item
\begin{enumerate}
\item[(i)] Let $\sigma _{n}^{2}(L_{\ell })=L_{\ell }(\psi ^{J})^{\prime
}\mho L_{\ell }(\psi ^{J})\nearrow +\infty $ as $n\rightarrow \infty $ for
each $\ell \in \mathcal{L}$. Let $\eta _{n}$ be a sequence of nonnegative
numbers such that $\eta _{n}=o(1)$. Let $\tilde{m}_{z}(x,t)=\psi
^{J}(x,t,z)^{\prime }\tilde{c}$ where $\tilde{c}=(\Psi ^{\prime }\Psi
)^{-}\Psi ^{\prime }\big(m_{Z_{1}}(X_{1},T_{1}),...,$\\$m_{Z_{n}}(X_{n},T_{n})%
\big)^{\prime }$ and $\sup_{\ell \in \mathcal{L}}\sqrt{n}|L_{\ell }(\tilde{m}%
_{z}(x,t))-L_{\ell }(m_{z}(x,t))|/\sigma _{n}(L_{\ell })=O_{p}(\eta _{n})$.

\item[(ii)] Let $u_n(L_\ell)(X_i, T_i, Z_i) = \psi^J(X_i, T_i, Z_i)^{\prime
-1}L_\ell(\psi^J)/\sigma_{n}(L_\ell)$ be the normalized sieve Riesz
representer.
Let $d_n(\ell_1, \ell_2) = \big(\mathbb{E}\big[(u_n(L_{\ell_1})(X_i, T_i,
$\\$
Z_i)-u_n(L_{\ell_2})(X_i, T_i, Z_i))^2\big]\big)^{1/2}$ be the semimetric on
$\mathcal{L}$.
Let $N(\mathcal{XT}, $\\$
d_n, \varsigma)$ be the $\varsigma$-covering number of $%
\mathcal{XT}$ with respect to $d_n$. There is a sequence of finite constant $%
c_n \gtrsim 1$ that could grow to infinity such that $1 + \int_0^\infty
\sqrt{\log N(\mathcal{XT}, d_n, \varsigma)} d\varsigma = O(c_n)$.

\item[(iii)] Let $\delta _{m,n}$ be a sequence of positive constants such
that $\Vert \hat{m}_{z}-m_{z}\Vert _{\infty }=O_{p}(\delta _{m,n})=o_{p}(1)$%
. 
Define $\delta _{V,n}$:=$\big( \zeta ^{(2+\delta )/\delta }\sqrt{(\log J)/n}%
\big) ^{\delta /(1+\delta )} +\delta _{m,n} +
$\\$
\zeta \sqrt{(\log J)/n}$. There
is a sequence of constant $r_{n}>0$ decreasing to zero slowly such that (a) $%
r_{n}c_{n}\lesssim 1$ and $\zeta J^{2}/(r_{n}^{3}\sqrt{n})=o(1)$, (b) $\zeta
\sqrt{(J\log J)/n}+\eta _{n}+\delta _{V,n}c_{n}=o(r_{n})$.
\end{enumerate}
\end{enumerate}

\label{ACC}
\end{customassumption}

\begin{customassumption}{A3}
Let $J\sqrt{(J\log J)/n}=o(1)$.
Let $B_{\infty ,\infty }^{p}$ denote the H\"{o}lder space of smoothness $p>0$
and $\Vert \cdot \Vert _{B_{\infty ,\infty }^{p}}$ denote its norm. Let $%
B_{\infty }(p,L)=\{m\in B_{\infty ,\infty }^{p}:\Vert m\Vert _{B_{\infty
,\infty }^{p}}\leq L\}$ denote a H\"{o}lder ball of smoothness $p>1$ and
radius $L\in (0,\infty )$. Let $m\in B_{\infty }(p,L)$ and $\Psi _{J}$ be
spanned by a B-spline basis of order $\gamma >p$ or a CDV wavelet basis of
regularity $\gamma >p$. \label{Astep1}
\end{customassumption}

Assumption~\ref{Astep1} ensures the uniform consistency of $\partial _{t}%
\hat{m}_{z}(x,t)=\partial \hat{m}_{z}(x,t)/\partial t$, which is used to
account for the Step 1 estimation error.

We show in Theorem~\ref{Tpiv} below that under Assumptions ~\ref{AQR}, \ref%
{ACC}, and \ref{Astep1}, the influence function of $\hat{\pi}(v)$ is given
by $R_{i}(v)/B(v)=(R_{1i}(v)+R_{2i}(v)+R_{3i}(v))/B(v)$, where $R_{1i}(v)$
captures the impact of Step 1, $R_{2i}(v)$ captures the impact of Step 2, $%
R_{3i}(v)$ is the influence function for the sample analogue estimator of $%
\hat{\pi}(v)$ (without accounting for the step 1 and step 2 estimation
errors) in Step 3, and $B(v)$ is from the normalization in the weighting
function. The exact formulas of $R_{ki}(v)$, $k=1,2,3$, are given in (\ref%
{AIFpiv}) in the online supplementary appendix. Let $\sigma _{n}^{2}(v)=%
\mathbb{E}\left[ R_{i}(v)^{2}\right] /B(v)^{2}$, which is the sieve variance
of $\sqrt{n}\hat{\pi}(v)$. Further let $\hat{\sigma}^{2}(v)$ be a uniformly
consistent estimator of $\sigma _{n}^{2}(v)$ in the sense that $\sup_{v\in 
\mathcal{V}_{\varrho }}|\sigma _{n}(v)/\hat{\sigma}(v)-1|=o_{p}(1)$ for a
closed set $\mathcal{V}_{\varrho }=\{v\in \mathcal{V}:\Pr (|\Delta
q(X,v)|>\varrho )>0\}$. For example, $\hat{\sigma}^{2}(v)$ can be estimated
by the sample analogue plug-in estimator, i.e., $\hat{\sigma}%
^{2}(v)=n^{-1}\sum_{i=1}^{n}\widehat{R}_{i}(v)^{2}/\widehat{B}(v)^{2}$,
where $\widehat{R}_{i}(v)$ and $\widehat{B}(v)$ are uniformly consistent
estimators of $R_{i}(v)$ and $B(v)$, respectively. We give the estimation
detail of $\hat{\sigma}^{2}(v)$ in Section~\ref{SecImplement} in the online
supplementary Appendix.

\begin{theorem}
Let Assumptions~\ref{AQR}, \ref{ACC}, and \ref{Astep1} hold. Let $\sqrt{n}%
(\varrho_n - \varrho) = o(1)$. Then $\sqrt{n}\big(\hat\pi(v) - \pi(v)\big)%
/\hat\sigma(v) = n^{-1/2}\sum_{i=1}^n R_i(v)/(B(v)\sigma_n(v)) + o_p(1) 
\overset{d}{\longrightarrow} \mathcal{N}(0,1)$ uniformly for $v\in \mathcal{V%
}_{\varrho}$. \label{Tpiv}
\end{theorem}

A $100(1-\alpha )\%$ confidence interval for $\pi (v)$ can be constructed as 
$\big[\hat{\pi}(v)-z_{1-\alpha }^{\ast }\hat{\sigma}(v)/\sqrt{n},\hat{\pi}%
(v)+z_{1-\alpha }^{\ast }\hat{\sigma}(v)/\sqrt{n}\big]$, where $z_{1-\alpha
}^{\ast }=\Phi ^{-1}(1-\alpha /2)$ is the $1-\alpha /2$ quantile of the
standard normal distribution, based on the asymptotically normal
approximation.

Similarly Theorem~\ref{TpiDR} shows that under Assumptions ~\ref{AQR}, \ref%
{ACC}, and \ref{Astep1}, the influence function of $\hat{\pi}^{DR}$ is given
by $R_{i}/B=(R_{1i}+R_{2i}+R_{3i})/B$. The exact formulas of $R_{ki}$, $%
k=1,2,3$, are given in (\ref{AIFpi}) in the online supplementary appendix.
Let $\sigma _{n}^{2}=\mathbb{E}\left[ R_{i}^{2}\right] /B^{2}$, which is the
sieve variance of $\sqrt{n}\hat{\pi}^{DR}$. Further let $\hat{\sigma}^{2}$
be a consistent estimator of $\sigma _{n}^{2}$ such that $|\sigma _{n}/\hat{%
\sigma}-1|=o_{p}(1)$. 

\begin{theorem}
Let Assumptions~\ref{AQR}, \ref{ACC}, and \ref{Astep1} hold. Let $\sqrt{n}%
(\varrho _{n}-\varrho )=o(1)$ and $\sqrt{n}l^{-1}=o(1)$. Then $\sqrt{n}\big(%
\hat{\pi}^{DR}-\pi ^{DR}\big)/\hat{\sigma}=n^{-1/2}\sum_{i=1}^{n}R_{i}/(B%
\sigma _{n})+o_{p}(1)\overset{d}{\longrightarrow }\mathcal{N}(0,1)$. \label%
{TpiDR}
\end{theorem}

Based on Theorem \ref{TpiDR}, a $100(1-\alpha )\%$ confidence interval for $%
\pi ^{DR}$ can be constructed as $\big[\hat{\pi}^{DR}-z_{1-\alpha }^{\ast }%
\hat{\sigma}/\sqrt{n},\hat{\pi}^{DR}+z_{1-\alpha }^{\ast }\hat{\sigma}/\sqrt{%
n}\big]$.

Note that our semiparametric estimation imposes certain functional form
assumptions. Causal interpretation of the estimated parameters require these
additional functional forms to hold. In theory, fully nonparametric
estimation and inference is possible. For example, in Step 1, one can use
the nonparametric QR series in Belloni et al. (2009), and in Step 2, one can
follow Chen and Christensen (2018) to estimate a fully nonparametric mean
regression. Our asymptotic theory for $\hat{\pi}(v)$ and $\hat{\pi}$ can
then be extended to the corresponding nonparametric estimators at the cost
of more complicated notations and stronger regularity conditions.

When monotonicity (along with other identifying assumptions) holds, $\pi
^{DR}=\tau ^{Wald\_X}$. So if the assumed semiparametric functional forms
are true or if both are non-parametrically estimated, the two estimators
converge to the same causal parameter and hence the corresponding estimates
should be similar in large samples. Seeing the estimates very different may
suggest that monotonicity does not hold (assuming other identifying
assumptions hold). When monotonicity does not hold, the usual Wald ratio
estimator, even when it is well-defined (or the denominator is not zero), is
not consistent, while our estimator can be consistent for a well-defined
causal parameter.

\section{\protect\large Extensions to a Multi-valued IV or Multiple IVs}

\label{SecExt}In this section we briefly discuss extensions of
identification, estimation and inference to the case of a multi-valued IV or
a vector of discrete IVs.\footnote{%
Mogstad et al.\ (2021) show that the LATE monotonicity may not be plausible
with multiple IVs for a binary treatment. This conclusion is generalizable
to a continuous treatment. While they seek to provide a causal
interpretation for the usual two stage least square (2SLS) estimand under a
weaker partial monotonicity condition (i.e., monotonicity holds with one IV
while holding other IVs fixed), we provide an estimand that is robust to the
failure of the LATE\ monotonicity assumption.} We first consider the basic
setup without covariates and then discuss the general setup with covariates.

Assume $T=g\left( T,\varepsilon \right) $ and $T=h\left( Z,U\right) $ as in
Section \ref{SecId}. Denote the support of $Z$ as $\mathcal{Z}=\left\{
z_{0},z_{1},...,z_{K}\right\} $. So e.g., if $Z=\left( Z_{1},Z_{2}\right) $,
where $Z_{1}\in \left\{ 0,1\right\} $ and $Z_{2}\in \left\{ 0,1\right\} $,
then one can let $z_{0}=\left( 0,0\right) $, $z_{1}=\left( 0,1\right) $, $%
z_{2}=\left( 1,0\right) $, and $z_{3}=\left( 1,1\right) $. Let $%
U_{k}=F_{T_{z_{k}}}\left( T_{z_{k}}\right) $ be the\ rank of the potential
treatment $T_{z_{k}}$ if $Z$ is exogenously set to be $z_{k}$. The observed
rank can be written as $U=\sum_{k=1}^{K}1\left( Z=z_{k}\right) U_{k}$. Let $%
T_{z_{k}}(u)$ be the $u$ quantile of the potential treatment $T_{z_{k}}$.
Further let $r_{k}=\Pr \left( Z=z_{k}\right) $, $p\left( Z\right) =\mathbb{E}%
\left[ T|Z\right] $, $p_{k}=\mathbb{E}\left[ T|Z=z_{k}\right] $, and $%
\overline{p}=\mathbb{E}\left[ T\right] $. Without loss of generality, assume
that the $K+1$ values of $Z$ is ordered such that $p_{k}\geq p_{k-1}$ for $%
k=1,...,K$, which may involve rearranging and is verifiable from the data.

We continue to use the same sets of assumptions when we consider either the
basic setup without covariates or the general setup with covariates, except
that the relevant assumptions need to be modified to accommodate the greater
support of $Z$, which is $\mathcal{Z}=\left\{ z_{0},z_{1},...,z_{K}\right\} $%
. For example, Assumption \ref{Cont_T} now requires that $T_{z_{k}}(u)$ is
strictly monotonic in $u$ for any $z_{k}\in \mathcal{Z}$, and that $%
U_{k}\sim Unif\left( 0,1\right) $ for $k=0,...,K$, and Assumption \ref{Inde}
independence now requires $Z\perp \left( U_{k},\varepsilon \right) $ for $%
k=0,...,K$. The same holds true for Assumptions \ref{Cont_T_extend} and \ref%
{Inde_extend}. Further Assumptions \ref{1st_stage} and \ref{Mono}, and \ref%
{RS}, and similarly Assumptions \ref{1st_extend} and \ref{RS_extend} need to
hold for each pair of IV values $z_{k}$ and $z_{k-1}$ for $k=1,...,K$. That
is, Assumption \ref{1st_stage} now requires that $T_{z_{k}}(u)\neq
T_{z_{k-1}}(u)$ for $k=1,...,K$ and at least some $u\in \left( 0,1\right) $.
Assumption \ref{Mono} monotonicity now states that $\Pr \left( T_{z_{k}}\geq
T_{z_{k-1}}\right) =1$, $k=1,...,K$. Assumption \ref{RS} now requires that $%
U_{k}|\varepsilon \sim U_{k-1}|\varepsilon $, $k=1,...,K$. The same holds
true for Assumption \ref{1st_extend} and Assumption \ref{RS_extend}. In
addition, Assumption \ref{CS} common support now requires $\Pr \left(
Z=z_{k}|X=x\right) \in \left( 0,1\right) $ for $k=0,...,K$ and any $x\in 
\mathcal{X}$.

Define the following estimand for each pair of the IV values $\left\{
z_{k-1},z_{k}\right\} $, $k=1,...K,$ 
\begin{equation*}
\tau _{k}(u)\text{:}\text{=}\frac{\mathbb{E}\left[ Y|Z=z_{k},U=u\right] -%
\mathbb{E}\left[ Y|Z=z_{k-1},U=u\right] }{\mathbb{E}\left[ T|Z=z_{k},U=u%
\right] -\mathbb{E}\left[ T|Z=z_{k-1},U=u\right] }
\end{equation*}%
if the denominator is not zero; otherwise, define $\tau _{k}(u)$:=$0$. Like
before, $T$\ and $U$ follow a one-to-one mapping given $Z=z_{k}$, so
conditioning on $U=u$ is the same as conditioning on $T=T_{z_{k}}(u)$.
Further given $Z\perp \left( U_{k},\varepsilon \right) $, we have $%
T_{z_{k}}(u)=q_{k}(u)$, where $q_{k}(u)=F_{T|Z}^{-1}\left( u|z_{k}\right) $
is the conditional $u$ quantile of $T$ given $Z=z_{k}$. Then $\tau _{k}(u)$
can be re-written as 
\begin{equation*}
\tau _{k}(u)=\frac{\mathbb{E}\left[ Y|Z=z_{k},T=q_{k}(u)\right] -\mathbb{E}%
\left[ Y|Z=z_{k-1},T=q_{k-1}(u)\right] }{q_{k}(u)-q_{k-1}(u)}.
\end{equation*}%
Following Theorem \ref{Thm_LATE_u}, $\tau _{k}(u)$ identifies an average
treatment effect at the $u$ quantile of treatment for units responding to
the IV change from $z_{k-1}$ to $z_{k}$.

Analogous to Proposition \ref{Prop_DR}, define a DR\ estimand for each pair
of IV\ values. In particular, let $\Delta q_{k}(u)=q_{k}(u)-q_{k-1}(u)$, $%
k=1,...,K$. The corresponding DR\ estimand is given by 
\begin{equation*}
\tau _{k}^{DR}\text{:=}\int_{0}^{1}\tau _{k}(u)w_{k}\left( u\right) du,
\end{equation*}%
where $w_{k}(u)=\frac{|\Delta q_{k}(u)|}{\int_{0}^{1}|\Delta q_{k}(u)|du}$. $%
\tau _{k}^{DR}$ identifies a weighted average of the average treatment
effect for all units that respond to the IV change from $z_{k-1}$ to $z_{k}$%
, under either monotonicity or rank similarity. Construct an aggregated DR
estimand as 
\begin{equation}
\tau ^{DR,K}\text{:=}\sum_{k=1}^{K}\lambda _{k}\tau _{k}^{DR},
\label{tau_DR_k}
\end{equation}%
where $\lambda _{k}$:=$\frac{\left( p_{k}-p_{k-1}\right)
\sum_{l=k}^{K}r_{_{l}}\left( p_{l}-\overline{p}\right) }{\sum_{k=1}^{K}%
\left( p_{k}-p_{k-1}\right) \sum_{l=k}^{K}r_{_{l}}\left( p_{l}-\overline{p}%
\right) }$. The weights $\lambda _{k}$ follow from Theorem 2 of Imbens and
Angrist (1994).

Note that $\lambda _{k}\geq 0$ and $\sum_{k=1}^{K}\lambda _{k}=1$, because
the IV values are ordered such that $p_{k}\geq p_{k-1}$. Therefore, $\tau
^{DR,K}$ is a convex combination of $\tau _{k}^{DR}$, $k=1,...,K$, and hence
has the DR property as well.\footnote{%
In theory, any convex combination of $\tau _{z_{k},z_{k-1}}^{DR}$, $%
k=1,...,K $, would have the DR\ property. Here our goal is to incorporate
the 2SLS or LATE-type estimand given by $\frac{Cov\left( Y,p\left( Z\right)
\right) }{Cov\left( T,p\left( Z\right) \right) }$ as a special case, which
leads to the particular choice of $\lambda _{k}$.} In particular, when
monotonicity holds, $\tau _{k}^{DR}$ reduces to the LATE Wald ratio $\tau
_{k}^{Wald}$:=$\frac{\mathbb{E}\left[ Y|Z=z_{k}\right] -\mathbb{E}\left[
Y|Z=z_{k-1}\right] }{\mathbb{E}\left[ T|Z=z_{k}\right] -\mathbb{E}\left[
T|Z=z_{k-1}\right] }$, and hence $\tau ^{DR,K}=\sum_{k=1}^{K}\lambda
_{k}\tau _{k}^{Wald}$. Further by Theorem 2 of Imbens and Angrist (1994), $%
\sum_{k=1}^{K}\lambda _{k}\tau _{k}^{Wald}=\frac{Cov\left( Y,p\left(
Z\right) \right) }{Cov\left( T,p\left( Z\right) \right) }$. Notice that $%
\tau _{k}^{Wald}$ in this case identifies a weighted average of LATEs for $%
Z\in \left\{ z_{k-1},z_{k}\right\} $ under monotonicity. Therefore, if
monotonicity holds, $\tau ^{DR,K}$ identifies a doubly weighted average of
LATEs, averaging over different compliers for a given pair of IV values and
over different pairs of IV values; otherwise, when rank similarity holds, $%
\tau ^{DR,K}$ identifies a doubly weighted average of the average treatment
effects at different treatment quantiles - the first averaging is over
different treatment quantiles for a given pair of IV values and the second
is over different pairs of IV values. Either way, $\tau ^{DR,K}$ identifies
a doubly weighted average of the average treatment effects for all the units
responding to IV changes.

Now consider the general setup where the IV independence and treatment rank
similarity are valid only conditional on covariates. One can incorporate
covariates as before for each pair of IV values. In particular for $%
k=1,...,K $, define the following estimand 
\begin{equation*}
\pi _{k}(x,v)\text{:=}\frac{\mathbb{E}\left[ Y|Z=z_{k},X=x,V=v\right] -%
\mathbb{E}\left[ Y|Z=z_{k-1}, X=x,V=v\right] }{\mathbb{E}\left[ T|Z=z_{k},
X=x,V=v\right] -\mathbb{E}\left[ T|Z=z_{k-1},X=x,V=v\right] }
\end{equation*}%
when the denominator is not zero; define $\pi _{k}(x,v)$:=$0$, otherwise.
Following Theorem \ref{Thm_LATE_v_extend}, $\pi _{k}\left( x,v\right) $
identifies an average treatment effect at the conditional (on $X=x$) $v$
quantile of treatment.

Further analogous to Proposition \ref{Prop_DR_extend}, define the DR
estimand for each pair of IV values, $z_{k-1}$ and $z_{k}$, as%
\begin{equation*}
\pi _{k}^{DR}\text{:=}\iint_{\left( 0,1\right) \mathcal{\times X}}\pi
_{k}(x,v)w_{k}(x,v)dvdx,
\end{equation*}%
where $w_{k}(x,v)=\frac{|\Delta q_{k}\left( x,v\right) |f(x)}{\iint_{\left(
0,1\right) \mathcal{\times X}}|\Delta q_{k}\left( x,v\right) |f(x)dvdx}$,
and $\Delta q_{k}\left( x,v\right) =q_{k}\left( x,v\right) -q_{k-1}\left(
x,v\right) $, and $q_{k}\left(x,v\right) =F_{T|Z,X}^{-1}\left( v|z_{k},
x\right) $ is the conditional $v$ quantile of $T$ given $Z=z_{k}$ and $X=x$.

Then define the aggregated\ DR\ estimand as 
\begin{equation*}
\pi ^{DR,K}\text{:=}\sum_{k=1}^{K}\lambda _{k}\pi _{k}^{DR},
\end{equation*}%
where $\lambda _{k}$ is defined as in (\ref{tau_DR_k}). When monotonicity
holds, $\pi ^{DR,K}$ identifies a doubly weighted average of LATEs;
otherwise when rank similarity holds, $\pi ^{DR,K}$ identifies a doubly
weighted average of the average treatment effects at different conditional
treatment quantiles. Note that the identified parameter in this case is
still the unconditional doubly weighted average, even though the instrument
validity holds only conditional on covariates.

One can estimate $\pi ^{DR,K}$ by $\hat{\pi}^{DR,K}=\sum_{k=1}^{K}\hat{%
\lambda}_{k}\hat{\pi}_{k}^{DR}$ given an $i.i.d.$\ sample $\{\left(
Y_{i},T_{i},X_{i},Z_{i}\right) \}_{i=1}^{n}$, where $\hat{\pi}_{k}^{DR}$ is
an estimator of $\pi _{k}^{DR}$ and $\hat{\lambda}_{k}$ is an estimator of $%
\lambda _{k}$. $\hat{\pi}_{k}^{DR}$ can be obtained similar to $\widehat{\pi 
}^{DR}$ proposed for a binary IV. $\hat{\lambda}_{k}$ can be estimated by a
simple sample analogue plug-in estimator. Let $D^{k}=1(Z=z_{k})$. One can
estimate $p_{k}=\mathbb{E}\left[ T|Z=z_{k}\right] $ by $\widehat{p}%
_{k}=\sum_{i=1}^{n}T_{i}D_{i}^{k}/\sum_{i=1}^{n}D_{i}^{k}$ for $k=0,1,...,K$%
, and estimate $\overline{p}$ by $\widehat{\overline{p}}=n^{-1}%
\sum_{i=1}^{n}T_{i}$. One can further estimate $r_{k}$ by $\widehat{r}%
_{k}=n^{-1}\sum_{i=1}^{n}D_{i}^{k}$ for $k=1,...,K$. Then the plug-in
estimator for $\lambda _{k}$ is $\hat{\lambda}_{k}=\frac{\left( \widehat{p}%
_{k}-\widehat{p}_{k-1}\right) \sum_{l=k}^{K}\widehat{r}_{_{l}}\left( 
\widehat{p}_{l}-\widehat{\overline{p}}\right) }{\sum_{k=1}^{K}\left( 
\widehat{p}_{k}-\widehat{p}_{k-1}\right) \sum_{l=k}^{K}\widehat{r}%
_{_{l}}\left( \widehat{p}_{l}-\widehat{\overline{p}}\right) }$, $k=1,...,K$

We provide the influence function for $\hat{\pi}^{DR,K}$, denoted as $R_{Ki}$%
, in eq.\ (\ref{IFm}) in the online supplementary appendix. The influence
function given in Theorem~\ref{TpiDR} is now indexed by $k$, i.e., $R_{i}/B$
defined in (\ref{AIFpi})\ is now $R_{i}^{k}/B^{k}$. Together with the
influence function of $\hat{\lambda}_{k}$, we can derive the influence
function of $\hat{\pi}^{DR,K}$. Define the sieve variance of $\sqrt{n}\hat{%
\pi}^{DR,K}$ as $\sigma _{Kn}^{2}=\mathbb{E}\left[ {R_{Ki}}^{2}\right] $.
Let $\hat{\sigma}_{K}^{2}$ be a consistent estimator of $\sigma _{Kn}^{2}$,
such that $|\sigma _{Kn}/\hat{\sigma}_{K}-1|=o_{p}(1)$. We have the
following asymptotics result for $\hat{\pi}^{DR,K}$.

\begin{theorem}
Let the conditions in Theorem~\ref{TpiDR} hold. Then $\sqrt{n}\big(\hat{\pi}%
^{DR, K}-\pi ^{DR,K}\big)/\hat{\sigma}_K$\newline
$=n^{-1/2}\sum_{i=1}^{n}R_{Ki}/\sigma_{Kn} +o_{p}(1)\overset{d}{%
\longrightarrow }\mathcal{N}(0,1)$. \label{TpiDRm}
\end{theorem}

\section{Empirical Analysis}

\label{SecEmp}In this section, we apply our doubly robust approach to
estimate the effects of night sleep on physical and psychological well-being
using data from a recent field experiment (Bessone et al. 2021). 452 adults
in Chennai, India participated in the experiment for a period of twenty
eight days. Baseline data were collected for the first eight days. Then
participants were randomized into three groups - a control group, a group
who were provided with (a) devices to improve their home-sleep environments,
and (b) information and verbal encouragement to increase their night sleep
(the Encouragement group) and a group who were provided with (a), (b) and
additional financial incentives to increase night sleep (the Encouragement +
Incentives group). The three groups were further cross-randomized with a nap
assignment that offered participants the opportunity for a daily half-hour
afternoon nap at their workplace. So all together, there are six groups -
control, encouragement, encouragement + incentives, naps, encouragement and
naps, encouragement + incentives and naps. Details on the study design can
be found in Bessone et al.\ (2021).

We use data from the first three non-nap assignment groups and take night
sleep as our treatment variable, i.e., $T=$ night sleep in hours, for two
reasons. First, night sleep is a primary form of sleep for most people.
Second, the control group has zero hours of nap, while our treatment
variable has to be absolutely continuous. For simplicity, we use as our
outcome the well-being index, a summary measure of physical and phycological
well-being, so $Y=$ well-being index.\footnote{%
Bessone et al.\ (2021) focus on the reduced-form impacts of the night sleep
and nap treatment assignments on a variety of work, well-being, cognition,
and economic preferences outcomes.}$^{,}$\footnote{%
The well-being index is constructed as a weighted average of a wide range of
standardized measures of psychological and physical well-being. Each
constituent measure is standardized by the control group's mean and standard
deviation. The weights are the inverse of the covariance matrix to ensure
that highly correlated measures receive less weights in the aggregation. The
measures of psychological well-being are happiness, sense of life
possibilities (Cantril Scale), life satisfaction, stress, and depression.
The measures of physical well-being are performance in a stationary biking
task, reported days of illness, self-reported pain, activities of daily
living, and blood pressure.} Some of the individual outcomes, like labor
supply etc.\ were recorded on a daily basis during the experimental period.
Analysis of these outcomes would require dealing with the panel structure of
the data. The well-being index is standardized by the baseline control
group's mean and standard deviation (std.\ dev.) as in Bessone et al.\
(2021), so the unit of measurement is standard deviations. Following Bessone
et al.\ (2021), our analysis controls for baseline measures of well-being
and night sleep. In a subset of analysis we additionally control for
participants' gender and age in four quartiles.

\begin{table}[tbph]
\begin{center}
{\footnotesize 
\begin{tabular}{lcccll}
\multicolumn{6}{c}{Table 1: Sample summary statistics} \\ \hline\hline
& (1) & (2) & (3) & $(2)-(1)$ & $(3)-(1)$ \\ \cline{2-6}
Baseline well-being & 0.00 (0.46) & 0.03 (0.40) & 0.09 (0.41) & 0.03 (0.07)
& 0.19 (0.07) \\ 
Baseline night sleep & 5.51 (0.90) & 5.60 (0.84) & 5.65 (0.79) & 0.09 (0.14)
& 0.14 (0.14) \\ 
Age in 1st quartile & 0.23 (0.43) & 0.25 (0.44) & 0.31 (0.47) & 0.02 (0.07)
& 0.08 (0.07) \\ 
Age in 2nd quartile & 0.27 (0.45) & 0.27 (0.45) & 0.20 (0.40) & -0.01 (0.07)
& -0.07 (0.07) \\ 
Age in 3rd quartile & 0.23 (0.43) & 0.27 (0.45) & 0.34 (0.48) & 0.03 (0.07)
& 0.10 (0.07) \\ 
Female & 0.68 (0.47) & 0.64 (0.48) & 0.64 (0.48) & -0.04 (0.08) & -0.04
(0.08) \\ 
Night sleep & 5.62 (0.80) & 5.99 (0.85) & 6.22 (0.95) & 0.37 (0.14) & 0.60
(0.14) \\ 
Well-being & -0.00 (0.41) & 0.14 (0.37) & 0.10 (0.37) & 0.15 (0.06) & 0.10
(0.06) \\ \cline{2-6}
Participants & 77 & 75 & 74 &  &  \\ \hline\hline
\multicolumn{6}{p{350pt}}{Note: Columns 1 - 3 report sample means and
standard deviations (in parentheses) of the three groups: (1) Control, (2)
Encouragement, (3) Encouragement + Incentives ; Columns 4 and 5 report the
mean differences and their standard errors.}%
\end{tabular}
\label{tab1} }
\end{center}
\end{table}

Our sample consists of 226 observations, including 77 from the control
group, 75 from the Encouragement group and 74 from the Encouragement +
Incentives group. Sample summary statistics are presented in Table 1. The
three experimental groups are well-balanced across all of the covariates.
Consistent with the results in Bessone et al.\ (2021), being assigned to
either the Encouragement group or the Encouragement + Incentives group
significantly increases night sleep on average. The increase in the
Encouragement + Incentives group is larger, as expected. Interestingly,
these simple mean comparisons show that being assigned to the Encouragement
group significantly increases well-being (by 0.15 std.\ dev.), while being
assigned to the Encouragement + Incentives group has no significant impacts
on well-being, even though it leads to a larger increase in the average
sleep time (0.60 vs.\ 0.37 hours).

Given the three experimental groups, we perform three sets of analysis. Let $%
Z_{1}$ be an indicator for whether one is assigned to the Encouragement
group, and $Z_{2}$ be an indicator for whether one is assigned to the
Encouragement + Incentives group. Define the three IV values based on the
values of $\left( Z_{1},Z_{2}\right) $, i.e., $z_{0}$:=$(0,0)$, $z_{1}$:=$%
(1,0)$ and $z_{2}$:=$(0,1)$. In the first set of analysis, we look at the
Encouragement group and the control group so that the IV is $Z=Z_{1}$. In
the second set of analysis, we look at the Encouragement + Incentives group
and the control group, so the IV is $Z=Z_{2}$. Note that these single IV
analyses condition on the other IV being zero, which is important (see,
discussion in Mogstad et al., 2021). In our third set of analysis, we use
data from all three groups, so that the IV is $Z=\left( Z_{1},Z_{2}\right) $
for $Z\in \left\{ z_{0},z_{1},z_{2}\right\} $. The first two sets of
analysis illustrate our proposed approach with a single binary IV, while the
the third analysis illustrates our extended result with a multi-valued IV.

For the first set of analysis, the monotonicity assumption requires (a)\
everyone is likely to increase their night sleep if they are assigned to the
Encouragement group instead of the control group; for the second set of
analysis, it requires (b) everyone is likely to increase their night sleep
if they are assigned to the Encouragement + Incentives group instead of the
control group. For the third set of analysis, the monotonicity assumption
requires (a) and additionally that everyone is likely to further increase
their night sleep if they are assigned to the Encouragement + Incentives
group instead of the Encouragement only group. Although we think these
conditions are plausible, they are not verifiable in principle.\footnote{%
The one-sided Kolmogorov-Smirnov (KS) test fails to reject first-order
stochastic dominance of the Encouragement group (or the Encouragement +
Incentives group) treatment over the Control group treatment at the 10\%
significance level. It also fails to reject the dominance of the
Encouragement + Incentives group treatment over the Encouragement only group
treatment at the 10\% significance level. When inspecting the empirical
treatment quantile curves for each comparison, we do not find quantile
crossing. However, as mentioned, stochastic dominance, $T_{1}(u)-T_{0}(u)%
\geq 0$ for all $u\in \left( 0,1\right) $, is a necessary but not sufficient
condition for monotonicity $\Pr \left( T_{1}-T_{0}\geq 0\right) =1$, and the
KS test is known to have low power for small samples, so we cannot conclude
that monotonicity holds in this case.} It is therefore useful to apply our
doubly robust approach. For comparison purposes, we also implement (i) the
usual linear 2SLS estimator, (ii) an estimator of $\tau ^{Wald\_X}$ in eq.\ (%
\ref{Wald_X}), where all the conditional means are assumed to be linear in
covariates and fully interacted with the relevant binary IV, as well as
(iii) a multi-valued IV extension of $\tau ^{Wald\_X}$, i.e., $\tau
^{Wald\_X,K}$:=$\sum_{k=1}^{K}\lambda _{k}\tau _{k}^{Wald\_X}$ for $K=2$,
where $\lambda _{k}$ is defined as in (\ref{tau_DR_k}) and $\tau
_{k}^{Wald\_X}$ is defined analogously to $\tau ^{Wald\_X}$ for the pair of
IV values, $z_{k-1}$ and $z_{k}$ for $k=1,2$.

\begin{table}[tbph]
\begin{center}
{\footnotesize 
\begin{tabular}{lllll}
\multicolumn{5}{c}{Table 2: Effects of per hour night sleep on well-being}
\\ \hline\hline
& 2SLS & Wald & DR & DR-2 \\ \hline
&  & \multicolumn{2}{c}{IV: Encouragement vs. Control ($Z_1$ )} &  \\ 
\cline{3-4}
(I) & 0.427 (0.195)** & 0.427 (0.236)* & 0.390 (0.225)* & 0.382 (0.220)* \\ 
(II) & 0.408 (0.187)** & 0.407 (0.236)* & 0.233 (0.128)* & 0.234 (0.127)* \\ 
&  & \multicolumn{2}{c}{IV: Incentives vs. Control ( $Z_2$)} &  \\ 
\cline{3-4}
(I) & 0.130 (0.109) & 0.131 (0.121) & 0.123 (0.133) & 0.122 (0.102) \\ 
(II) & 0.111 (0.107) & 0.111 (0.122) & 0.077 (0.103) & 0.078 (0.132) \\ 
&  & \multicolumn{2}{c}{Two IVs: $(Z_{1}, Z_{2})$} &  \\ \cline{3-4}
(I) & 0.151 (0.107) & 0.174 (0.229) & 0.165 (0.158) & 0.157 (0.156) \\ 
(II) & 0.144 (0.104) & 0.149 (0.226) & 0.099 (0.122) & 0.097 (0.121) \\ 
\hline\hline
\multicolumn{5}{p{310pt}}{Note: (I) controls for baseline measure of
well-being and that of night sleep, and (II) additionally controls for
participants gender and age in four quartiles. 2SLS - linear 2SLS estimate;
Wald - estimates of $\tau^{Wald\_X}$ in eq. (14) or a multiple IV extension
of it, where the conditional mean functions of $Y$ and $T$ are assumed to be
linear in covariates and fully interacted with IV $Z$ (see details in the
main text); DR - doubly robust IV estimates based on the estimator in
Section 4; DR-2 - doubly robust IV estimates, where the trimming parameter
is set to be 3 times the baseline value. Standard errors are in the
parenthesis. ** Significant 5$\%$; * Significant at 10$\%$.}%
\end{tabular}
\label{tab2} }
\end{center}
\end{table}

Table 2 reports estimates from the three sets of analysis in three panels.
Column 1 reports estimates by the usual linear 2SLS estimator. Column 2
reports estimates of $\tau ^{Wald\_X}$ (top and middle panels) and estimates
of $\tau ^{Wald\_X,2}$ (bottom panel).\ The linear 2SLS estimator is a
special case of the estimator of $\tau ^{Wald\_X}$ (or $\tau ^{Wald\_X,2}$).
Both estimators similarly require monotonicity for causal interpretation but
the former additionally assumes homogeneity of the instrument and treatment
effects in covariates. Column 3 reports estimates by our doubly robust
estimation proposed in Section \ref{SecEst} (top and middle panels) and
estimates by the multi-valued IV extension of the doubly robust estimation
discussed in Section \ref{SecExt} (bottom panel). Lastly, Column 4 reports
similar doubly robust estimates to those reported in Column 3. The
difference is that in Column 3 the trimming parameter $\mathfrak{\varrho }%
_{n}$ is set to be the baseline value specified in Section \ref{SecEst},
while in Column 4 $\mathfrak{\varrho }_{n}$ is set to be three times of the
baseline value. In all the doubly robust estimation, the polynomial order of
the power series of $T$ is chosen to be one, considering the relatively
small sample sizes. We report bootstrapped standard errors based on 200
bootstrap replications for estimates in Columns 2-4, since bootstrapping is
straightforward and is computationally convenient.

Note that each instrument is associated with a different group of
individuals responding to it. We found interesting treatment effect
heterogeneity across the different groups responding to the two instruments.
The local estimates using the encouragement assignment $Z_{1}$ as an IV\
range from 0.38 to 0.43 std.\ dev.\ when controlling for baseline sleep and
baseline well-being, which are significant at the $5\%$ or $10\%$ level. So
for the group that respond to the encouragement instrument, increased night
sleep has marginally significant impacts on well-being. These estimates
reduce to 0.23-0.41 std.\ dev., which are still significant at the $5\%$ or $%
10\%$ level, when additionally controlling for participants' gender and age.
In contrast, the local estimates using the encouragement + incentives
assignment $Z_{2}$ as an IV\ are smaller (yet still positive) but are not
statistically significant, meaning that for the groups that respond to the
encouragement + incentives instrument, increased night sleep does not
translate into better well-being.

\begin{table}[htbp]
\begin{center}
{\footnotesize 
\begin{tabular}{llllll}
\multicolumn{6}{c}{Table 3: Effects of per hour night sleep on well-being:}
\\ 
\multicolumn{6}{c}{breakdown of the combined two IV estimates} \\ 
\hline\hline
& (I) & (II) &  & (I) & (II) \\ \cline{2-3}\cline{5-6}
$\pi^{DR}_{1}$ & 0.399 (0.187)** & 0.233 (0.157) & $\tau_{1}^{Wald\_X}$ & 
0.440 (0.243)** & 0.404 (0.248) \\ 
$\pi^{DR}_{2}$ & -0.297 (0.189) & -0.167 (0.120) & $\tau_{2}^{Wald\_X}$ & 
-0.354 (0.357) & -0.355 (0.485) \\ 
Wt. avg. & 0.165 (0.158) & 0.099 (0.122) & Wt. avg. & 0.174 (0.229) & 0.149
(0.226) \\ \hline\hline
\multicolumn{6}{p{350pt}}{Note: (I) controls for baseline well-being and
baseline night sleep; (II) additionally controls for participants' gender
and age in four quartiles. $\pi^{DR}_{1}$ and$\tau_{1}^{Wald\_X}$ compares
the Encouragement group with Control; $\pi^{DR}_{1}$ and $\tau_{2}^{Wald\_X}$
compares the Encouragement + Incentives group with the Encouragement only
group. Wt. avg. is the weighted average of $\pi^{DR}_{1}$ and $\pi^{DR}_{1}$
(or $\tau_{1}^{Wald\_X}$ and $\tau_{2}^{Wald\_X}$ ), where the weights $%
\lambda_{1} = 0.664$ (std. err. $=$ 0.239) and $\lambda_{2} = 0.336$ (std.
err.$=$ 0.239). ** Significant 5$\%$.}%
\end{tabular}
\label{tab3} }
\end{center}
\end{table}

The estimates using the two instruments $Z_{1}$ and $Z_{2}$ jointly lie
between the estimates using each of the two instruments separately, which
are not statistically significant. Recall that the doubly robust estimator
estimates $\pi ^{DR,2}$:=$\sum_{k=1}^{2}\lambda _{k}\pi _{k}^{DR}$ and the
Wald estimator estimates $\tau ^{Wald\_X,2}$:=$\sum_{k=1}^{2}\lambda
_{k}\tau _{k}^{Wald\_X}$, where $\pi _{1}^{DR}$ and $\tau _{1}^{Wald\_X}$
utilize the IV variation from $z_{0}$ to $z_{1}$ (comparing the
Encouragement group to the control) while $\pi _{2}^{DR}$ and $\tau
_{2}^{Wald\_X}$ utilize the IV variation from $z_{1}$ to $z_{2}$ (comparing
the Encouragement + Incentives group to the Encouragement group). Table 3
reports a detailed breakdown of the joint IV estimates. Estimates of $\pi
_{1}^{DR}$ and $\tau _{1}^{Wald\_X}$ are positive, while estimates of $\pi
_{2}^{DR}$ and $\tau _{2}^{Wald\_X}$ are always negative, even though they
are not statistically significant. Consistent with the above analysis, these
results once again suggest that those who respond to the additional
financial incentives do not experience improved well-being.

Across all analysis, our doubly robust estimates are similar to the
estimates of $\tau ^{Wald\_X}$ or $\tau ^{Wald\_X,2}$ and the 2SLS
estimates. The doubly robust estimates come with slightly inflated standard
errors compared with the 2SLS estimates. The inflated standard errors
reflect partly the tradeoff between robustness and efficiency. The
similarity of the point estimates between our doubly robust estimator and
the 2SLS estimator is reassuring. In this case, the doubly robust approach
serves as a valuable tool to corroborate the usual 2SLS estimates, so that
the 2SLS estimates can be relied upon with greater confidence.

Compared with the analysis in Bessone et al.\ (2021), which focuses on
reduced-form analysis and uses different IVs jointly in one regression, we
analyze each IV separately and when using the two IVs jointly, we give a
detailed breakdown of the overall estimates. Our analysis yields the new
finding that those individuals who slept longer due to the better sleep
environment and verbal encouragement experienced improved mental and
physical well-being, while those who slept more due to the financial
incentives did not retain such benefits. This result is largely in line with
the reduced-form estimates in Bessone et al.\ (2021, see, e.g., Table III).

\section{Conclusion}

\label{SecCon} Many empirical applications feature a continuous endogenous
variable (treatment) and a binary or discrete IV. In this paper, we propose
nonparametric doubly robust identification of the causal effects of a
continuous treatment with a binary or discrete instrument.

We consider the two commonly imposed restrictions on the first-stage
instrument effect heterogeneity: the LATE-type monotonicity vs.\ treatment
rank similarity. Both assumptions can be used to identify causal effects of
treatment in non-separable models, which accommodate arbitrary treatment
effect heterogeneity and individuals sell-selection of different treatment
levels. These assumptions are not nested. Both assumptions are not
verifiable. We first show that with a continuous treatment, both can yield
weighted average effects for the units that respond to the instrument
change. In practice, it is not ideal to choose estimands based on, say, some
pre-testing results. We further develop doubly robust estimands that are
robust to failure of either one, so that one does not have to rely on
pre-testing. When the LATE-type monotonicity holds, they reduce to the
LATE-type estimands; otherwise, they continue to be valid under treatment
rank similarity. Further, when treatment rank similarity holds, we can
identify treatment effect heterogeneity at different (conditional) treatment
quantiles.{\small \ }Based on our nonparametric identification results, we
propose convenient semiparametric estimators and establish consistency and
asymptotic normality of the proposed estimators. While our primary focus is
on a binary instrument, we extend all of the identification, estimation and
asymptotic results to the case with a multi-valued IV or a vector of
discrete IVs, with or without covariates.

The usefulness of our proposed approach is illustrated in an empirical
analysis estimating the impacts of night sleep on well-being, using data
from a recent field experiment. We show that the group of individuals who
increased night sleep due to information and verbal encouragement had
improved psychological and physical well-being, while those who slept more
due to the additional financial incentives did not experience such positive
effects. In this case, our doubly robust estimation yields estimates that
are similar to the usual linear 2SLS estimates across different sets of
analysis, which further establishes the credibility of the IV/2SLS estimates.

It is worth mentioning that we seek robust identification of some
unconditional weighted average effects. When monotonicity fails, such
weighted average effects average over units experiencing positive treatment
changes and those experiencing negative treatment changes, which may not be
ideal. However, it is well-known that individual types are not identified;
therefore, point identification of (weighted) average effects separately for
each individual type is not possible without further assumptions. An
interesting direct of future research is then to develop partial
identification results.

\newpage

\renewcommand{\theequation}{S.\arabic{equation}} \renewcommand{%
\thesection}{S} \setcounter{equation}{0} \setcounter{page}{1}

\begin{center}
{\Large Online supplementary appendix for \\[0pt]
Nonparametric Doubly Robust Identification of Causal Effects of a Continuous
Treatment using Discrete Instruments}\\[5pt]
{\large Yingying Dong\ and\ Ying-Ying Lee$^{\dagger}$ \let\thefootnote\relax%
\footnotetext{$^{\dagger}$Yingying Dong and Ying-Ying Lee, Department of
Economics, University of California Irvine, yyd@uci.edu and
yingying.lee@uci.edu.} } 
\end{center}

\medskip

In this supplementary appendix, Section \ref{SecIdenProof} provides proofs
for the identification results presented in Sections \ref{SecId} and \ref%
{SecGen}. Section~\ref{Secpixv} presents the inference theory for $\pi (x,v)$%
. Section~\ref{ASecInfPf} presents the proofs of the inference results
presented in Section \ref{SecAsy}. Section~\ref{SecImplement} provides more
details on computing the standard errors.

\subsection{Proofs: Identification}

\label{SecIdenProof}

\paragraph{Proof of Lemma~\protect\ref{L_LATE}:}

By definition, 
\begin{eqnarray*}
\tau ^{Wald} &=&\frac{\mathbb{E}\left[ g\left( T_{1},\varepsilon \right) |Z=1%
\right] -\mathbb{E}\left[ g\left( T_{0},\varepsilon \right) |Z=0\right] }{%
\mathbb{E}\left[ T_{1}|Z=1\right] -\mathbb{E}\left[ T_{0}|Z=0\right] } \\
&=&\frac{\mathbb{E}\left[ g\left( T_{1},\varepsilon \right) -g\left(
T_{0},\varepsilon \right) \right] }{\mathbb{E}\left[ T_{1}-T_{0}\right] } \\
&=&\frac{\mathbb{E}\left[ \left\{ g\left( T_{1},\varepsilon \right) -g\left(
T_{0},\varepsilon \right) \right\} \cdot 1\left( T_{1}-T_{0}>0\right) \right]
}{\mathbb{E}\left[ \left\{ T_{1}-T_{0}\right\} \cdot 1\left(
T_{1}-T_{0}>0\right) \right] } \\
&=&\frac{\iint_{\mathcal{T}_{c}}\int \left\{ g\left( t_{1},e\right) -g\left(
t_{0},e\right) \right\} F_{\varepsilon |T_{0},T_{1}}\left(
de|t_{0},t_{1}\right) F_{T_{0},T_{1}}\left( dt_{0},dt_{1}\right) }{\iint_{%
\mathcal{T}_{c}}\left\{ t_{1}-t_{0}\right\} F_{T_{0},T_{1}}\left(
dt_{0},dt_{1}\right) } \\
&=&\iint_{\mathcal{T}_{c}}w_{t_{0},t_{1}}\left\{ \int \frac{g\left(
t_{1},e\right) -g\left( t_{0},e\right) }{t_{1}-t_{0}}F_{\varepsilon
|T_{0},T_{1}}\left( de|t_{0},t_{1}\right) \right\} F_{T_{0},T_{1}}\left(
dt_{0},dt_{1}\right) \\
&=&\iint_{\mathcal{T}_{c}}w_{t_{0},t_{1}}\mathbb{E}\left[ \frac{%
Y_{t_{1}}-Y_{t_{0}}}{t_{1}-t_{0}}|T_{0}=t_{0},T_{1}=t_{1}\right]
F_{T_{0},T_{1}}\left( dt_{0},dt_{1}\right) \\
&=&\iint_{\mathcal{T}_{c}}w_{t_{0},t_{1}}LATE(t_{0},t_{1})F_{T_{0},T_{1}}%
\left( dt_{0},dt_{1}\right),
\end{eqnarray*}%
where the first equality follows from the models for $Y$\ and $T$ without
covariates as specified in eq.s (\ref{Y_eq}) and (\ref{T_eq2}),
respectively, the second equality follows from Assumption \ref{Inde}, the
third equality follows from Assumption \ref{Mono}, the fourth equality
follows from the law of iterated expectations, and the fifth to the last
equalities follow from rearranging and our notation $w_{t_{0},t_{1}}=\frac{%
t_{1}-t_{0}}{\iint_{\mathcal{T}_{c}}\left( t_{1}-t_{0}\right)
F_{T_{0},T_{1}}\left( dt_{0},dt_{1}\right) }$ and $\mathcal{T}_{c}=\left\{
(t_{0},t_{1})\in \mathcal{T}_{0}\times \mathcal{T}_{1}\text{ : }%
t_{1}-t_{0}>0\right\} $. Under monotonicity, $w_{t_{0},t_{1}}\geq 0$ and $%
\iint_{\mathcal{T}_{c}}w_{t_{0},t_{1}}F_{T_{0},T_{1}}(dt_{0},dt_{1})=1$, so $%
\tau ^{Wald}$ identifies a weighted average of $LATE(t_{0},t_{1})$:=$\mathbb{%
E}\left[ \frac{Y_{t_{1}}-Y_{t_{0}}}{t_{1}-t_{0}}|T_{0}=t_{0},T_{1}=t_{1}%
\right] $ for $\left( t_{0},t_{1}\right) \in \mathcal{T}_{c}$.

Further, when $g\left( T,\varepsilon \right) $ is continuously
differentiable in $T$, 
\begin{eqnarray*}
\tau ^{Wald} &=&\frac{\mathbb{E}\left[ \int_{T_{0}}^{T_{1}}\frac{\partial
g\left( t,\varepsilon \right) }{\partial t}dt\right] }{\mathbb{E}\left[
\int_{T_{0}}^{T_{1}}1dt\right] } \\
&=&\frac{\mathbb{E}\left[ \int_{\mathcal{T}}\frac{\partial g\left(
t,\varepsilon \right) }{\partial t}1\left( T_{0}\leq t\leq T_{1}\right) dt%
\right] }{\mathbb{E}\left[ \int 1\left( T_{0}\leq t\leq T_{1}\right) dt%
\right] } \\
&=&\frac{\int_{\mathcal{T}}\mathbb{E}\left[ \frac{\partial g\left(
t,\varepsilon \right) }{\partial t}|T_{0}\leq t\leq T_{1}\right] \Pr \left(
T_{0}\leq t\leq T_{1}\right) dt}{\int \Pr \left( T_{0}\leq t\leq
T_{1}\right) dt} \\
&=&\int_{\mathcal{T}}\mathbb{E}\left[ \frac{\partial g\left( t,\varepsilon
\right) }{\partial t}\Bigg|T_{0}\leq t\leq T_{1}\right] \widetilde{w}dt,
\end{eqnarray*}%
where $\widetilde{w}=\frac{\Pr \left( T_{0}\leq t\leq T_{1}\right) }{\int_{%
\mathcal{T}}\Pr \left( T_{0}\leq t\leq T_{1}\right) dt}$, the first equality
follows from Assumption \ref{Mono} and differentiability of $g\left(
T,\varepsilon \right) $ in $T$, the second to the last equalities follow
from the law of iterated expectations and interchanging the order of
integration when standard regularity conditions hold.

\paragraph{Proof of Lemmas \protect\ref{L_CV} and \protect\ref{L_CCV}:}

By $Z\perp \left( U_{z},\varepsilon \right) $ specified in Assumption \ref%
{Inde}, $Z\perp U_{z}|\varepsilon $. That is, $U_{0}|\varepsilon \sim
U_{0}|\left( \varepsilon ,Z=0\right) $ and $U_{1}|\varepsilon \sim
U_{1}|\left( \varepsilon ,Z=1\right) $. Further by Assumption \ref{RS}, $%
U_{0}|\varepsilon \sim U_{1}|\varepsilon $. Together they imply $%
U_{0}|\left( \varepsilon ,Z=0\right) \sim U_{1}|\left( \varepsilon
,Z=1\right) $, i.e., $U|\left( \varepsilon ,Z=1\right) \sim U|\left(
\varepsilon ,Z=0\right) $, so that $U\perp Z|\varepsilon $. Further by
Assumption \ref{Inde}, $Z\perp \varepsilon $. Therefore, $Z\perp \left(
U,\varepsilon \right) $, and hence $Z\perp \varepsilon |U$. It further
implies $T\perp \varepsilon |U$, since $T=h\left( Z,U\right) $.

Replacing the above proof of Lemma \ref{L_CV} by conditioning on $X$ in each
step proves Lemma \ref{L_CCV}.

\paragraph{Proof of Theorem~\protect\ref{Thm_LATE_v_extend}:}

Similar to the derivation of Lemma \ref{L_CV}, one can show $Z\perp \epsilon
|\left( V,X\right) $ under Assumptions \ref{Inde_extend} and \ref{RS_extend}%
. In particular, Assumption \ref{Inde_extend} states $Z\perp \left(
V_{z},\epsilon \right) |X$, which implies $Z\perp V_{z}|\left( X,\epsilon
\right) $, i.e., $V_{z}|\left( X,\epsilon ,Z=z\right) \sim V_{z}|\left(
X,\epsilon \right) $, and hence $V|\left( X,\epsilon ,Z=z\right) \sim
V_{z}|\left( X,\epsilon \right) $. In addition, Assumption \ref{RS_extend}
states $V_{1}|\left( X,\epsilon \right) \sim V_{0}|\left( X,\epsilon \right) 
$. Then, $V|\left( X,\epsilon ,Z=0\right) \sim V|\left( X,\epsilon
,Z=1\right) $, i.e., $Z\perp V|\left( X,\epsilon \right) $. Further by
Assumption \ref{Inde_extend}, $Z\perp \epsilon |X$. Therefore, $Z\perp
\left( V,\epsilon \right) |X$, and hence $Z\perp \epsilon |\left( V,X\right) 
$.

Consider now the two terms in the numerator of $\pi (x,v)$: 
\begin{eqnarray*}
\mathbb{E}\left[ Y|Z=z,X=x,V=v\right] &=&\mathbb{E}\left[ G\left(
T_{z}(x,v),x,\epsilon \right) |Z=z,X=x,V=v\right] \\
&=&\mathbb{E}\left[ G\left( T_{z}(x,v),x,\epsilon \right) |X=x,V=v\right] \\
&=&\mathbb{E}\left[ Y_{T_{z}(x,v)}|X=x,V=v\right] \\
&=&\int G\left( T_{z}(x,v),x,e\right) F_{\epsilon |X,V}\left( de|x,v\right),
\end{eqnarray*}%
where the first equality follows from our models (\ref{Y_general_eq})\ and (%
\ref{T_general_eq}), the second equality follows from the condition $Z\perp
\epsilon |\left( V,X\right) $ shown above, and the third equality follows
from the definition of potential outcomes.

Consider next the two terms in the denominator of $\pi (x,v)$. By eq. (\ref%
{T_general_eq}), 
\begin{equation*}
\mathbb{E}\left[ T|Z=z,X=x,V=v\right] =T_{z}\left( x,v\right) .
\end{equation*}

Together they prove the theorem.

\paragraph{Proof of Lemma \protect\ref{L_LATE_X}:}

First notice%
\begin{eqnarray*}
&&\mathbb{E}\left[ Y|Z=1,X=x\right] -\mathbb{E}\left[ Y|Z=0,X=x\right] \\
&=&\mathbb{E}\left[ G\left( T_{1},X,\epsilon \right) |Z=1,X=x\right] -%
\mathbb{E}\left[ G\left( T_{0},X,\epsilon \right) |Z=0,X=x\right] \\
&=&\mathbb{E}\left[ G\left( T_{1},X,\epsilon \right) |X=x\right] -\mathbb{E}%
\left[ G\left( T_{0},X,\epsilon \right) |X=x\right] ,
\end{eqnarray*}%
where the first equality follows from our models of $Y$\ and $T$, equations (%
\ref{Y_general_eq})\ and (\ref{T_general_eq}), respectively, while the
second equality follows from Assumption \ref{Inde_extend}.

Consider now the numerator of $\tau ^{LATE\_X}$: 
\begin{eqnarray*}
&&\int_{\mathcal{X}}\left\{ \mathbb{E}\left[ Y|Z=1,X=x\right] -\mathbb{E}%
\left[ Y|Z=0,X=x\right] \right\} f_{X}(x) dx \\
&=&\int_{\mathcal{X}}\left\{ \mathbb{E}\left[ G\left( T_{1},X,\epsilon
\right) |X=x\right] -\mathbb{E}\left[ G\left( T_{0},X,\epsilon \right) |X=x%
\right] \right\} f_{X}(x) dx \\
&=&\mathbb{E}\left[ \left( G\left( T_{1},X,\epsilon \right) -G\left(
T_{0},X,\epsilon \right) \right) \right] \\
&=&\mathbb{E}\left[ G\left( T_{1},X,\epsilon \right) -G\left(
T_{0},X,\epsilon \right) \cdot 1\left( T_{1}-T_{0}<0\right) \right] \\
&=&\iint_{\mathcal{T}_{c}}\iint \left\{ G\left( t_{1},x,e\right) -G\left(
t_{0},x,e\right) \right\} F_{X,\epsilon |T_{0},T_{1}}\left(
dx,de|t_{0},t_{1}\right) F_{T_{0},T_{1}}\left( dt_{0},dt_{1}\right),
\end{eqnarray*}%
where the first equality follows from the derivation above, the second
equality follows from the law of total expectation, and the third equality
follows from Assumption \ref{Mono}, and the last equality follows from
iterated expectations.

Similarly, the numerator of $\tau ^{Wald\_X}$ can be derived as follows 
\begin{eqnarray*}
&&\int_{\mathcal{X}}\left\{ \mathbb{E}\left[ T_{1}|Z=1,X=x\right] -\mathbb{E}%
\left[ T_{0}|Z=0,X=x\right] \right\} f_{X}(x) dx \\
&=&\mathbb{E}\left[ \left( T_{1}-T_{0}\right) \cdot 1\left(
T_{1}-T_{0}<0\right) \right] \\
&=&\iint_{\mathcal{T}_{c}}\left\{ t_{1}-t_{0}\right\} F_{T_{0},T_{1}}\left(
dt_{0},dt_{1}\right).
\end{eqnarray*}

Therefore, 
\begin{align*}
&\tau ^{Wald\_X} \\
&=\frac{\iint_{\mathcal{T}_{c}}\iint \{ G( t_{1},x,e) -G( t_{0},x,e)\}
F_{X,\varepsilon |T_{0},T_{1}}\left( dx,de|t_{0},t_{1}\right)
F_{T_{0},T_{1}}\left( dt_{0},dt_{1}\right) }{\iint_{\mathcal{T}_{c}}\left\{
t_{1}-t_{0}\right\} F_{T_{0},T_{1}}\left( dt_{0},dt_{1}\right) } \\
&=\iint_{\mathcal{T}_{c}}w_{t_{0},t_{1}}\Big\{ \iint \frac{G( t_{1},x,e) -G(
t_{0},x,e) }{t_{1}-t_{0}}F_{X,\varepsilon |T_{0},T_{1}}(dx,de|t_{0},t_{1}) %
\Big\} F_{T_{0},T_{1}}( dt_{0},dt_{1}) \\
&=\iint_{\mathcal{T}_{c}}w_{t_{0},t_{1}}\mathbb{E}\left[ \frac{%
Y_{t_{1}}-Y_{t_{0}}}{t_{1}-t_{0}}|T_{1}=t_{1},T_{0}=t_{0}\right]
F_{T_{0},T_{1}}\left( dt_{0},dt_{1}\right) \\
&=\iint_{\mathcal{T}_{c}}w_{t_{0},t_{1}}LATE(t_{0},t_{1})
F_{T_{0},T_{1}}\left( dt_{0},dt_{1}\right).
\end{align*}

\paragraph{Proof of Proposition~\protect\ref{Prop_DR_extend}:}

When Assumption \ref{Mono} monotonicity holds, 
\begin{equation*}
\pi ^{DR}=\frac{\iint \pi (x,v)\Delta q(x,v)f_{X}(x)dvdx}{\iint \Delta
q(x,v)f_{X}(x)dvdx}.
\end{equation*}%
Plug in the expression of $\pi (x,v)$ and $\Delta q(x,v)$, and notice $%
V=V_{z}$ when $Z=z$, for $z=0,1$. The numerator of $\pi ^{DR}$ is $\int_{%
\mathcal{X}}\{\int_{0}^{1}\{\mathbb{E}\left[ Y|Z=1,X=x,V_{1}=v\right] -%
\mathbb{E}\left[ Y|Z=0,X=x,V_{0}=v\right] \}dv\}f_{X}(x)dx$. Consider the
two terms involved in the difference. For $z=0,1$, we have 
\begin{align*}
& \int_{\mathcal{X}}\left\{ \int_{0}^{1}\mathbb{E}\left[ Y|Z=z,X=x,V_{1}=v%
\right] dv\right\} f_{X}(x)dx \\
& =\int_{\mathcal{X}}\left\{ \int_{0}^{1}\mathbb{E}\left[ G(T_{z}(x,v),x,%
\epsilon )|X=x,V_{1}=v\right] dv\right\} f_{X}(x)dx \\
& =\int_{\mathcal{X}}\mathbb{E}\left[ G\left( T_{z}\left( x,V_{1}\right)
,x,\epsilon \right) |X=x\right] f_{X}(x)dx \\
& =\int_{\mathcal{X}}\mathbb{E}\left[ G\left( T_{z}\left( x,V_{1}\right)
,x,\epsilon \right) |Z=1,X=x\right] f_{X}(x)dx \\
& =\int_{\mathcal{X}}\mathbb{E}\left[ Y|Z=z,X=x\right] f_{X}(x)dx,
\end{align*}%
where the first equality follows from the models of $Y$ given by (\ref%
{Y_general_eq}) and Assumption \ref{Inde_extend}, which implies $Z\perp
\epsilon |\left( V_{z},X\right) $, the second equality follows from
averaging over the conditional distribution of $V_{z}$ given $X$, which is $%
Unif\left( 0,1\right) $ by construction, the third equality follows from
Assumption \ref{Inde_extend}, which states $Z\perp \left( V_{z},\epsilon
\right) |X$, the last equality follows from the models of $Y$ given by (\ref%
{Y_general_eq}).

Now consider the numerator of\ $\pi ^{DR}$. It is given by $\int_{\mathcal{X}%
}\{\int_{0}^{1}\{\mathbb{E}[Y|Z=1,X=x,V_{1}=v]-\mathbb{E}[Y|Z=0,X=x,V_{0}=v]%
\}dv\}f_{X}(x)dx$. Consider the two terms involved in the difference. For $%
z=0,1$, we have 
\begin{align*}
& \int_{\mathcal{X}}\left\{ \int_{0}^{1}\{\mathbb{E}\left[ T|Z=z,X=x,V_{z}=v%
\right] dv\right\} f_{X}(x)dx \\
& =\int_{\mathcal{X}}\left\{ \int_{0}^{1}\mathbb{E}\left[
T_{z}(x,v)|X=x,V_{z}=v\right] dv\right\} f_{X}(x)dx \\
& =\int_{\mathcal{X}}\mathbb{E}\left[ T_{z}\left( x,V_{z}\right) |X=x\right]
f_{X}(x)dx \\
& =\int_{\mathcal{X}}\mathbb{E}\left[ T_{z}\left( x,V_{z}\right) |Z=z,X=x%
\right] f_{X}(x)dx \\
& =\int_{\mathcal{X}}\mathbb{E}\left[ T|Z=z,X=x\right] f_{X}(x)dx,
\end{align*}%
where the first equality follows from the models of $T$ given by (\ref%
{T_general_eq}) and Assumption \ref{Inde_extend}, which implies $Z\perp
\epsilon |\left( V_{z},X\right) $, the second equality follows from
averaging over the conditional distribution of $V_{z}$ given $X$, which is $%
Unif\left( 0,1\right) $ by construction, the third equality follows from
Assumption \ref{Inde_extend}, which states $Z\perp \left( V_{z},\epsilon
\right) |X$ and further implies $Z\perp V_{z}|X$, the last equality follows
from the model of $T$ given by eq.\ (\ref{T_general_eq}).

Together we have 
\begin{eqnarray*}
\pi ^{DR} &=&\frac{\int_{\mathcal{X}}\left\{ \mathbb{E}\left[ Y|Z=1,X=x%
\right] -\mathbb{E}\left[ Y|Z=0,X=x\right] \right\} f_{X}(x)dx}{\int_{%
\mathcal{X}}\left\{ \mathbb{E}\left[ T|Z=1,X=x\right] -\mathbb{E}\left[
T|Z=0,X=x\right] \right\} f_{X}(x)dx} \\
&=&\tau ^{Wald\_X}.
\end{eqnarray*}%
Then by Lemma \ref{L_LATE_X}, $\pi ^{DR}$ identifies a weighted average of $%
LATE\left( t_{0},t_{1}\right) $ for $(t_{0},t_{1})\in \mathcal{T}_{c}$ under
Assumption \ref{Mono} monotonicity.

Otherwise, when Assumption \ref{Mono} monotonicity does not hold, but
Assumption \ref{RS_extend} conditional treatment rank similarity holds, 
\begin{align*}
\pi ^{DR}\text{:=}\iint \pi(x,v) w(x,v) dvdx,
\end{align*}%
where $w(x,v) \geq 0$ and $\iint w(x,v) dvdx=1$. So $\pi ^{DR}$ is a
weighted average of $\pi(x,v) $, the conditional average treatment effect
given $X=x$ and $V=v$, by Theorem \ref{Thm_LATE_v_extend}.

\subsection{Inference for $\protect\pi(x,v)$}

\label{Secpixv}

A $100(1-\alpha)\%$ confidence interval for $\pi(x,v)$ is constructed as $%
\big[\hat \pi(x,v) - z_{1-\alpha}^\ast \hat \sigma(x,v)/\sqrt{n}, \hat
\pi(x,v) + z_{1-\alpha}^\ast \hat \sigma(x,v)/\sqrt{n}\big]$, where the
critical value $z_{1-\alpha}^\ast$ can be $\Phi^{-1}(1-\alpha/2)$ by the
asymptotically normal approximation. 
The sieve variance estimator for $\hat\pi(x,v)$ is $\hat\sigma^2(x,v) =
\Delta \hat\psi(x,v)^{\prime }\hat\mho \Delta \hat\psi(x,v)/\Delta \hat
T(x,v)^2$, where $\Delta\hat\psi(x,v) = \psi^J(x, \hat T_1(x,v), 1) -
\psi^J(x, \hat T_0(x,v), 0)$.

\begin{theorem}
Let Assumptions~\ref{AQR}-\ref{Astep1} hold. Then $\sqrt{n}(\hat\pi(x,v) -
\pi(x,v))/\hat\sigma(x,v) $\newline
$\overset{d}{\longrightarrow}\mathcal{N}(0,1)$ uniformly for $(x,v)\in \Uppi%
_\varrho = \{(x,v)\in \mathcal{X}\times\mathcal{V}: |\Delta T(x,v)| \geq
\varrho\}$. \label{TPL}
\end{theorem}

For the uniform confidence interval over $(x,v)\in \Uppi_\varrho$, the
critical value $z_{1-\alpha}^\ast$ is simulated from the bootstrap sieve $t$%
-statistic $\mathbb{Z}_n^\ast(x,v)$ for $(x,v) \in \Uppi_\varrho$: Let $%
\varpi_1,..., \varpi_n$ be i.i.d. random variables independent of the data
with mean zero, unit variance, and finite third moment, e.g., $\mathcal{N}%
(0,1)$. Let 
\begin{align*}
\mathbb{Z}^\ast_{n}(x,v) = \frac{\Delta\hat\psi(x,v)^{\prime }\hat G^{-1} }{%
\Delta\hat T(x,v)\hat\sigma(x,v)\sqrt{n}} \sum_{i=1}^n \psi^J(x, T, Z_i)
\hat e_i\varpi_i.
\end{align*}
Calculate $\mathbb{Z}_n^\ast(x,v)$ for a large number of independent draws
of $\varpi_1,..., \varpi_n$. Then the critical value $z_{1-\alpha}^\ast$ is
the $(1-\alpha)$ quantile of $\sup_{(x,v)\in \Uppi_\varrho}|\mathbb{Z}%
_n^\ast(x,v)|$ over the draws. Theorem 4.1 in Chen and Christensen (2018)
implies the result on the consistency of the sieve score bootstrap. $%
\sup_{s\in\mathcal{R}} \Big|
\mathbb{P}\Big( \sup_{(x,v)\in \Uppi_\varrho} |\sqrt{n}\big(\hat\pi(x,v) -
\pi(x,v)\big)/\hat\sigma(x,v) | \leq s \Big) - \mathbb{P}^\ast\Big( %
\sup_{(x,v)\in \Uppi_\varrho} | \mathbb{Z}_n^\ast(x,v) | \leq s \Big) \Big| %
= o_p(1), $ 
where $\mathbb{P}^\ast$ denotes a probability measure conditional on the
data $\{Y_i, T_i, X_i, Z_i\}_{i=1}^n$.

\subsection{Proofs: Estimation and Inference}

\label{ASecInfPf} The proofs use the results in Angrist, Chernozhukov, and
Ferna\'{n}dez-Val (2006) (ACF, henceforth) and Chen and Christensen (2018)
(CC, henceforth). To simplify exposition, we collect notations used in the
proofs below. We suppress the subscripts $i, z$ and dependence on $v$, when
there is no confusion.

\paragraph{Notation:}

\begin{align*}
\phi_i(v) &= \vartheta(v)^{-1}\big( 1(T_i \leq S_{i}^{\prime }a(v)) - v %
\big) S_{i} \\
S_{1i} &= (1, X_i^{\prime }, 1, X_i^{\prime })^{\prime }, S_{0i} = (1,
X_i^{\prime }, 0, {\mathbf{0}}_{(d_x\times 1)}^{\prime })^{\prime }, \Delta
S_i = S_{1i} - S_{0i} \\
\partial_t m_z(X, q_{z}(X,v)) &= \frac{\partial}{\partial t} m_z(X,
t)|_{t=q_z(X,v)} \\[3pt]
q_{zi} &= q_z(X_i, v), \hat q_{zi} = \hat q_z(X_i, v) \\
\Delta q_i &=\Delta q(X_i,v)= q_{1i} - q_{0i} = (S_{1i} - S_{0i})^{\prime
}a(v) = \Delta S_i^{\prime }a(v) \\
\Delta \hat q_i &=\Delta \hat q(X_i,v)= \hat q_{1i} - \hat q_{0i} = (S_{1i}
- S_{0i})^{\prime }\hat a(v) = \Delta S_i^{\prime }\hat a(v) \\[3pt]
\Delta\psi_i &= \Delta\psi(X_i,v) = \psi^J(X_i, q_1(X_i,v), 1) - \psi^J(x,
q_0(X_i,v), 0) \\
\Delta\hat\psi_i &= \Delta\hat\psi(X_i,v) = \psi^J(X_i, \hat q_1(X_i,v), 1)
- \psi^J(X_i, \hat q_0(X_i,v), 0) \\[3pt]
\Delta m_i &= \Delta m(X_i, v) = m_1(X_i, q_1(X_i,v)) - m_0(X_i, q_0(X_i,v))
\\
\Delta\hat m_i &= \Delta\hat m(X_i,v) = \hat m_1(X_i, \hat q_1(X_i,v)) -
\hat m_0(X_i, \hat q_0(X_i,v)) = \Delta\hat\psi_i^{\prime }\hat c \\
\Delta\check m_i &= \Delta\check m(X_i,v) = \hat m_1(X_i, q_1(X_i,v)) - \hat
m_0(X_i, q_0(X_i,v)) = \Delta\psi_i^{\prime }\hat c \\[3pt]
\chi_i &= \chi(X_i, v) = 1(|\Delta q(X_i, v)| \geq \varrho|) \\
\chi_i^\pm &= \chi^\pm(X_i, v) = 1(\pm\Delta q(X_i, v) \geq \varrho|)
\end{align*}

Lemma~\ref{Ltrim} is for estimating the trimming function.

\begin{lemma}
Let Assumption~\ref{AQR} hold. Let $\sqrt{n}(\varrho_n-\varrho) = o(1)$ and $%
\sqrt{n}l^{-1} = o(1)$. Then

\begin{enumerate}
\item 
\begin{align*}
&\frac{1}{\sqrt{n}}\sum_{i=1}^n \frac{1}{l} \sum_{v\in V^{(l)}} \Delta
m(X_i, v) \left(\hat\chi^+(X_i,v) - \chi^+(X_i,v) \right) \\
&=\frac{1}{\sqrt{n}}\sum_{i=1}^n \int_0^1 \frac{\partial}{\partial \alpha} 
\mathbb{E}\left[ \Delta m(X,v)1(\Delta S^{\prime }\alpha\geq \varrho) \right]%
^{\prime }\big|_{\alpha= a(v)} \phi_i(v) dv + o_p(1).
\end{align*}

\item 
\begin{align*}
&\frac{1}{\sqrt{n}}\sum_{i=1}^n \frac{1}{l} \sum_{v\in V^{(l)}} \Delta
q(X_i, v) \left(\hat\chi^+(X_i,v) - \chi^+(X_i,v) \right) \\
&=\frac{1}{\sqrt{n}}\sum_{i=1}^n \int_0^1 \frac{\partial}{\partial \alpha} 
\mathbb{E}\left[\Delta q(X,v)1(\Delta S^{\prime }\alpha\geq \varrho)\right]%
^{\prime }\big|_{\alpha= a(v)} \phi_i(v) dv + o_p(1).
\end{align*}
\end{enumerate}

\label{Ltrim}
\end{lemma}

Step 1 is $O_{p}(n^{-1/2})$, so the estimation error of $\chi $ is of first
order asymptotically by Lemma~\ref{Ltrim}. The rate condition on $\sqrt{n}%
(\varrho _{n}-\varrho )=o(1)$ means that using $\varrho _{n}$ rather than $%
\varrho $ is first-order asymptotically ignorable.

Lemma~\ref{Lint} is for the approximation error from the numerical
integration.

\begin{lemma}
Let a function $f(x,v)$ be of bounded variation in $v\in\mathcal{V}$,
uniformly in $x \in\mathcal{X}$. Then 
\begin{align*}
\sup_{x\in\mathcal{X}} \big|l^{-1}\sum_{v\in V^{(l)}} f(x,v) 1(\Delta q(x,v)
> \varrho) - \int_0^1 f(x,v) 1(\Delta q(x,v) > \varrho) dv\big| = O(l^{-1}).
\end{align*}
\label{Lint}
\end{lemma}

The inference theory for ${\pi}(v)$ follows analogously to that of ${\pi}%
^{DR}$, but without integrating over $v$. Therefore we first present the
proof of Theorem~\ref{TpiDR} for $\pi^{DR}$.

\paragraph{Proof of Theorem~\protect\ref{TpiDR}:}

Define $A_+$ and $A_-$ as $A_\pm = \int_0^1\int_\mathcal{X} \Delta m(x,v)
\chi^\pm(x,v) $\newline
$f_X(x) dx dv$. So $A = A_+ - A_- = 
\int_0^1\int_\mathcal{X} \Delta m(x,v)/\Delta q(x,v)\big(
\Delta q(x,v) 1(\Delta q(x,v) \geq \varrho) - \Delta q(x,v) 1(\Delta q(x,v)
\leq -\varrho) \big) f_X(x)dx dv = \int_0^1\int_\mathcal{X} \pi(x,v) |\Delta
q(x,v)| 1(|\Delta q(x,v)| $\newline
$\geq \varrho) f_X(x) dx dv. $

Define $B_+$ and $B_-$ as $B_\pm = \int_0^1\int_\mathcal{X} \Delta q(x,v)
\chi^\pm(x,v)f_X(x)dx dv$. By a similar argument as $A$, we can show that $B
= B_+- B_-$. Therefore, $\pi^{DR} = A/B$ and $\pi^{DR}_\pm = A_\pm/B_\pm$.
Linearize $\hat\pi^{DR}-\pi^{DR} = (\hat A- A)/B - (\hat B-B)\pi/B +
O_p\left(|\hat A-A| |\hat B - B|/B^2 + |\hat B - B|^2/B^2\right) $.

The proof is focused on $\hat A_+$, the estimator of $A_+$. The same
arguments apply to $\hat B_+$, the estimator of $B_+$. The same arguments
apply to $\hat\pi_-^{DR}$ and hence $\hat\pi^{DR}$.

Write $\hat\pi^{DR}_+ = \hat A_+/\hat B_+$, where 
\begin{align*}
\hat A_+ &= \frac{1}{n}\sum_{i=1}^n \frac{1}{l}\sum_{v\in V^{(l)}} \Delta
\hat m(X_i, v ) \hat\chi^+(X_i,v), \\
\hat B_+ &= \frac{1}{n}\sum_{i=1}^n \frac{1}{l}\sum_{v\in V^{(l)}} \Delta
\hat q(X_i, v ) \hat\chi^+(X_i,v).
\end{align*}

In the following, we suppress the subscripts of $+$ and superscripts of $DR$
for expositional simplicity. Linearize $\hat\pi-\pi = (\hat A- A)/B - (\hat
B-B)\pi/B + O_p\left(|\hat A-A| |\hat B - B|/B^2 + |\hat B - B|^2/B^2
\right) $. 

Let $\tilde A = n^{-1}\sum_{i=1}^n l^{-1}\sum_{v\in V^{(l)}} \Delta \hat
m(X_i, v ) \chi(X_i,v)$ for a known trimming function. Decompose $\hat A - A
= \hat A - \tilde A + \tilde A - A$. The estimation error in $\Delta \hat m$%
. 
\begin{align}
\tilde A - A &= \frac{1}{n}\sum_{i=1}^n \frac{1}{l}\sum_{v\in V^{(l)}}
\left( \Delta \hat m(X_i, v ) - \Delta m(X_i, v)\right) \chi(X_i,v)
\label{Atilde1} \\
&\ \ \ + \frac{1}{n}\sum_{i=1}^n \frac{1}{l}\sum_{v\in V^{(l)}} \Delta
m(X_i, v ) \chi(X_i,v) - A.  \label{Atilde2}
\end{align}
By Lemma~\ref{Lint} and assuming $\sqrt{n}l^{-1} = o(1)$, (\ref{Atilde2}) is 
$n^{-1}\sum_{i=1}^n R_{A3i} + o_p(n^{-1/2})$, where $R_{A3i} = \int_0^1
\Delta m(X_i, v ) \chi^+(X_i,v) dv - A_+$.


We focus on (\ref{Atilde1}) next. Decompose $\Delta\hat m_i - \Delta m_i = %
\big(\Delta\hat m_i - \Delta\check m_i \big)+ \big(\Delta\check m_i - \Delta
m_i\big)$. The first part is for Step 1 estimation error, and the second
part is for Step 2 estimation error.

\paragraph{Step 1}

Theorem 3 in ACF shows that $\hat a(v) - a(v) = n^{-1}\sum_{i=1}^n
\phi_i(v)+ o_p(n^{-1/2})$ uniformly over $v\in\mathcal{V}$ and converges in
distribution to a zero mean Gaussian process indexed by $v$. Decompose 
\begin{align*}
&\Delta\hat m_i - \Delta\check m_i \\
&= m_1(X_i, \hat q_{1i}) - m_1(X_i, q_{1i}) - (m_0(X_i, \hat q_{0i}) -
m_0(X_i, q_{0i})) + so1 \\
&= \partial_t m_1(X_i, q_{1i}) (\hat q_{1i} - q_{1i}) - \partial_t m_0(X_i,
q_{0i}) (\hat q_{0i} - q_{0i}) + so1 + so2 \\
&= \partial_t m_1(X_i, q_{1i}) S_{1i}(\hat a(v) - a(v)) - \partial_t
m_0(X_i, q_{0i}) S_{0i}(\hat a(v) - a(v)) + so1 + so2,
\end{align*}
where (We suppress the subscript $i$ for simplicity.) 
\begin{align*}
so1 &= \hat m_1(\hat q_1) - m_1(\hat q_1) - \big(\hat m_0(\hat q_0) -
m_0(\hat q_0)\big) - \big( \hat m_1(q_1) - m_1(q_1)\big) \\
&\ \ \ + \big(\hat m_0(q_0) - m_0(q_0)\big) \\
&= O_p\left( \|\partial_t\hat m _z- \partial_tm_z\|_\infty\|\hat q_z -
q_z\|_\infty \right), \\
so2 &= O_p\left( \partial_t^2 m_1 \big(\hat q_1 - q_1\big)^2 + \partial_t^2
m_0 \big(\hat q_0 - q_0\big)^2 \right) = O_p(\|\hat q_z - q_z\|_\infty^2),
\end{align*}
as $\partial_t^2 m_z$ is uniformly bounded by Assumption~\ref{Astep1}. ACF
and Corollary 3.1(ii) in CC implies that $so1 + so2 = O_p(\|\hat q_z
-q_z\|_\infty\|\partial_t \hat m_z-\partial_t m_z\|_\infty + \|\hat q_z -
q_z\|_\infty^2) = O_p(n^{-1/2} (J^{-(p-1)} + J\sqrt{(J \log J)/n} )+ n^{-1})
= o_p(n^{-1/2})$ uniformly over $v\in\mathcal{V}$, by assuming $J\sqrt{(J
\log J)/n}=o(1)$ and $p>1$.

Then 
\begin{align*}
&\sqrt{n} \frac{1}{n}\sum_{i=1}^n \frac{1}{l}\sum_{v \in V^{(l)}}
(\Delta\hat m_i - \Delta\check m_i )\chi_i \\
&= \frac{1}{n}\sum_{i=1}^n \frac{1}{l}\sum_{v \in V^{(l)}} (\partial_t
m_1(X_i, q_{1i}) S_{1i}^{\prime }\chi_i \sqrt{n}(\hat a(v) - a(v)) \\
&\ \ \ - \partial_t m_0(X_i, q_{0i}) S_{0i}^{\prime }\chi_i \sqrt{n}(\hat
a(v) - a(v)) + o_p(1) \\
&= \frac{1}{l}\sum_{v \in V^{(l)}} \mathbb{E}\left[\left(\partial_t m_1(X_i,
q_{1i}) S_{1i} - \partial_t m_0(X_i, q_{0i}) S_{0i} \right)\chi_i\right]%
^{\prime }\sqrt{n}(\hat a(v) - a(v)) + o_p(1) \\
&= \frac{1}{\sqrt{n}}\sum_{j=1}^n \frac{1}{l}\sum_{v \in V^{(l)}} \mathbb{E}%
\left[\left(\partial_t m_1(X_i, q_{1i}) S_{1i} - \partial_t m_0(X_i, q_{0i})
S_{0i} \right)\chi_i\right]^{\prime }\phi_j(v) + o_p(1) \\
&= \frac{1}{\sqrt{n}}\sum_{j=1}^n \int_0^1 \mathbb{E}\left[\left(\partial_t
m_1(X_i, q_{1i}) S_{1i} - \partial_t m_0(X_i, q_{0i}) S_{0i} \right)\chi_i%
\right]^{\prime }\phi_j(v) dv + o_p(1),
\end{align*}
where the third equality is by ACF, and the last equality is by Lemma~\ref%
{Lint} and $\sqrt{n}l^{-1} = o(1)$.

For the second equality, let $\mathcal{F}=\big\{1(\Delta S_{i}^{\prime
}a>\varrho ),a\in \mathcal{B}\big\}$ that is a VC subgraph class and hence a
bounded Donsker class. Then $\mathcal{F}( \partial
_{t}m_{1}(X_{i},S_{1i}^{\prime }a)S_{1i}-\partial
_{t}m_{0}(X_{i},S_{0i}^{\prime }a)S_{0i}) $ is also bounded Donsker with a
square-integrable envelop $2\sup_{z,x,t}|\partial _{t}m_{z}(x,t)|\max_{j\in
\{1,2,...,d_{x}\}}|X_{j}|$ by Theorem 2.10.6 in Van der Vaart and Wellner
(1996). So $n^{-1}\sum_{i=1}^{n}(\partial
_{t}m_{1}(X_{i},q_{1i})S_{1i}-\partial _{t}m_{0}(X_{i},q_{0i})S_{0i})\chi
_{i}= \mathbb{E}[( \partial _{t}m_{1}(X_{i},q_{1i})S_{1i}-\partial
_{t}m_{0}(X_{i},q_{0i})S_{0i}) \chi _{i}] +o_{p}(n^{-1/2})$ uniformly in $%
v\in \mathcal{V}$.

\paragraph{Step 2}

We show the stochastic equicontinuity, $n^{-1}\sum_{i=1}^n(\Delta\check
m_i-\Delta m_i)\chi_i = \mathbb{E}\left[(\Delta\check m_i-\Delta m_i)\chi_i%
\right] + so3 $, where $so3 = o_p(n^{-1/2})$ uniformly in $v\in\mathcal{V}$.

Let $\Delta \tilde m_i = \Delta \psi_i^{\prime }\tilde c$ and $\Delta \check
m_i = \Delta \psi_i^{\prime }\hat c$. Then decompose $so3 = so31 + so32$ to
the ``standard deviation" term $so31$ and the ``bias" term $so32$, 
\begin{align}
so3 &= \frac{1}{n}\sum_{i=1}^n\chi_i \left(\Delta \check m_i - \Delta \tilde
m_i\right) - \int_\mathcal{X} \chi_i\left(\Delta \check m_i- \Delta \tilde
m_i\right) F_X(dX_i)  \tag{so31}  \label{SEvar} \\
& \ \ \ + \frac{1}{n}\sum_{i=1}^n \chi_i\left(\Delta \tilde m_i - \Delta
m_i\right) - \int_\mathcal{X}\chi_i \left(\Delta \tilde m_i - \Delta
m_i\right) F_X(dX_i).  \tag{so32}  \label{SEbias}
\end{align}

Let $\sqrt{n}so31 = Q_J^{\prime }(\hat c - \tilde c)$, where $Q_J = \sqrt{n}(%
\frac{1}{n}\sum_{i=1}^n \chi_i\Delta \psi_i - \int_\mathcal{X} \chi_i\Delta
\psi_i F_X(dX_i))$. By $var(Q_J) = \mathbb{E}[\chi_i\Delta
\psi_i\Delta\psi_i^{\prime }]$ and the Jensen's inequality, $\mathbb{E}%
[\|Q_J\|] \leq $\newline
$O(\sqrt{\mathbb{E}[\|\Delta\psi_i\|^2]} ) 
= O(\zeta )$. As given in the proof of Lemma 3.1 in CC, $\|\hat c - \tilde
c\|_{\ell^\infty} = O_p(\sqrt{\log J/(n\lambda_{min}(G))})$, where the
minimum eigenvalue $\lambda_{min}(G) > 0$.\footnote{%
By Lemma A.1 in CC, $s_{JK}^{-1}\asymp \pi_J =1$ for the exogenous case.}
Then $\mathbb{E}[|\ref{SEvar}|] = O(n^{-1/2}\zeta \sqrt{\log
J/(n\lambda_{min}(G))})$ by the Cauchy-Schwartz inequality. The Markov's
inequality implies $so31 = O_p(n^{-1}\zeta \sqrt{\log J/\lambda_{min}(G)}) =
o_p(n^{-1/2})$ implied by Assumption~\ref{ACC}.5.

$var(\sqrt{n} so32) = O(\mathbb{E}\left[\chi_i \left(\Delta\tilde m_i -
\Delta m_i\right)^2\right]) = O( \|m - \Pi_J m\|_\infty^2 )$, where $\Pi_J m
= \arg\min_{h\in\Psi_J}\|m - h\|_{L^2(X,T,Z)}$, by Theorem 3.1 (i) in CC. 
The Markov's inequality yields $so32=O_p(n^{-1/2}\|m - \Pi_J m\|_\infty ) =
O_p(n^{-1/2}J^{-p} ) = o_p(1)$ by the results in the proof of Corollary 3.1
in CC.

\medskip

By Lemma~\ref{Lint} and assuming $\sqrt{n}l^{-1} = o(1)$, $l^{-1}\sum_{v \in
V^{(l)}}\mathbb{E}\left[(\Delta\check m_i-\Delta m_i)\chi_i\right] =
\int_0^1 \mathbb{E}\left[\Delta\check m_i \chi_i\right] dv - A +
o_p(n^{-1/2})$.

Note that $A$ is based on a linear functional of $m$, $L(m) = \int_0^1 \int_%
\mathcal{X} m_z(x, q_z(x,v))$\newline
$1(\Delta q(x,v) > \varrho)F_X(dx) dv$. So we use the results on linear
functionals of a sieve estimator in CC. 
Let $\sigma_{A2n}^2 = \mathbb{E}\left[R_{A2i}^2\right]$, where $R_{A2i} = 
\mathcal{D}^{+\prime} G^{-1}\psi^J(X_i, T_i, Z_i) e_i$ and $\mathcal{D}^{+}
= \int_{0}^{1}\mathbb{E}\left[\Delta\psi^{J}(X,v)\chi ^{+}(X,v)\right]dv$,
with a consistent estimator $\hat \sigma_{A2}^2$. Lemma 4.1 in CC provides 
\begin{align*}
\left|\frac{\sqrt{n}}{\hat \sigma_{A2}} \left(\int_0^1 \mathbb{E}\left[%
\Delta\check m_i \chi_i\right] dv-A\right) - \frac{1}{\sigma_{A2n} \sqrt{n}}
\sum_{i=1}^n R_{A2i} \right| = o_p(1).
\end{align*}

The estimation error from the trimming function $\hat{A}-\tilde{A}%
=n^{-1}\sum_{i=1}^{n}l^{-1} $\newline
$\sum_{v\in V^{(l)}}\Delta m(X_{i},v)\big(\hat{\chi}(X_{i},v)-\chi (X_{i},v)%
\big)+o_{p}(1)$ by $n^{-1}\sum_{i=1}^{n}l^{-1}\sum_{v\in V^{(l)}} $\newline
$\big(\Delta \hat{m}(X_{i},v)-\Delta m(X_{i},v)\big)\big(\hat{\chi}%
(X_{i},v)-\chi (X_{i},v)\big)=O_{p}\Big(\Vert \Delta \hat{m}-\Delta m\Vert
_{\infty }\Vert \hat{q}_z-q_z\Vert _{\infty }\Big)=o_{p}(n^{-1/2})$.
Together with Lemma~\ref{Ltrim}(i), $\big|\sqrt{n}(\hat{A}%
-A)-n^{-1/2}\sum_{i=1}^{n}R_{Ai}\big|=o_{p}(1)$, where $%
R_{Ai}=R_{A1i}+R_{A2i}+R_{A3i}$ with 
\begin{align*}
R_{A1i}& =\int_{0}^{1}\bigg(\mathbb{E}\left[ \left( \partial
_{t}m_{1}(X,q_{1})S_{1}-\partial _{t}m_{0}(X,q_{0})S_{0}\right) \chi
^{+}(X,v)\right] \\
& \ \ \ +\frac{\partial }{\partial \alpha }\mathbb{E}\left[ \Delta
m(X,v)1(\Delta S^{\prime }\alpha \geq \varrho )\right] \big|_{\alpha =a(v)}%
\bigg)^{\prime }\phi _{i}(v)dv,
\end{align*}

By the similar arguments as for $A$ in (\ref{Atilde1}) and (\ref{Atilde2}), 
\begin{align}
\tilde B - B &= \frac{1}{n}\sum_{i=1}^n \frac{1}{l}\sum_{v\in V^{(l)}}
\left( \Delta \hat q(X_i, v ) - \Delta q(X_i, v)\right) \chi(X_i,v)
\label{Btilde1} \\
&\ \ \ + \frac{1}{n}\sum_{i=1}^n \frac{1}{l}\sum_{v\in V^{(l)}} \Delta
q(X_i, v ) \chi(X_i,v) - B.  \label{Btilde2}
\end{align}
By Lemma~\ref{Lint}, (\ref{Btilde2}) is $n^{-1}\sum_{i=1}^n \int_0^1 \Delta
q(X_i, v ) \chi(X_i,v) dv - B + o_p(n^{-1/2})$. 
(\ref{Btilde1}) is 
\begin{align*}
&\frac{1}{n}\sum_{i=1}^n \frac{1}{l}\sum_{v\in V^{(l)}} \Delta S_i^{\prime
}\left( \hat a(v) - a(v)\right) \chi_i \\
&= \frac{1}{n}\sum_{j=1}^n \frac{1}{l}\sum_{v\in V^{(l)}} \frac{1}{n}%
\sum_{i=1}^n \chi_i \Delta S_i^{\prime }\phi_j(v) + o_p(n^{-1/2}) \\
&= \frac{1}{n}\sum_{j=1}^n \frac{1}{l}\sum_{v\in V^{(l)}} \mathbb{E}\left[
\chi_i \Delta S_i\right]^{\prime }\phi_j(v) + o_p(n^{-1/2})
\end{align*}
\begin{align*}
&= \frac{1}{n}\sum_{j=1}^n \int_0^1 \mathbb{E}\left[ \Delta S_i^{\prime
}\chi_i \right]\phi_j(v) dv + o_p(n^{-1/2}),
\end{align*}
where the first equality by ACF, and the third equality by Lemma~\ref{Lint}.
For the second equality, let $\mathcal{F} = \big\{1(\Delta S_i^{\prime }a >
\varrho), a \in\mathcal{B} \big\}$ that is a VC subgraph class and hence a
bounded Donsker class. Then $\mathcal{F}\Delta S$ is Donsker with a
square-integrable envelop $\max_{j \in \{1,2,...,d_x\}} |X_j|$ by Theorem
2.10.6 in Van der Vaart and Wellner (1996). So $n^{-1}\sum_{i=1}^n
\chi_i\Delta S_i - \mathbb{E}\left[ \chi_i\Delta S_i\right] = o_p(1)$
uniformly over $v \in \mathcal{V}$

Together with Lemma~\ref{Ltrim}(ii), we obtain $\big| \sqrt{n}(\hat B - B) -
n^{-1/2}\sum_{i=1}^n R_{Bi} \big| = o_p(1)$, where $R_{Bi} = R_{B1i} +
R_{B3i}$ with 
\begin{align*}
R_{B1i} &= \int_0^1 \left( \mathbb{E}\left[ \Delta S^{\prime +}(X,v) \right]
+ \frac{\partial}{\partial \alpha} \mathbb{E}\left[ \Delta q(X,v)1(\Delta
S^{\prime }\alpha\geq \varrho) \right]^{\prime }\big|_{\alpha= a(v)}
\right)\phi_i(v) dv \\
R_{B3i} &= \int_0^1 \Delta q(X_i, v ) \chi^+(X_i,v) dv - B.
\end{align*}

By a linearization for $\hat\pi_+^{DR}$, $\hat\pi_+^{DR} - \pi_+^{DR} = 
\frac{\hat A_+}{\hat B_+} - \frac{A_+}{B_+} = \frac{\hat A_+ - A_+}{B_+} - 
\frac{\pi_+^{DR}}{B_+}(\hat B_+-B_+) + o_p(n^{-1/2})$. Therefore, we define $%
R_{i}^+ = R_{Ai} - \pi_+^{DR} R_{Bi} = R_{1i}^+ + R_{2i}^+ + R_{3i}^+$,
where $R_{1i}^+ = R_{A1i} - \pi_+^{DR} R_{B1i}$, $R_{2i}^+ = R_{A2i}$, and $%
R_{3i}^+ = R_{A3i} - \pi_+^{DR} R_{B3i}$. That is, 
\begin{align*}
R_{1i}^{+}& =\int_{0}^{1}\bigg(\mathbb{E}\left[ \left( \partial
_{t}m_{1}(X,q_{1}(X,v))S_{1}-\partial _{t}m_{0}(X,q_{0}(X,v))S_{0}-\pi
_{+}^{DR}\Delta S\right) \chi ^{+}(X,v)\right] \\
& \ \ \ +\frac{\partial }{\partial \alpha }\mathbb{E}\left[ \left( \Delta
m(X,v)-\pi _{+}^{DR}\Delta q(X,v)\right) 1(\Delta S^{\prime }\alpha \geq
\varrho )\right] \big|_{\alpha =a(v)}\bigg)^{\prime }\phi _{i}(v)dv, \\
R_{2i}^{+}& =\mathcal{D}^{+^{\prime }}G^{-1}\psi
^{J}(X_{i},T_{i},Z_{i})e_{i},\mbox{with}\ \mathcal{D}^{+} = \int_{0}^{1}%
\mathbb{E}\left[\Delta\psi^{J}(X,v)\chi ^{+}(X,v)\right]dv, \\
R_{3i}^{+}& =\int_{0}^{1}\left( \Delta m(X_{i},v)-\pi _{+}^{DR}\Delta
q(X_{i},v)\right) \chi ^{+}(X_{i},v)dv.
\end{align*}
Then we obtain $\hat\pi_+^{DR} - \pi_+^{DR} = n^{-1}\sum_{i=1}^n \big( %
R_{Ai}- \pi_+^{DR}R_{Bi}\big)/B_+ + o_p(n^{-1/2}) = n^{-1}\sum_{i=1}^n
R_i^+/B_+ + o_p(n^{-1/2})$.

\paragraph{Asymptotic normality}

We suppress the subscripts of $+$ and superscripts of $DR$ for expositional
simplicity. Because $R_{2i}$ depends on $(Y_i, T_i, X_i)$, $R_{1i}$ depends
on $(T_i, X_i)$, and $R_{3i}$ depends on $X_i$, the law of iterated
expectations yields $\sigma_n^2 = \big(\mathbb{E}\big[R_{1i}^2\big] + 
\mathbb{E}\big[R_{2i}^2\big] + \mathbb{E}\big[R_{3i}^2\big]\big)/B^2 =
(\sigma_1^2+ \sigma_{2n}^2+ \sigma_3^2)/B^2$.

We will show the Bahadur representation that 
\begin{align}
&\left| \frac{\sqrt{n}(\hat \pi - \pi)}{\hat \sigma} - \frac{1}{\sqrt{n}}%
\sum_{i=1}^n \frac{R_i}{B\sigma_n} \right|  \notag \\
&\leq \left| \frac{\sqrt{n}(\hat \pi - \pi)}{\sigma_n} - \frac{1}{\sqrt{n}}%
\sum_{i=1}^n \frac{R_i}{B\sigma_n} \right| + \left| \frac{\sqrt{n}(\hat \pi
- \pi)}{\sigma_n} \left(\frac{\sigma_n}{\hat\sigma}-1\right) \right| = o_p(1)
\label{ALRso0}
\end{align}
by (i) $n^{-1/2}\sum_{i=1}^n R_i/(B\sigma_n)\overset{d}{\rightarrow}\mathcal{%
N}(0,1)$, and (ii) $|\sigma_n/\hat\sigma -1| = o_p(1)$, as shown below.

\paragraph{(i)}

Asymptotic normality will follow from the Lyapunov central limit theorem
with the third absolute moment, $n^{-1/2}\mathbb{E}[|R_i|^3]/(B\sigma_n)^3
\rightarrow 0$, since $\{R_i\}_{i=1}^n$ are independent across $i$, with
mean zero and variance 1. By the assumed conditions, it is straightforward
to show that $n^{-1/2}\mathbb{E}[|R_{1i}|^3]/(B\sigma_1)^{3} \rightarrow 0$.
We show below that $n^{-1/2}\mathbb{E}[|R_{2i}|^3]/(B\sigma_{2n})^{3}
\rightarrow 0$. Then it implies that all the cross-product terms $n^{-1/2}%
\mathbb{E}[|R_{1i} R_{2i}R_{3i}|]/(B\sigma_n)^{3} \rightarrow 0$ and $%
n^{-1/2}\mathbb{E}[|R_{ji}^2R_{ki}|]/(B\sigma_n)^{3} \rightarrow 0$ for $%
j,k=1,2,3$, $j\neq k$.

Denote as $\psi_i = \psi^J(X_i, T_i, Z_i)$. By Assumption~\ref{ACC}.2(ii), 
\begin{align}
\sigma_{2n}^2 &= \mathbb{E}\left[R_{2i}^2\right]/B^2 = \mathbb{E}\left[ (%
\mathcal{D}^{\prime} G^{-1} \psi_i)^2 e_i^2\right]/B^2  \notag \\
&\geq \mathbb{E}\left[ (\mathcal{D}^{\prime} G^{-1} \psi_i)^2 \right]%
\underline{\sigma}^2/B^2 = \mathcal{D}^{\prime} G^{-1}\mathcal{D}\underline{%
\sigma}^2/B^2.  \label{20.52}
\end{align}
By the Schwarz inequality, (\ref{20.52}), and Assumption~\ref{ACC}.3(ii), 
\begin{align}
\frac{(\mathcal{D}^{\prime} G^{-1} \psi_i)^2}{\sigma_{2n}^2} \leq \frac{(%
\mathcal{D}^{\prime} G^{-1} \mathcal{D}^{\prime })(\psi_i^{\prime} G^{-1}
\psi_i)}{\sigma_{2n}^2} \leq \frac{\zeta^2}{\underline{\sigma}^2}.
\label{20.521}
\end{align}
Then by (\ref{20.52}), (\ref{20.521}), and Assumption~\ref{ACC}.2(iii), 
\begin{align}
\frac{1}{\sqrt{n}}\mathbb{E}\left[\frac{|R_{2i}|^3}{B^3\sigma_{2n}^{3}}%
\right] &= \frac{1}{\sqrt{n}}\mathbb{E}\left[\frac{|\mathcal{D}^{\prime}
G^{-1} \psi_i e_i|^3}{B^3\sigma_{2n}^{3}}\right]  \notag \\
&= \frac{1}{\sqrt{n}}\mathbb{E}\left[\frac{(\mathcal{D}^{\prime} G^{-1}
\psi_i)^2 }{B^3\sigma_{2n}^2}\frac{|\mathcal{D}^{\prime} G^{-1} \psi_i| }{%
\sigma_{2n}} \mathbb{E}\left[|e_i|^3|X_i,T_i,Z_i\right] \right]  \notag \\
&\leq \frac{\zeta}{\sqrt{n}B^3\underline{\sigma}^3} \sup_{x,t,z} \mathbb{E}%
\left[|e_i|^3|X_i=x, T_i=t, Z_i=z\right] = O\left(\frac{\zeta}{\sqrt{n}}%
\right) = o(1).  \notag
\end{align}

\paragraph{(ii)}

It is straightforward that $\hat\sigma_1^2 = n^{-1}\sum_{i=1}^n \hat
R_{1i}^2/\hat B^2 \overset{p}{\longrightarrow} \sigma_1^2 = \mathbb{E} \left[%
R_{1i}^2\right]/B^2$ and $\hat\sigma_3^2 \overset{p}{\longrightarrow}
\sigma_3^2$. The same arguments in Lemma G.4 in CC give $|\sigma_{2n}/\hat%
\sigma_{2} -1 | = O_p(\delta_{V,n}) = o_p(1)$. So $|\sigma_n/\hat\sigma -1|
= o_p(1)$.

\medskip

By (i) that $n^{-1/2}\sum_{i=1}^n R_{i}/(B\sigma_n) = O_p(1)$ and (ii), the
second term $\Big|
\frac{\sqrt{n}(\hat \pi - \pi)}{\hat\sigma} \Big(\frac{\hat\sigma}{\sigma_n}%
-1\Big)
\Big| = O_p(1)o_p(1) = o_p(1)$. We then obtain the Bahadur representation.
The asymptotic normality follows from the result~(i).

\medskip

Therefore, we obtain that when $B_{+}>0$, $\sqrt{n}\big(\hat{\pi}%
_{+}^{DR}-\pi _{+}^{DR}\big)/\hat{\sigma}_{n+}=n^{-1/2}%
\sum_{i=1}^{n}R_{i}^{+}$\newline
$/(B_+\sigma _{n+})+o_{p}(1)\overset{d}{\longrightarrow }\mathcal{N}(0,1)$,
where $\hat{\sigma}_+^2$ is a consistent estimator of $\sigma_{n+}^{2}=%
\mathbb{E}\left[R_{i}^{+2}\right]/B_{+}^{2}$.

\bigskip

For $\pi^{DR}_-$, define 
\begin{align*}
R_{1i}^{-}& =\int_{0}^{1}\bigg(\mathbb{E}\left[ \left( \partial
_{t}m_{1}(X,q_{1}(X,v))S_{1}-\partial _{t}m_{0}(X,q_{0}(X,v))S_{0}-\pi
_{-}^{DR}\Delta S\right) \chi^{-}(X,v)\right] \\
& \ \ \ +\frac{\partial }{\partial \alpha }\mathbb{E}\left[ \left( \Delta
m(X,v)-\pi _{+}^{DR}\Delta q(X,v)\right) 1(\Delta S^{\prime }\alpha \leq -
\varrho )\right] \big|_{\alpha =a(v)}\bigg)^{\prime }\phi _{i}(v)dv.
\end{align*}
Define $R^-_{i}$ as $R^+_i$ by replacing $+$ with $-$ in all the components
in $R^+_i$. By the same arguments for $\pi^{DR}_+$, we obtain that when $%
B_{-} >0$, $\sqrt{n}\big(\hat{\pi}_{-}^{DR}-\pi _{-}^{DR}\big)/\hat{\sigma}%
_{-}=n^{-1/2}\sum_{i=1}^{n}R_{i}^{-}/(B_-\sigma _{n-})+o_{p}(1)\overset{d}{%
\longrightarrow }\mathcal{N}(0,1)$, where $\hat{\sigma}_-^2$ is a consistent
estimator of $\sigma_{n-}^{2}=\mathbb{E}\left[R_{i}^{-2}\right]/B_{-}^{2}$,
such that $|\sigma_{n-}/\hat\sigma_- - 1| = o_p(1)$.

\bigskip

For $\pi^{DR}$, the same linearization yields $\hat\pi^{DR}-\pi^{DR} = (\hat
A- A)/B - (\hat B-B)\pi^{DR}/B + O_p\big(|\hat A-A| |\hat B - B|/B^2 + |\hat
B - B|^2/B^2\big) $. Let $R_i = R_i^+ - R_i^- = R_{1i} + R_{2i} + R_{3i}$,
where $R_{li}=R_{li}^+ - R_{li}^-$ for $l=1,2,3$ by replacing $\pi^{DR}_+$
and $\pi^{DR}_-$ with $\pi^{DR}$. Specifically, let $\text{sgn}(x,v) =
1(\Delta q(x,v) \geq \varrho) - 1(\Delta q(x,v) \leq -\varrho)$, 
\begin{align}
R_{1i} & =\int_{0}^{1}\Big(\mathbb{E}\big[ ( \partial
_{t}m_{1}(X,q_{1}(X,v))S_{1}-\partial
_{t}m_{0}(X,q_{0}(X,v))S_{0}-\pi^{DR}\Delta S) \text{sgn}(X,v)\big]  \notag
\\
& \ \ +\frac{\partial }{\partial \alpha }\mathbb{E}[ \left( \Delta
m(X,v)-\pi ^{DR}\Delta q(X,v)\right) (1(\Delta S^{\prime }\alpha \geq
\varrho )  \notag \\
&\ \ - 1(\Delta S^{\prime }\alpha \leq -\varrho ) )] \big|_{\alpha =a(v)}%
\Big)^{\prime }\phi _{i}(v)dv,  \notag \\
&\mbox{with}\ \phi_i(v) = \vartheta(v)^{-1}\big( 1(T_i \leq S_{i}^{\prime
}a(v)) - v \big) S_{i},  \notag \\
&\ \ \ \ \ \ \ S_{1i} = (1, X_i^{\prime }, 1, X_i^{\prime })^{\prime },
S_{0i} = (1, X_i^{\prime }, 0, {\mathbf{0}}_{(d_x\times 1)}^{\prime
})^{\prime}, \Delta S_i = S_{1i} - S_{0i},  \notag \\
R_{2i} &= \mathcal{D}^{\prime}G^{-1}\psi ^{J}(X_{i},T_{i},Z_{i})e_{i}, 
\notag \\
&\mbox{with}\ \mathcal{D}=\int_{0}^{1}\mathbb{E}\left[(\psi
^{J}(X,q_{1}(X,v),1) - \psi ^{J}(X,q_{0}(X,v),0)) \text{sgn}(X,v)\right]dv. 
\notag \\
R_{3i}&= \int_{0}^{1}\left( \Delta m(X_{i},v)-\pi^{DR}\Delta
q(X_{i},v)\right) \text{sgn}(X_{i},v)dv,  \notag \\
B&= \int_{0}^{1}\int_{\mathcal{X}}|\Delta q(x,v)| 1(|\Delta q(x,v)| \geq
\varrho) f(x)dxdv.  \label{AIFpi}
\end{align}

\paragraph{Proof of Theorem~\protect\ref{Tpiv}:}


The proof follows exactly the same arguments in the proof of Theorem~\ref%
{TpiDR} and Lemma~\ref{Ltrim} by removing all ``$\int_0^1\cdots dv$" and ``$%
l^{-1}\sum_{v \in V^{(l)}}$". We can derive the influence function of $%
\hat\pi(v)$ to be $R_{i}(v)/B(v)$ defined as the influence function of $\hat{%
\pi}^{DR}$ given in (\ref{AIFpi}) by removing all $\int_0^1\cdots dv$.
Specifically, as $\pi^{DR}$, define $\pi_+(v)$ over units experiencing
positive changes for $v\in \mathcal{V}_{+\varrho} = \{v\in\mathcal{V}:
P(\Delta q(X,v) \geq \varrho) > 0\}$. Define $B_+(v) = \int_{\mathcal{X}}
\Delta q(x,v)\chi^+(x, v) f(x)dx$, so $B_+ = \int_0^1 B_+(v) dv$. The
influence function of $\hat\pi_+(v)$ is $R^+_{i}(v)/B_+(v) = (R^+_{1i}(v) +
R^+_{2i}(v) + R^+_{3i}(v))/B_+(v)$, where 
\begin{align}
R_{1i}^+(v) &= \bigg( \frac{\partial}{\partial \alpha} \mathbb{E}\left[
\left(\Delta m(X,v) -\pi_+(v) \Delta q(X,v) \right) 1(\Delta S^{\prime
}\alpha \geq \varrho) \right]\big|_{\alpha= a(v)}  \notag \\
&\ \ + \mathbb{E}\big[(\partial_t m_1(X, q_{1}(X,v))S_{1} - \partial_t
m_0(X, q_{0}(X,v)) S_{0}  \notag \\
&\ \ - \pi_+(v)\Delta S)\chi^+(X,v)\big] \bigg)^{\prime }\phi_i(v),  \notag
\\
R_{2i}^+(v) &= \mathcal{D}^{+^{\prime }}(v) G^{-1}\psi^J(X_i, T_i, Z_i) e_i, %
\mbox{with}\ \mathcal{D}^+(v) = \mathbb{E}\left[\Delta\psi^J(X, v)
\chi^+(X,v)\right],  \notag \\
R_{3i}^+(v) &= \left(\Delta m(X_i, v ) - \pi_+(v) \Delta q(X_i, v )
\right)\chi^+(X_i,v).  \label{AIFpiv}
\end{align}

Similarly consider $\pi_-(v)$ over units experiencing negative changes for $%
v\in \mathcal{V}_{-\varrho} = \{v\in\mathcal{V}: P(-\Delta q(X,v) \geq
\varrho) > 0\}$. Let $B(v)=B_+(v) - B_-(v)$, where $B_-(v) = \int_{\mathcal{X%
}} \Delta q(x,v)\chi^-(x, v) f(x)dx$. Let $R_i(v) = R_i^+(v) - R_i^-(v)$,
and the influence function of $\hat\pi(v)$ is $R_{i}(v)/B(v)$.

Define $\sigma^2(v) = \mathbb{E}\left[R_i(v)^2\right]/B(v)^2$. The unknown
elements are estimated following the same procedure as $\hat\pi^{DR}$ by
removing ``$l^{-1}\sum_{v\in V^{(l)}}$." For example, $\hat{\mathcal{D}}%
^+(v) = n^{-1}\sum_{i=1}^n \Delta\hat\psi_i\hat\chi^+(X_i,v)$.

\paragraph{Proof of Theorem~\protect\ref{TPL}:}

We first show that the estimation error of $\hat q_z(x,v)$ in Step 1 is of
smaller order than the estimation error in Step 2, i.e., the first-order
asymptotic distribution of $\hat \pi(x,v)$ is as if $q_z(x,v)$ was known.
Under Assumption~\ref{AQR}, Theorem 3 in ACF implies that $\sup_{(x,v)\in%
\mathcal{X} \times \mathcal{V}}|\hat q_z(x,v) - q_z(x,v)| = O_p(n^{-1/2})$.
The Step 2 series least squares estimator converges at a nonparametric rate
shower than $\sqrt{n}$. 
Therefore the first-order asymptotic distribution of $\hat\pi(x,v)$ is
dominated by Step~2 $\Delta\check m(x,v)$.

\paragraph{Step 1}

When $T_{zi}$ is observed, i.e., there is no Step 1 estimation error, define 
$\check\pi(x,v) = \Delta\check m(x,v)/\Delta q(x,v)$. Decompose $%
\hat\pi(x,v)-\check\pi(x,v) = \frac{\Delta\hat m}{\Delta \hat q} -\frac{%
\Delta\check m}{\Delta q} = \Big(
\frac{\Delta\hat m}{\Delta \hat q} -\frac{\Delta\check m}{\Delta \hat q}%
\Big)
+ \Big( \frac{\Delta\check m}{\Delta \hat q} -\frac{\Delta\check m}{\Delta q}%
\Big)$. The second part is for Step 1 in the denominator: 
$\frac{\Delta\check m}{\Delta \hat q} -\frac{\Delta\check m}{\Delta q} = 
\frac{\Delta m}{\Delta q^2}(\Delta q - \Delta \hat q) + so1$. The first part
is for Step 1 in the argument in the numerator, 
\begin{align*}
&\frac{\Delta\hat m}{\Delta \hat q} -\frac{\Delta\check m}{\Delta \hat q} \\
&= \frac{1}{\Delta q}(\Delta\hat m - \Delta\check m) + so2 \\
&= \frac{1}{\Delta q}(m_1(x, \hat q_{1}) - m_1(x, q_{1}) - (m_0(x, \hat
q_{0}) - m_0(x, q_{0}))) + so2 + so3 \\
&= \frac{1}{\Delta q}(\partial_t m_1(x, q_{1}) (\hat q_{1} - q_{1}) -
\partial_t m_0(x, q_{0}) (\hat q_{0} - q_{0}) ) + so2 + so3 + so4,
\end{align*}
where 
\begin{align*}
so1 &= \frac{\Delta\check m}{\Delta\hat q \Delta q} (\Delta q - \Delta \hat
q) - \frac{\Delta m}{\Delta q^2} (\Delta q - \Delta \hat q) = (\Delta q -
\Delta \hat q)\frac{1}{\Delta q}\left( \frac{\Delta \check m}{\Delta \hat q}
- \frac{\Delta m}{\Delta q}\right), \\
so2 &= \Delta\hat m\left(\frac{1}{\Delta\hat q} - \frac{1}{\Delta q}\right)
+ \Delta\check m\left(\frac{1}{\Delta q} - \frac{1}{\Delta \hat q}\right) =
\left( \Delta\hat m - \Delta \check m\right)\left(\frac{1}{\Delta\hat q} - 
\frac{1}{\Delta q}\right), \\
so3 &= \frac{1}{\Delta q}\Big\{ \hat m_1(x, \hat q_1) - m_1(x, \hat q_1) - %
\big(\hat m_0(x, \hat q_0) - m_0(x, \hat q_0)\big) - \big( \hat m_1(x, q_1)
\\
&\ \ \ - m_1(x, q_1)\big) + \big(\hat m_0(x, q_0) - m_0(x, q_0)\big) \Big\}
\\
&= O_p\left( (\partial_t\hat m_1(x, q_1) - \partial_tm_1(x, q_1))(\hat q_1 -
q_1) \right), \\
so4 &= O_p\left( \partial_t^2 m_1 \big(\hat q_1 - q_1\big)^2 + \partial_t^2
m_0 \big(\hat q_0 - q_0\big)^2 \right) = O_p(\|\hat q_z - q_z\|_\infty^2).
\end{align*}
Thus $so1 + so2 + so3 + so4 = O_p(\|\hat T - T\|_\infty^2 + \|\hat T
-T\|_\infty\|\partial_t \check m-\partial_t m\|_\infty) = O_p( n^{-1} +
n^{-1/2} (J^{-(p-1)} + J\sqrt{(J4 \log J)/n}) ) = o_p(n^{-1/2})$ uniformly
over $(x,v)\in \Uppi_\varrho$, by Corollary 3.1(ii) in CC and assuming $J%
\sqrt{(J \log J)/n}=o(1)$ and $p>1$. Therefore, 
\begin{align}
&\sqrt{n}\big(\hat\pi(x,v)-\check\pi(x,v) \big)  \notag \\
&= \sqrt{n}\left\{ \frac{\Delta m}{\Delta q^2}(\Delta q - \Delta \hat q) + 
\frac{1}{\Delta q}(\partial_t m_1(x, q_{1}) (\hat q_{1} - q_{1}) -
\partial_t m_0(x, q_{0}) (\hat q_{0} - q_{0}) )\right\}  \notag \\
&\ \ + o_p(1)  \notag \\
&= \left\{ -\frac{\pi(x,v)}{\Delta q}(S_{1} - S_{0}) + \frac{1}{\Delta q}%
(\partial_t m_1(x, q_{1}) S_{1} - \partial_t m_0(x, q_{0}) S_{0})
\right\}^{\prime }\sqrt{n}(\hat a(v)-a(v))  \notag \\
&\ \ + o_p(1)  \notag \\
&= \left\{ -\frac{\pi(x,v)}{\Delta q}\Delta S + \frac{1}{\Delta q}%
(\partial_t m_1(x, q_{1}) S_{1} - \partial_t m_0(x, q_{0}) S_{0})
\right\}^{\prime }\frac{1}{\sqrt{n}}\sum_{j=1}^n \phi_j(v) + o_p(1)
\label{Estep1}
\end{align}
by Theorem 3 in ACF and $\|\hat\pi-\check\pi\|_\infty = O_p(n^{-1/2})$. 

\paragraph{Step 2}

Define $\mathcal{Z}_n \sim \mathcal{N}(0, \mho)$, $\sigma_n^2(x,v) =
\Delta\psi(x,v)^{\prime}\mho \Delta\psi(x,v)/\Delta q(x,v)^2$, and 
\begin{align*}
\mathbb{Z}_n^\pi(x,v) = \frac{\Delta\psi(x,v)^{\prime }}{\Delta
q(x,v)\sigma_n(x,v)} \mathcal{Z}_n.
\end{align*}
Lemma 4.1 in CC provides uniform Bahadur representation and uniform Gaussian
process strong approximation 
%
\begin{align*}
&\sup_{(x,v)\in \Uppi_\varrho}\left| \frac{\sqrt{n}\left(\hat\pi(x,v) -
\pi(x,v)\right)}{\hat\sigma(x,v)} - \mathbb{Z}_n^\pi(x,v) \right| = o_p(1).
\end{align*}

\paragraph{Proof of Lemma~\protect\ref{Ltrim}:}

Since $\Delta S_{i}^{\prime }=(0,\mathbf{0}_{(d_{x}\times 1)}^{\prime
},1,X_{i}^{\prime })^{\prime }$, let $\Delta S_{i}^{\prime }a-\varrho
=\Delta S_{i}^{\prime }\beta $, where $\beta =(a_{0}(v),a_{1}^{\prime }(v),
a_{2}(v)-\varrho ,a_{3}^{\prime }(v))^{\prime }$. Let $\hat{\beta}=(\hat{a}%
_{0}(v),\hat{a}_{1}^{\prime }(v),\hat{a}_{2}(v)-\varrho _{n},\hat{a}%
_{3}^{\prime }(v))^{\prime }$.

We show that $(v,\beta )\mapsto \mathbb{G}_{n}[\Delta m_{i}\chi _{i}]=\sqrt{n%
}\sum_{i=1}^{n}\big(\Delta m_{i}\chi _{i}-\mathbb{E}\left[ \Delta m_{i}\chi
_{i}\right] \big)$ is stochastic equicontinuous over $\mathcal{V}\times 
\mathcal{B}$, with respect to the $L_{2}(P)$ pseudometric $\rho
((v_{1},\beta _{1}),(v_{2},\beta _{2}))^{2}=\mathbb{E}\big[\big(\Delta
m(X_{i},v_{1})(1(\Delta S_{i}^{\prime }\beta _{1}\geq 0)-\Delta
m(X_{i},v_{2})1(\Delta S_{i}^{\prime }\beta _{2}\geq 0)\big)^{2}\big]$.

Following the proof of Theorem 3 in Section A.1.2 in the appendix of ACF,
let $\mathcal{F} =\big\{1(\Delta S_i^{\prime }\beta > 0), \beta \in\mathcal{B%
} \big\}$ that is a VC subgraph class and hence a bounded Donsker class. $%
\mathcal{F}\Delta m(X, v)$ is Donsker with a square-integrable envelop $%
|\Delta m(X, v)|$ by Theorem 2.10.6 in Van der Vaart and Wellner (1996).

By stochastic equicontinuity of $(v,\beta) \mapsto \mathbb{G}_n[\Delta m_i
\chi_i]$, $n^{-1/2}\sum_{i=1}^n\Delta m_i\big(\hat \chi_i - \chi_i\big)
=\sqrt{n}\mathbb{E}\big[ \Delta m_i\big(\hat \chi_i - \chi_i\big)
\big] + o_{p^*}(1) = \frac{\partial}{\partial \alpha} \mathbb{E}\big[
\Delta m(X_i,v) 1(\Delta S_i^{\prime }\alpha \geq 0) \big]^{\prime }\big|%
_{\alpha= \beta(v)} \times \sqrt{n}(\hat \beta(v) - \beta(v)) + o_{p^*}(1)$
uniformly over $v\in\mathcal{V}$, which follows from $\|\hat \beta(v) -
\beta(v)\| = o_{p^*}(1)$, and resulting convergence with respect to the
pseudometric $\sup_{v\in\mathcal{V}}\rho((v,\hat \beta(v)), (v,\beta(v)))^2
= o_p(1)$. The latter is from $\rho((v,\beta), (v,B))^2 = \mathbb{E}\big[%
\Delta m(X_i, v)^2 (1(\Delta S_i^{\prime }\beta \geq 0) - 1(\Delta
S_i^{\prime 2}\big]
= O\big(
\frac{\partial}{\partial \alpha}\mathbb{E}\big[\Delta m(X_i, v)^2 1(\Delta
S_i^{\prime }\alpha\geq 0) \big]\big|_{\alpha= \beta}^{\prime } $\newline
$(B- \beta) \big)$ for $\beta, B\in\mathcal{B}$, which we show below.

We can rewrite $1(\Delta S_i^{\prime }\beta \geq 0) - 1(\Delta S_i^{\prime
}B\geq 0) = 1(\Delta S_i^{\prime }\beta \geq 0, \Delta S_i^{\prime }B< 0) -
1(\Delta S_i^{\prime }\beta < 0, \Delta S_i^{\prime }B\geq 0)$, and hence $%
\big(1(\Delta S_i^{\prime }\beta \geq 0) - 1(\Delta S_i^{\prime }B\geq 0) %
\big)^2 = 1(\Delta S_i^{\prime }\beta \geq 0, \Delta S_i^{\prime }B< 0) +
1(\Delta S_i^{\prime }\beta < 0, \Delta S_i^{\prime }B\geq 0)$. By symmetry,
we focus on the second term. We can write $1( \Delta S_i^{\prime }\beta < 0,
\Delta S_i^{\prime }B\geq 0) = (1(\Delta S_i^{\prime }B\geq 0) -1(\Delta
S_i^{\prime }\beta \geq 0) ) 1(\Delta S_i^{\prime }(B-\beta) \geq 0)$. Then $%
\mathbb{E}\big[\Delta m(X_i, v)^2 (1(\Delta S_i^{\prime }B\geq 0) -1(\Delta
S_i^{\prime }\beta \geq 0) ) 1(\Delta S_i^{\prime }(B-\beta) \geq 0) \big] %
\leq \mathbb{E}\big[\Delta m(X_i, v)^2 (1(\Delta S_i^{\prime }B\geq 0)
-1(\Delta S_i^{\prime }\beta \geq 0) ) \big]
= \frac{\partial}{\partial \alpha}\mathbb{E}\big[\Delta m(X_i, v)^2 1(\Delta
S_i^{\prime }\alpha\geq 0) \big]\big|_{\alpha=\bar \beta}^{\prime }(B-
\beta), $ where $\bar \beta$ is between $\beta$ and $B$ by the mean value
theorem.

$n^{-1/2}\sum_{i=1}^{n}l^{-1}\sum_{v\in V^{(l)}}\Delta m_{i}\big(\hat{\chi}%
_{i}-\chi _{i}\big)=l^{-1}\sum_{v\in V^{(l)}}\frac{\partial }{\partial
\alpha }\mathbb{E}\big[\Delta m(X_{i},v) 
1(\Delta S_{i}^{\prime }\alpha \geq 0)\big]^{\prime }\big|_{\alpha =\beta
(v)} \sqrt{n}(\hat{\beta}(v)-\beta (v))+o_{p^{\ast
}}(1)=n^{-1/2}\sum_{j=1}^{n}\int_{0}^{1}\frac{\partial }{\partial \alpha }%
\mathbb{E}\big[\Delta m(X_{i},v) 
1(\Delta S_{i}^{\prime }\alpha \geq \varrho )\big]^{\prime }\big|_{\alpha
=a(v)}\phi _{j}(v)dv 
+\int_{0}^{1}\frac{\partial }{\partial a_{2}(v)}\mathbb{E}\big[\Delta
m(X_{i},v)1(\Delta S_{i}^{\prime }a(v)\geq \varrho )\big]dv\sqrt{n}(\varrho
_{n}-\varrho )+o_{p^{\ast }}(1)$ by Lemma~\ref{Lint}.

The same arguments yield the result in 2.\ by replacing $\Delta m$ with $%
\Delta q$.

\paragraph{Proof of Lemma~\protect\ref{Lint}:}

Let $\mathcal{V}(x)=\{v\in \mathcal{V}:\Delta q(x,v)>\varrho \}$. The
approximation error of Riemann sum is $\sup_{x\in \mathcal{X}}\big|%
l^{-1}\sum_{v\in V^{(l)}\cap \mathcal{V}(x)}f(x,v)-\int_{\mathcal{V}%
(x)}f(x,v)dv\big|=O\big(\sup_{x\in \mathcal{X}}l^{-1}\sum_{v_{j}\in V^{(l)}}%
\big(\sup_{v\in (v_{j-1},v_{j})}f(x,v)-\inf_{v\in (v_{j-1},v_{j})}f(x,v)\big)%
\big)
$\newline
$=O\big(\sup_{x\in \mathcal{X}} l^{-1}\sup_{P\in \mathcal{P}%
}\sum_{j=0}^{n_{P}}\big|f(x,v_{j})-f(x,v_{j-1})\big|\big)=O(l^{-1}),$ where
the set of all partitions ${\mathcal{P}}=\left\{ P=\{v_{0},\dots
,v_{n_{P}}\}\subset \mathcal{V}\right\} $.

\paragraph{Proof of Theorem~\protect\ref{TpiDRm}:}

Decompose $\hat \pi^{DR, K} - \pi^{DR, K} = \sum_{k=1}^K \hat \lambda_k \hat
\pi_k - \lambda_k \pi_k = \sum_{k=1}^K (\hat \lambda_k -\lambda_k) \pi_k +
\lambda_k(\hat \pi_k - \pi_k) + O_p((\hat \lambda_k
-\lambda_k)(\hat\pi_k-\pi_k)).$

Let $n_k = \sum_{i=1}^n D^k_i$. By the proof of Theorem~\ref{TpiDR}, $%
\sum_{k=1}^K \lambda_k(\hat \pi_k - \pi_k) = \sum_{k=1}^K \lambda_k
(n_k+n_{k-1})^{-1} 
\sum_{i=1}^n (D^k_i+D^{k-1}_i) R^k_i/B^k + o_p(n^{-1/2}) = n^{-1}
\sum_{i=1}^n \sum_{k=1}^K $\newline
$\lambda_k \frac{(D^k_i+D^{k-1}_i)R^k_i}{(p_k+p_{k-1})B^k} + o_p(n^{-1/2})$,
where $R^k_i = R^k_{1i} + R^k_{2i} + R_{3i}$, 
\begin{align}
R^k_{1i} =&\ \int_{0}^{1}\bigg(\mathbb{E}[ ( \partial
_{t}m_{1}(X,q_{1}(X,v))S_{1}-\partial
_{t}m_{0}(X,q_{0}(X,v))S_{0}-\pi^{DR}\Delta S)  \notag \\
&\text{sgn}(X,v)(D^k+D^{k-1})] +\frac{\partial }{\partial \alpha }\mathbb{E}%
\big[ \big( \Delta m(X,v)-\pi ^{DR}\Delta q(X,v)\big) \big(1(\Delta
S^{\prime }\alpha \geq \varrho )  \notag \\
& - 1(\Delta S^{\prime }\alpha \leq -\varrho ) \big) (D^k+D^{k-1})\big] \big|%
_{\alpha =a(v)}\bigg)^{\prime } \phi^k_{i}(v)dv/(p_k + p_{k-1}),  \notag \\
&\mbox{with}\ \phi^k_i(v) = \vartheta_k(v)^{-1}\big( 1(T_i \leq
S_{i}^{\prime }a_k(v)) - v \big) S_{i},  \notag \\
&\vartheta_k(v) = \mathbb{E}\left[f_{T|X,Z}(S^{\prime }a_k(v)|X,Z)SS^{\prime
}(D^k + D^{k-1})\right]/(p_k + p_{k-1}),  \notag \\
&S_{1i} = (1, X_i^{\prime }, 1, X_i^{\prime })^{\prime }, S_{0i} = (1,
X_i^{\prime }, 0, {\mathbf{0}}_{(d_x\times 1)}^{\prime})^{\prime}, \Delta
S_i = S_{1i} - S_{0i},  \notag
\end{align}
\begin{align}
R^k_{2i} =&\ \mathcal{D}_k^{\prime}G_k^{-1}\psi ^{J}(X_{i},T_{i},Z_{i})e_{i},
\notag \\
& \mbox{with}\ G_k = \mathbb{E}\left[e^2\psi^J(X,T,Z)\psi^J(X,T,Z)^\prime
(D^k+D^{k-1}) \right]/(p_k+p_{k-1}),  \notag \\
&\mathcal{D}_k=\int_{0}^{1}\mathbb{E}\big[(\psi ^{J}(X,q_{1}(X,v),1) - \psi
^{J}(X,q_{0}(X,v),0)) \text{sgn}(X,v)  \notag \\
& \ \ \ \ \ \ \ \times (D^k+D^{k-1})\big]dv/(p_k+p_{k-1}),  \notag \\
R_{3i}=&\ \int_{0}^{1}\left( \Delta m(X_{i},v)-\pi^{DR}\Delta
q(X_{i},v)\right) \text{sgn}(X_{i},v)dv,  \notag \\
B^k=&\ \int_{0}^{1}\mathbb{E}\left[|\Delta q(X,v)| 1(|\Delta q(X,v)| \geq
\varrho) (D^k+D^{k-1}) \right]dv/(p_k+p_{k-1}).  \label{AIFpim}
\end{align}

\bigskip

Next we analyze $\sum_{k=1}^K (\hat \lambda_k -\lambda_k) \pi_k$. Let $%
\mathsf{A}_k = Q_kP_k$, where $Q_k = q_k - q_{k-1}$ and $P_k = \sum_{l=k}^K
p_l(q_l - \mathbb{E}\left[T\right])$. Let $\lambda_k = \mathsf{A}_k/\mathsf{B%
}$, where $\mathsf{B} = \sum_{k=1}^K \mathsf{A}_k$. So $\pi^{DR, K} =
\sum_{k=1}^K \pi_k\mathsf{A}_k/\mathsf{B}$. Then $\sum_{k=1}^K (\hat
\lambda_k -\lambda_k) \pi_k = \sum_{k=1}^K \big\{ (\hat{\mathsf{A}}_k - 
\mathsf{A}_k)/\mathsf{B} - (\hat{\mathsf{B}} - \mathsf{B})\mathsf{A}_k/%
\mathsf{B}^2 \big\}\pi_k + o_p(n^{-1/2}) = \sum_{k=1}^K (\hat{\mathsf{A}}_k
- \mathsf{A}_k)(\pi_k - \pi^{DR, K})/\mathsf{B} + o_p(n^{-1/2}) $.

Decompose $\hat{\mathsf{A}}_k - \mathsf{\mathsf{A}}_k = (\hat Q_k - Q_k)P_k
+ (\hat P_k - P_k)Q_k + o_p(n^{-1/2})$. It is straightforward to show that $%
\hat q_k - q_k = n^{-1}\sum_{i=1}^n \big\{ (T_i D^k_i - \mathbb{E}\left[%
T_iD^k_i\right])/p_k - (D^k_i- p_k)q_k/p_k \big\} + o_p(n^{-1/2}) =
n^{-1}\sum_{i=1}^n (T_i - q_k) D^k_i/p_k + o_p(n^{-1/2})$.

$\hat P_k - P_k = \sum_{l=k}^K\Big\{
(\hat p_l - p_l)(q_l - \mathbb{E}\left[T\right]) + p_l\big(\hat q_l - q_l -
\bar T + \mathbb{E}\left[T\right]\big)
\Big\} + o_p(n^{-1/2}) = n^{-1}\sum_{i=1}^n \sum_{l=k}^K\Big\{
(D_{li} - p_l)(q_l - \mathbb{E}\left[T\right]) + p_l\big( (T_i -
q_l)D_{li}/p_l - T_i + \mathbb{E}\left[T\right]\big)\Big\}
+ o_p(n^{-1/2}) = n^{-1}\sum_{i=1}^n \sum_{l=k}^K \big(D_{li} - p_l\big)\big(%
T_i-\mathbb{E}\left[T\right] \big) - P_k + o_p(n^{-1/2}). $

Therefore $\sum_{k=1}^K (\hat \lambda_k -\lambda_k) \pi_k = \sum_{k=1}^K (%
\hat{\mathsf{A}}_k - \mathsf{A}_k)(\pi_k - \pi^{DR, K})/\mathsf{B} +
o_p(n^{-1/2}) = n^{-1}\sum_{i=1}^n \sum_{k=1}^K R_{4ki} + o_p(n^{-1/2})$,
where 
\begin{align}
R^k_{4i} =&\ \bigg\{ \left( (T_i - q_k) \frac{D^k_i}{p_k} - (T_i - q_{k-1}) 
\frac{D^{k-1}_i}{p_{k-1}} \right) \sum_{l=k}^K p_l(q_l - \mathbb{E}\left[T%
\right]) + (q_k-q_{k-1})  \notag \\
&\left(T_i-\mathbb{E}\left[T\right] \right) \sum_{l=k}^K \big(D_{li} - p_l%
\big) \bigg\}\frac{\pi_k - \pi^{DR, K}}{\sum_{k=1}^K (q_k - q_{k-1})
\sum_{l=k}^K p_l(q_l - \mathbb{E}\left[T\right]).}.  \label{IFm4}
\end{align}
By $R_i^k$ given in (\ref{AIFpim}) and $R_{4i}^k$ given in (\ref{IFm4}), we
obtain the influence function 
\begin{align}
R_{Ki} =\sum_{k=1}^K \lambda_k \frac{(D^k_i+D^{k-1}_i)R^k_i}{(p_k+p_{k-1})B^k%
} + R^k_{4i}.  \label{IFm}
\end{align}

Asymptotic normality follows the same arguments in the proof of Theorem~\ref%
{TpiDR} with the following modifications. The law of iterated expectations
yields $\sigma_{Kn}^2 = \sigma_{K1}^2+ \sigma_{K2n}^2+ \sigma_{K3}^2$, where 
$\sigma_{K1}^2 = \mathbb{E}\left[\left( \sum_{k=1}^K \lambda_k \frac{%
(D^k_i+D^{k-1}_i)R^k_{1i}}{(p_k+p_{k-1})B^k} + R^k_{4i} \right)^2\right]$, $%
\sigma_{K2n}^2 = \mathbb{E}\left[\left( \sum_{k=1}^K \lambda_k \frac{%
(D^k_i+D^{k-1}_i)R^k_{2i}}{(p_k+p_{k-1})B^k} \right)^2\right] $, and $%
\sigma_{K3}^2 = \mathbb{E}\left[\left( \sum_{k=1}^K \lambda_k \frac{%
(D^k_i+D^{k-1}_i)R_{3i}}{(p_k+p_{k-1})B^k} \right)^2\right]$.

\subsection{Variance Estimation}

\label{SecImplement}

For convenience, we first collect the relevant notations and then discuss
implementation details.

\subsubsection{Notation:}

Let $\phi_i(v) = \vartheta(v)^{-1}\big( 1(T_i \leq S_{i}^{\prime }a(v)) - v %
\big) S_{i}$. Let the trimming function $\chi ^{+}(x,v)=1(\Delta q(x,v)\geq
\varrho )$. Let $S_{1i}=(1,X_{i}^{\prime },1,X_{i}^{\prime })^{\prime }$, $%
S_{0i}=(1,X_{i}^{\prime },0,{\mathbf{0}}_{(d_{x}\times 1)}^{\prime
})^{\prime }$, $\Delta S_{i}=S_{1i}-S_{0i}$, $\partial _{t}m_{z}(X,q_{z})=%
\frac{\partial }{\partial t}m_{z}(X,t)|_{t=q_{z}(X,v)}$. 

\begin{align*}
R_{1i}^{+}& =\int_{0}^{1}\bigg(\mathbb{E}\left[ \left( \partial
_{t}m_{1}(X,q_{1}(X,v))S_{1}-\partial _{t}m_{0}(X,q_{0}(X,v))S_{0}-\pi
_{+}^{DR}\Delta S\right) \chi ^{+}(X,v)\right] \\
& \ \ \ +\frac{\partial }{\partial \alpha }\mathbb{E}\left[ \left( \Delta
m(X,v)-\pi _{+}^{DR}\Delta q(X,v)\right) 1(\Delta S^{\prime }\alpha \geq
\varrho )\right] \big|_{\alpha =a(v)}\bigg)^{\prime }\phi _{i}(v)dv, \\
R_{2i}^{+}& =\mathcal{D}^{+^{\prime }}G^{-1}\psi
^{J}(X_{i},T_{i},Z_{i})e_{i},\mbox{with}\ G\equiv \mathbb{E}\left[\psi
^{J}(X,T,Z)\psi ^{J}(X,T,Z)^{\prime }\right]=\mathbb{E}\left[\Psi ^{\prime
}\Psi /n\right], \\
&\ \ \ \mathcal{D}^{+}=\mathcal{D}_{1}^{+}-\mathcal{D}_{0}^{+},\mathcal{D}%
_{z}^{+}=\int_{0}^{1}\mathbb{E}\left[\psi ^{J}(X,q_{z}(X,v),z)\chi ^{+}(X,v)%
\right]dv, \\
R_{3i}^{+}& =\int_{0}^{1}\left( \Delta m(X_{i},v)-\pi _{+}^{DR}\Delta
q(X_{i},v)\right) \chi ^{+}(X_{i},v)dv, \\
B_{+}& =\int_{0}^{1}\int_{\mathcal{X}}\Delta q(x,v)\chi ^{+}(x,v)f(x)dxdv.
\end{align*}

Let $\chi^-(x, v) = 1(\Delta q(x,v) < -\varrho)$ and $B_- = \int_0^1\int_{%
\mathcal{X}} \Delta q(x,v)\chi^-(x, v) f(x)dxdv$. Let $B=B_+ - B_-$.

For $\pi^{DR}_-$, define $R^-_{i}$ as $R^+_i$ by replacing $+$ with $-$ in
all the components in $R^+_i$.

For $\pi ^{DR}$, define $R_{i}=R_{1i}+R_{2i}+R_{3i}$, where $%
R_{ki}=R_{ki}^{+}-R_{ki}^{-}$ for $k=1,2,3$ except that one needs to replace 
$\pi _{+}^{DR}$ and $\pi _{-}^{DR}$ with $\pi ^{DR}$ in $R_{ki}^{+}$ and $%
R_{ki}^{-}$, $k=1,3$. 

\subsubsection{Implementation}

We estimate $\sigma^2$ by the sample analogue plug-in estimator, i.e., $%
\hat\sigma^2 = \hat\sigma^2_1+ \hat\sigma^2_2 + \hat\sigma^2_3$, where $%
\hat\sigma^2_{k} = n^{-1}\sum_{i=1}^n \hat R_{ki}^2/\hat B^2$, $\hat B$ and $%
\hat R_{ki}$ are consistent estimators for $B$ and $R_{ki}$ for $k=1,2,3$,
respectively, given in (\ref{AIFpi}):

For $\hat R_{1i}$, $\partial_t\hat m_z$ is directly computed from Step 2.
From the linear quantile regression literature, it is standard $\hat{%
\vartheta}(v) = n^{-1}\sum_{i=1}^n \hat f_{T|X,Z}(S_i^{\prime }\hat
a(v)|X_i, Z_i) S_iS_i^{\prime }$. The derivative $\frac{\partial}{\partial
\alpha}{\mathbb{E}}\big[
\Delta m(X,v) 1(\Delta S_i^{\prime }\alpha \geq \varrho) \big]\big|_{\alpha=
a(v)}$ may be estimated by a numerical differentiation, i.e., $%
n^{-1}\sum_{i=1}^n \Delta \hat m(X_i,v) \big(1(\Delta S_i^{\prime }(\hat
a(v)+\iota/2) \geq \varrho_n) - 1(\Delta S_i^{\prime }(\hat a(v)-\iota/2)
\geq \varrho_n) \big)\big/\iota $ for some small $\iota > 0$.

For $\hat R_{2i}$, 
let $\hat e_i = Y_i - \psi^J(X_i, T_i, Z_i)^{\prime }\hat c$, $\hat \Omega =
n^{-1}\sum_{i=1}^n \hat e_i^2 \psi^J(X_i, T_i, Z_i)\psi^J(X_i, $\newline
$T_i, Z_i)^{\prime }$, $\hat G = \Psi^{\prime }\Psi/n$, and $\hat\mho = \hat
G^{-1}\hat\Omega \hat G^{-1}$. Let $\hat{\mathcal{D}}^+ = n^{-1}\sum_{i=1}^n
l^{-1}\sum_{v\in V^{(l)}}\hat\psi^J_i $\newline
$\hat\chi^+(X_i,v)$. Let $\hat{\mathcal{D}} = \hat{\mathcal{D}}^+ - \hat{%
\mathcal{D}}^-$. Then $\hat \sigma_{2n}^{2} = \hat{\mathcal{D}}^{\prime
}\hat\mho\hat{\mathcal{D}}$, $\hat \sigma_{+2}^{2} = \hat{\mathcal{D}^+}%
^{\prime }\hat\mho\hat{\mathcal{D}^+}$, and $\hat\sigma_{-2}^{2} = \hat{%
\mathcal{D}^-}^{\prime }\hat\mho\hat{\mathcal{D}^-}$.

$\hat R_{3i}^+ = l^{-1}\sum_{v\in V^{(l)}} \left(\Delta\hat m(X_i,v)
-\hat\pi^{DR}_+ \Delta \hat q(X_i, v)\right)\hat\chi^+(X_i,v)$, and $\hat
B_+ $ is analogous.

\medskip

Consider $\varrho =0$. In practice, one may choose $\varrho _{n}=1.96\times
\min_{v\in {V}^{(l)},\{X_{i}\}_{i=1,...,n}} $\newline
$se(\Delta \hat{T}(X_{i},v))/\log (n)$. This procedure includes
insignificant estimates of \newline
$\Delta \hat{T}(X_{i},v)$ (at the $5\%$ significance level).\footnote{%
Step 1 is $O_{p}(n^{-1/2})$, so the estimation error of $\chi $ is of first
order asymptotically by Lemma~\ref{Ltrim}. The rate condition on $\sqrt{n}%
(\varrho _{n}-\varrho )=o(1)$ means that using $\varrho _{n}$ rather than $%
\varrho $ is first-order asymptotically ignorable.}

\end{document}